\documentclass[a4paper,11pt]{article}
\pdfoutput=1 
\usepackage{jheppub} 
\usepackage{bm} 
\usepackage{array}
\usepackage{multirow} 

\usepackage[T1]{fontenc}

\usepackage[pdftex]{color}

\title{\boldmath 
Inclusive production of $J/\psi$, $\psi(2S)$, and $\Upsilon$ states in pNRQCD
}
\preprint{TUM-EFT 170/22}

\author[a,b,c]{Nora~Brambilla,} \author[a,d,e]{Hee~Sok~Chung,}
\author[a]{Antonio~Vairo,} \author[a]{and Xiang-Peng~Wang}

\affiliation[a]{Physik Department, Technische Universit\"at M\"unchen,\\
James-Franck-Strasse 1, 85748 Garching, Germany} 
\affiliation[b]{Institute for Advanced Study, 
Technische Universit\"at M\"unchen,\\ 
Lichtenbergstrasse 2 a, 85748 Garching, Germany} 
\affiliation[c]{Munich Data Science Institute, 
Technische Universit\"at M\"unchen, \\ 
Walther-von-Dyck-Strasse 10, 85748 Garching, Germany} 
\affiliation[d]{Excellence Cluster ORIGINS, 
Boltzmannstrasse 2, 85748 Garching, Germany} 
\affiliation[e]{Department of Physics, Korea University, Seoul 02841, Korea}

\emailAdd{nora.brambilla@tum.de} \emailAdd{heesok.chung@tum.de}
\emailAdd{antonio.vairo@tum.de} \emailAdd{xiangpeng.wang@tum.de}

\abstract{
Under some assumptions on the hierarchy of relevant energy scales,
we compute the nonrelativistic QCD (NRQCD) long-distance matrix elements 
(LDMEs) for inclusive production of $J/\psi$, $\psi(2S)$, and $\Upsilon$ states
based on the potential NRQCD (pNRQCD) effective field theory. 
Based on the pNRQCD formalism, we obtain expressions for the LDMEs in terms of
the quarkonium wavefunctions at the origin and universal gluonic correlators,
which do not depend on the heavy quark flavor or the radial excitation. 
This greatly reduces the number of nonperturbative unknowns and substantially
enhances the predictive power of the nonrelativistic effective field theory formalism. 
We obtain improved determinations of the LDMEs for $J/\psi$, $\psi(2S)$, and
$\Upsilon$ states thanks to the universality of the gluonic correlators, and
obtain phenomenological results for cross sections and polarizations at large
transverse momentum that agree well with measurements at the LHC. 
}

\begin{document} 
\maketitle
\flushbottom

\section{Introduction}
\label{sec:intro}

Understanding the production mechanism of heavy quarkonia remains a formidable
challenge in QCD phenomenology~\cite{Brambilla:2004wf, Brambilla:2010cs,
Bodwin:2013nua, Brambilla:2014jmp}. 
Investigation of many unexplored areas of QCD 
rely heavily on heavy quarkonium production rates, especially the $S$-wave
spin-triplet states including $J/\psi$, $\psi(2S)$, and $\Upsilon$. 
Much theoretical effort has been made in the nonrelativistic QCD (NRQCD)
effective field theory~\cite{Caswell:1985ui}, which provides a factorization
formalism where the production cross section is given by sums of products of
short-distance coefficients and long-distance matrix elements
(LDMEs)~\cite{Bodwin:1994jh}. 
NRQCD utilizes the separation of the scale of the heavy quark mass $m$ from the
scales of the momentum $mv$ and the energy $mv^2$ of the heavy quark $Q$ and
antiquark $\bar Q$, where $v$ is the velocity of $Q$ or $\bar Q$ inside the
quarkonium. 
The LDMEs encode the physics below the scale $m$ and 
correspond to the nonperturbative 
probability for a $Q$ and a $\bar Q$ in a nonrelativistic
state to evolve into a heavy quarkonium, while the short-distance coefficients
correspond to cross sections for production of a $Q \bar Q$. 
Because it has not been known how to compute an important class of LDMEs from
first principles, they have usually been determined phenomenologically by
comparing to cross section measurements. 
This approach has not led to a satisfactory description of the quarkonium
production mechanism: different analyses lead to inconsistent sets of LDMEs,
while none of the LDME determinations can give a comprehensive description of
important observables such as total and differential cross sections and
polarizations at different colliders~\cite{Chung:2018lyq}. 
This raises questions on the range of 
applicability of the NRQCD factorization approach and the validity of existing
LDME determinations. 

Naturally, a first-principles calculation of the LDMEs would substantially 
enhance our understanding of the quarkonium production mechanism and the 
predictive power of the NRQCD factorization formalism. 
It has long been known that color-singlet LDMEs, which correspond to the
probability for a $Q \bar Q$ in a color-singlet state to evolve into a
quarkonium, can be computed in lattice QCD or potential models, or obtained
phenomenologically from decay rates of heavy quarkonia. 
On the other hand, it has not been known how to compute color-octet LDMEs,
which often give rise to dominant contributions to heavy quarkonium production
rates. 

This unfortunate situation has recently been improved by analyses based on the
potential NRQCD (pNRQCD) effective field theory, which utilizes the separation
of scales $mv$ and $mv^2$~\cite{Pineda:1997bj, Brambilla:1999xf,
Brambilla:2004jw, Brambilla:2000gk, Pineda:2000sz, Brambilla:2001xy,
Brambilla:2002nu}. 
In this approach, color-octet LDMEs are given by
the product of the quarkonium wavefunction at the origin, which can be obtained
by solving a Schr\"odinger equation, times universal gluonic correlators, 
which are vacuum expectation values of gluonic
operators~\cite{Brambilla:2020ojz, Brambilla:2021abf}. 
The pNRQCD calculation of the LDMEs is valid for non-Coulombic, strongly 
coupled quarkonia, including charmonium and excited bottomonium states.
Not only the gluonic correlators are much more amenable to nonperturbative
determinations on the lattice than LDMEs themselves, but their universality 
reveals more symmetries and enhances the predictive power of the NRQCD 
factorization formalism. That is, even without the knowledge of the values of
the gluonic correlators, the phenomenological determination of the LDMEs
improves due to the universality of the gluonic correlators. 
This approach has been successfully applied to production of $\chi_c$ and
$\chi_b$ in refs.~\cite{Brambilla:2020ojz, Brambilla:2021abf}, 
and has recently been applied to production of $S$-wave heavy quarkonia, 
including $J/\psi$, $\psi(2S)$, and excited $\Upsilon$ states in 
ref.~\cite{Brambilla:2022rjd}. 
The analysis of $S$-wave quarkonia in ref.~\cite{Brambilla:2022rjd} led to
improved determinations of LDMEs, which, in turn, led to satisfactory descriptions of 
cross section and polarization of  $J/\psi$, $\psi(2S)$, $\Upsilon(2S)$, and 
$\Upsilon(3S)$ at the LHC. It is expected that the pNRQCD result for the
$S$-wave quarkonia will play a pivotal r\^{o}le in understanding the production
mechanism of $S$-wave heavy quarkonia. 

In this paper, we describe in detail the pNRQCD calculation of the LDMEs for
$S$-wave states that were first presented in ref.~\cite{Brambilla:2022rjd}. 
We also improve the phenomenological analysis and update the results for their cross section and polarization at the LHC,
and provide results for $J/\psi$ photoproduction, $\eta_c$ production, 
associated production of $J/\psi + Z$ and $J/\psi + W$, 
and our prediction for $J/\psi$
production at the Electron-Ion Collider. 

The paper is organized as follows. We introduce the NRQCD factorization
formalism and definitions of the LDMEs in section~\ref{sec:NRQCDfac}. 
Detailed calculations of $S$-wave LDMEs in pNRQCD are presented in
section~\ref{sec:LDMEs}, 
followed by phenomenological results in section~\ref{sec:pheno}.
We conclude in section~\ref{sec:summary}.

\section{NRQCD factorization formula}
\label{sec:NRQCDfac}

The inclusive production cross section of a $S$-wave heavy quarkonium 
${\cal Q}$ can be written in NRQCD in the form~\cite{Bodwin:1994jh}
\begin{equation}
\label{eq:fac_general}
\sigma_{{\cal Q}+X} = \sum_N \hat{\sigma}_{Q \bar{Q}(N)} 
\langle {\cal O}^{\cal Q} (N) \rangle, 
\end{equation}
where $\langle {\cal O}^{\cal Q} (N) \rangle$ is a NRQCD long-distance matrix
element (LDME) that corresponds to the probability for a $Q \bar Q$ pair in a
color and angular momentum state $N$ to produce a quarkonium ${\cal Q}+{\rm
anything}$, and $\hat{\sigma}_{Q \bar{Q}(N)}$ is the corresponding 
short-distance coefficient (SDC). 
The notation $\langle \cdots \rangle$ stands for the expectation value on the
QCD vacuum $|\Omega\rangle$. 
The operators ${\cal O}^{\cal Q} (N)$ have the schematic form 
\begin{equation}
\label{eq:operator_generic}
{\cal O}^{\cal Q} (N) = 
\chi^\dag {\cal K}_N \psi {\cal P}_{{\cal Q}
(\bm{P}=\bm{0})} \psi^\dag {\cal K}'_N \chi, 
\end{equation}
where $\psi$ and $\chi$ are Pauli spinor fields that annihilate and create a
heavy quark and antiquark, respectively, ${\cal K}_N$ and ${\cal K}'_N$ are
products of covariant derivatives, gluon field operators, 
and spin and color matrices. The projector ${\cal P}_{{\cal Q} (\bm{P})} = 
a^\dag_{{\cal Q} (\bm{P})} a_{{\cal Q} (\bm{P})}$ projects onto
states that include the quarkonium ${\cal Q}$ with three momentum $\bm{P}$. 
For polarization-summed cross sections, the projection operator
is summed over the polarizations of ${\cal Q}$. For polarized
cross sections, the projection operator only projects onto states that
contain ${\cal Q}$ with specific polarization. Throughout this paper, we take
the operators to be summed over all possible polarizations of the produced
quarkonium, unless the polarization of the quarkonium is specified. 
That is, we take ${\cal P}_{{\cal Q} (\bm{P})} = \sum_\lambda {\cal P}_{{\cal Q}(\lambda, \bm{P})}$,
where the sum is over all possible polarizations of ${\cal Q}$. 

If the $\chi^\dag {\cal K}_N \psi$ and $\psi^\dag {\cal K}'_N \chi$ 
on the right-hand side of eq.~(\ref{eq:operator_generic})
transform as color octets under SU(3), then 
$\langle {\cal O}^{\cal Q} (N) \rangle$ is a 
color-octet LDME; if $\chi^\dag {\cal K}_N \psi$ and $\psi^\dag {\cal K}'_N
\chi$ are color singlets, 
then $\langle {\cal O}^{\cal Q} (N) \rangle$ is a color-singlet LDME. 
By using the vacuum-saturation approximation, which is accurate up to
corrections of relative order $v^4$, a color-singlet LDME 
of the form 
$\langle \Omega | \chi^\dag {\cal K}_N \psi {\cal P}_{{\cal Q}
(\bm{P}=\bm{0})} \psi^\dag {\cal K}'_N \chi | \Omega \rangle $
can be related to its decay counterpart, which is the expectation value on the
quarkonium state at rest given by 
$\langle {\cal Q} | \psi^\dag {\cal K}'_N \chi \chi^\dag {\cal K}_N \psi | {\cal Q} \rangle$. 
The vacuum-saturation approximation does not apply for the color-octet LDMEs,
so unlike the color-singlet case, color-octet production LDMEs cannot be 
related to color-octet decay LDMEs. 

For the NRQCD factorization formula in eq.~(\ref{eq:fac_general}) to hold, 
the SDCs must be perturbatively calculable. That is, the infrared (IR) 
divergences that appear in perturbative QCD must either cancel or be absorbed 
into the LDMEs. Arguments for proof of NRQCD factorization 
have been given in an expansion in powers of 
$m/p_T$ to relative order $m^2/p_T^2$, where $m$ is the heavy quark mass and 
$p_T$ the transverse momentum of the quarkonium~\cite{Nayak:2005rt,
Nayak:2005rw, Nayak:2006fm, Kang:2014tta}. 
Hence, eq.~(\ref{eq:fac_general}) is expected to hold for values of $p_T$ much
larger than the quarkonium mass. 

Once a power counting is established, here we will assume the one in~\cite{Bodwin:1994jh}, 
the LDMEs have known scalings in $v$, and the sum in eq.~(\ref{eq:fac_general})
is organized in powers of $v$. In practice, the sum is truncated at a desired
accuracy in $v$.
For production of a $S$-wave spin-triplet ($^3S_1$) heavy quarkonium $V$, 
the following color-singlet LDME contributes at leading order in $v$:
\begin{equation}
\langle {\cal O}^{V} (^3S_1^{[1]}) \rangle
= \langle \Omega|\chi^\dag \sigma^i \psi 
{\cal P}_{V (\bm{P}=\bm{0})} 
\psi^\dag \sigma^i \chi|\Omega \rangle. 
\end{equation}
This corresponds to the probability for a color-singlet $Q \bar Q$ in a $^3S_1$
state to evolve into $V$. At relative order $v^2$, the following color-singlet
LDME appears:
\begin{equation}
\langle \Omega|\chi^\dag 
\left( -\frac{i}{2} \overleftrightarrow{\bm{D}} \right)^2 
\sigma^i \psi 
{\cal P}_{V (\bm{P}=\bm{0})} \psi^\dag 
\sigma^i \chi|\Omega \rangle 
+ {\rm c.c.},
\end{equation}
where ${\rm c.c.}$ stands for complex conjugation of the preceding terms, 
$\bm{D} = \bm{\nabla} -i g \bm{A}$ is the covariant derivative, 
$\bm{A}$ is the gluon field, and 
$\chi^\dag \overleftrightarrow{\bm{D}} \psi
= \chi^\dag \bm{D} \psi - (\bm{D} \chi)^\dag \psi$. 
To relative order $v^4$ accuracy, there are contributions from the 
color-octet LDMEs defined by 
\begin{subequations}
\begin{eqnarray}
\langle {\cal O}^{V} (^3S_1^{[8]}) \rangle &=& 
\langle \Omega|\chi^\dag \sigma^i T^a  \psi \Phi^{\dag ab}_\ell (0)
{\cal P}_{V (\bm{P}=\bm{0})} 
\Phi^{bc}_\ell (0) \psi^\dag \sigma^i T^c \chi|\Omega \rangle, \\
\langle {\cal O}^{V} (^1S_0^{[8]}) \rangle &=& 
\langle \Omega|\chi^\dag T^a  \psi \Phi^{\dag ab}_\ell (0)
{\cal P}_{V (\bm{P}=\bm{0})} 
\Phi^{bc}_\ell (0) \psi^\dag T^c \chi|\Omega \rangle, \\
\langle {\cal O}^{V} (^3P_0^{[8]}) \rangle &=& 
\frac{1}{3}
\langle \Omega|\chi^\dag 
\left( -\frac{i}{2} \overleftrightarrow{\bm{D}} \cdot \bm{\sigma} \right) 
T^a \psi \Phi^{\dag ab}_\ell (0)
\nonumber \\ && \hspace{5ex} \times 
{\cal P}_{V (\bm{P}=\bm{0})} 
\Phi^{bc}_\ell (0) \psi^\dag 
\left( -\frac{i}{2} \overleftrightarrow{\bm{D}} \cdot \bm{\sigma} \right) 
T^c \chi|\Omega \rangle, \\
\langle {\cal O}^{V} (^3P_1^{[8]}) \rangle &=& 
\frac{1}{2}
\langle \Omega|\chi^\dag 
\left( -\frac{i}{2} \overleftrightarrow{\bm{D}} \times \bm{\sigma} \right)^i
T^a \psi \Phi^{\dag ab}_\ell (0)
\nonumber \\ && \hspace{5ex} \times 
{\cal P}_{V (\bm{P}=\bm{0})} 
\Phi^{bc}_\ell (0) \psi^\dag 
\left( -\frac{i}{2} \overleftrightarrow{\bm{D}} \times \bm{\sigma} \right)^i
T^c \chi|\Omega \rangle, \\
\langle {\cal O}^{V} (^3P_2^{[8]}) \rangle &=& 
\langle \Omega|\chi^\dag 
\left( -\frac{i}{2} \overleftrightarrow{\bm{D}}^{(i} \bm{\sigma}^{j)} \right)
T^a \psi \Phi^{\dag ab}_\ell (0)
\nonumber \\ && \hspace{5ex} \times 
{\cal P}_{V (\bm{P}=\bm{0})} 
\Phi^{bc}_\ell (0) \psi^\dag 
\left( -\frac{i}{2} \overleftrightarrow{\bm{D}}^{(i} \bm{\sigma}^{j)} 
\right) T^c \chi|\Omega \rangle, 
\end{eqnarray}
\end{subequations}
where $\Phi_\ell(x) = P \exp \left[ -i g \int_0^\infty d \lambda \, 
\ell \cdot A^{\rm adj} (x+\ell \lambda) \right]$ is a Wilson line 
in the adjoint representation in the $\ell$ direction, 
and $T^{(ij)} = \frac{1}{2} (T^{ij}+T^{ji}) - \frac{1}{3} T^{ii}$ is the
symmetric traceless part of a rank two tensor. 
The gauge-completion Wilson line $\Phi_\ell(0)$ is necessary in order to ensure
the gauge invariance of color-octet LDMEs~\cite{Nayak:2005rt, Nayak:2005rw,
Nayak:2006fm}.
We note that there are also color-singlet LDMEs at relative orders $v^3$ and
$v^4$, which involve Pauli matrices, covariant derivatives, and gauge field
strengths between the quark and antiquark fields. 
They can be obtained from their decay counterparts, which are listed in 
refs.~\cite{Bodwin:2002cfe, Brambilla:2006ph, Brambilla:2008zg}. 

It has been known that, for inclusive production of a spin-triplet $S$-wave
quarkonium at large $p_T$, the SDCs for the color-octet channels are enhanced by powers of $\alpha_s$ compared to the
color-singlet channels~\cite{Braaten:1994vv, Cho:1995vh, Cho:1995ce}. 
This can be understood from the fact that, due to conservation of color and
angular momentum, an energetic gluon can produce a $Q \bar Q$ in a
color-singlet $^3S_1$ state from order $\alpha_s^3$, 
while a color-octet $Q \bar Q$ can be produced from order $\alpha_s$. 
Hence, in practice, the contributions from the color-octet channels can 
be larger than the color-singlet contribution appearing at leading order in $v$, even though
the color-octet LDMEs are suppressed by powers of $v$. 
Because of this, in phenomenological studies of inclusive production of heavy quarkonia,
the color-octet channels were customarily considered as leading order contributions,
together with the color-singlet contribution appearing at leading order in $v$. 
That is, the inclusive cross section of $V$ is written at leading order as
\begin{eqnarray}
\label{eq:fac_jpsi}
\sigma_{V+X} &=& 
\hat{\sigma}_{Q \bar{Q}(^3S_1^{[1]})} 
\langle {\cal O}^V(^3S_1^{[1]}) \rangle 
+ \hat{\sigma}_{Q \bar{Q}(^3S_1^{[8]})} 
\langle {\cal O}^V(^3S_1^{[8]}) \rangle 
\nonumber\\ && 
+ \hat{\sigma}_{Q \bar{Q}(^1S_0^{[8]})} 
\langle {\cal O}^V(^1S_0^{[8]}) \rangle 
+ \sum_{J=0,1,2} \hat{\sigma}_{Q \bar{Q}(^3P_J^{[8]})} 
\langle {\cal O}^V(^3P_J^{[8]}) \rangle. 
\end{eqnarray}
If we use the heavy-quark spin symmetry relations for the color-octet $^3P_J$ LDMEs given by 
$\langle \Omega | {\cal O}^V(^3P_J^{[8]}) | \Omega \rangle
= (2 J+1) \times \langle \Omega | {\cal O}^V(^3P_0^{[8]}) | \Omega \rangle$, 
which are accurate up to corrections of relative order $v^2$, 
then eq.~(\ref{eq:fac_jpsi}) describes at leading order the inclusive production of $V$ 
with four nonperturbative LDMEs. 
This {\it four-LDME phenomenology} has long been a standard for 
NRQCD-based description of spin-triplet $S$-wave heavy
quarkonia~\cite{Leibovich:1996pa, Beneke:1996yw, Beneke:1998re, Braaten:1999qk,
Chung:2009xr, Chung:2010iq,
Ma:2010jj, Butenschoen:2010rq, Ma:2010yw, Chao:2012iv, Butenschoen:2012px,
Gong:2012ug, Butenschoen:2011yh, Butenschoen:2009zy, Butenschoen:2011ks,
Butenschoen:2012qh, Bodwin:2014gia, 
Shao:2014yta, Bodwin:2015iua, Butenschoen:2012qr, Bodwin:2015yma,
Braaten:2000cm, 
Wang:2012is, Gong:2013qka, Han:2014kxa}. 

Since the SDCs can be computed perturbatively as series in $\alpha_s$, 
the determination of the four LDMEs in eq.~(\ref{eq:fac_jpsi}) directly leads 
to a description of $^3S_1$ heavy quarkonium cross sections. 
Because the color-singlet LDME can be related to its decay counterpart by using
the vacuum-saturation approximation, it can be obtained from quarkonium decay
rates, or can be evaluated using lattice QCD or potential
models~\cite{Bodwin:1994jh, Eichten:1995ch, Bodwin:1996tg, Bodwin:2001mk, 
Bodwin:2007fz}. 
On the other hand, since it has not been known how to compute color-octet LDMEs 
from first principles, they have usually been determined phenomenologically by
comparing eq.~(\ref{eq:fac_jpsi}) with cross section measurements. 
So far, this approach has not led to a satisfactory description of 
the production mechanism of $J/\psi$, $\psi(2S)$, and $\Upsilon$. 
One major problem is that, if we only employ the $p_T$-differential cross 
section measurements at $p_T$ much larger than the quarkonium mass, 
which are mainly available from hadron collider experiments, 
the phenomenological approach cannot strongly 
constrain all three color-octet LDMEs~\cite{Ma:2010jj, Bodwin:2014gia}. 
This happens because the $p_T$ shape of the cross 
section is in general given by a linear combination of leading power (LP) and
next-to-leading power (NLP) contributions, which behave like 
$d \sigma^{\textrm{LP}}/dp_T^2 \sim 1/p_T^4$ and 
$d \sigma^{\textrm{NLP}}/dp_T^2 \sim 1/p_T^6$,
respectively at the parton level~\cite{Kang:2014tta}. 
Hence in the hadroproduction-based phenomenological approach, only 
certain linear combinations of LDMEs are well determined. 
As we will see in the following sections, computation of the LDMEs in pNRQCD
leads to expressions involving quarkonium wavefunctions at the origin and
universal gluonic correlators, which reveal more symmetries and reduce the
number of nonperturbative unknowns. This leads to stronger constraints on the 
LDMEs compared to existing hadroproduction based approaches.

\section{\boldmath $S$-wave LDMEs in pNRQCD}
\label{sec:LDMEs} 

In this section, we compute the NRQCD LDMEs that appear in the NRQCD
factorization formula for inclusive production of $^3S_1$ quarkonium $V$ in
eq.~(\ref{eq:fac_jpsi}) by using the techniques developed in
refs.~\cite{Brambilla:2000gk, Brambilla:2002nu, Brambilla:2020ojz,
Brambilla:2021abf}. 
This lets us write a color-singlet or color-octet LDME in terms of heavy
quarkonium wavefunctions at the origin and its derivatives, times universal
coefficients that can be written in terms of vacuum expectation values of gluonic operators. 
The result follows from using quantum-mechanical perturbation theory (QMPT)
on the NRQCD Hamiltonian expanded in powers of $1/m$: 
\begin{equation}
H_{\rm NRQCD} = H_{\rm NRQCD}^{(0)} + \frac{1}{m} H_{\rm NRQCD}^{(1)} + \cdots,
\end{equation}
where 
\begin{eqnarray}
H_{\rm NRQCD}^{(0)} &=& \frac{1}{2} \int d^3x \left( \bm{E}^a \cdot \bm{E}^a + 
\bm{B}^a \cdot \bm{B}^a \right)
- \sum_{k=1}^{n_f} \int d^3x \, \bar{q}_k \, i \bm{D} \!\!\!\!/ \, q_k, 
\nonumber \\
H_{\rm NRQCD}^{(1)} &=& 
-\frac{1}{2} \int d^3x \, \psi^\dag \bm{D}^2 \psi 
- \frac{c_F}{2} \int d^3 x \, \psi^\dag \bm{\sigma} \cdot g \bm{B} \psi 
\nonumber \\ && 
+\frac{1}{2} \int d^3x \, \chi^\dag \bm{D}^2 \chi 
+ \frac{c_F}{2} \int d^3 x \, \chi^\dag \bm{\sigma} \cdot g \bm{B} \chi. 
\end{eqnarray}
Here $\bm{E}^a$ and $\bm{B}^a$ are the chromoelectric and chromomagnetic
fields, $q_k$ is the light quark field with flavor $k$, 
$c_F$ is a short-distance coefficient given in the $\overline{\rm MS}$ scheme
by $c_F = 1 + [ C_F + C_A (1+\log \Lambda/m)] \alpha_s/(2 \pi) +
O(\alpha_s^2)$~\cite{Eichten:1990vp, Czarnecki:1997dz, Grozin:2007fh}, 
where $C_A = N_c$, $C_F = (N_c^2-1)/(2 N_c)$, $N_c =3$ being the number of colors. 
The normalized eigenstates of $H_{\rm NRQCD}$ in the $Q \bar Q$ sector are 
labeled as $|\underline{\rm n}; \bm{x}_1, \bm{x}_2 \rangle$, where $\bm{x}_1$ 
and $\bm{x}_2$ are the positions of the heavy quark and antiquark, 
respectively. Here $\rm n=0$ is the ground state. 
The eigenstates have the expansion 
\begin{equation}
\label{eq:stateexpansion}
|\underline{\rm n} ; \bm{x}_1, \bm{x}_2 \rangle
=
|\underline{\rm n} ; \bm{x}_1, \bm{x}_2 \rangle^{(0)} 
+
\frac{1}{m}
|\underline{\rm n} ; \bm{x}_1, \bm{x}_2 \rangle^{(1)} 
+ \cdots, 
\end{equation}
where $|\underline{\rm n} ; \bm{x}_1, \bm{x}_2 \rangle^{(0)}$ is an eigenstate 
of $H_{\rm NRQCD}^{(0)}$ with eigenvalue $E_n^{(0)} (\bm{x}_1, \bm{x}_2)$. 
Expressions for $|\underline{\rm n} ; \bm{x}_1, \bm{x}_2 \rangle^{(1)}$ in
terms of $|\underline{\rm n} ; \bm{x}_1, \bm{x}_2 \rangle^{(0)}$ and 
$E_n^{(0)} (\bm{x}_1, \bm{x}_2)$ can be found in refs.~\cite{Brambilla:2000gk, Brambilla:2002nu}.
Since the scales that appear in NRQCD are $mv$, $\Lambda_{\rm QCD}$, and 
$mv^2$, the expansion in powers of $1/m$ in the calculation of the LDMEs 
corresponds to an expansion in powers of $v$ and $\Lambda_{\rm QCD}/m$. 

For a given NRQCD LDME $\langle \Omega | {\cal O}^{\cal Q}(N)|\Omega\rangle$, we have the following pNRQCD expression 
\begin{eqnarray}
\label{eq:masterformula}
\langle \Omega | {\cal O}^{\cal Q}(N)|\Omega\rangle &=& 
\frac{1}{\langle \bm{P}=\bm{0}|\bm{P}=\bm{0} \rangle} 
\int d^3x_1 d^3x_2 d^3x_1' d^3x_2' \, \phi_{\cal Q}^{(0)} (\bm{x}_1-\bm{x}_2) 
\\ && \times 
\left[ -V_{{\cal O}(N)} (\bm{x}_1, \bm{x}_2; \bm{\nabla}_1, \bm{\nabla}_2)
\delta^{(3)} (\bm{x}_1-\bm{x}_1') \delta^{(3)} (\bm{x}_2-\bm{x}_2') 
\right] \phi_{\cal Q}^{(0)\dagger} (\bm{x}_1'-\bm{x}_2') ,
\nonumber 
\end{eqnarray}
where $V_{{\cal O}(N)} (\bm{x}_1, \bm{x}_2; \bm{\nabla}_1, \bm{\nabla}_2)$ 
is the contact term given by 
\begin{eqnarray}
&& \hspace{-8ex} 
-V_{{\cal O}(N)} (\bm{x}_1, \bm{x}_2; \bm{\nabla}_1, \bm{\nabla}_2)
\delta^{(3)} (\bm{x}_1-\bm{x}_1') \delta^{(3)} (\bm{x}_2-\bm{x}_2')
\nonumber \\ &=& 
\sum_{n \in {\mathbb S}} \int d^3x \langle \Omega | \left( \chi^\dag {\cal K}_N
\psi \right) (\bm{x}) | \underline{\rm n}; \bm{x}_1, \bm{x}_2 \rangle
\langle \underline{\rm n}; \bm{x}_1' \bm{x}_2'| \left( \psi^\dag {\cal K}'_N
\chi \right) (\bm{x}) | \Omega \rangle, 
\end{eqnarray}
when the operator ${\cal O}^{\cal Q}(N)$ takes the form given in eq.~(\ref{eq:operator_generic}).
Here, the sum over $n$ is restricted to only include states in ${\mathbb S}$, 
which are made up of states where the $Q$ and $\bar Q$ are in a color-singlet state
in the static limit when located at the same point.
This restriction is necessary in order to have nonzero overlap with quarkonium states. 
Equation~(\ref{eq:masterformula}) is accurate up to corrections of relative
order $1/N_c^2$~\cite{Brambilla:2020ojz,Brambilla:2021abf}. 

In order to compute the contact term in the QMPT, we expand the states $| \underline{\rm n}; \bm{x}_1, \bm{x}_2 \rangle$ 
according to eq.~(\ref{eq:stateexpansion}), and make explicit the heavy quark
and antiquark content of the states $| \underline{\rm n}; \bm{x}_1, \bm{x}_2
\rangle^{(0)}$ by using 
\begin{equation}
\label{eq:QQexplicit}
| \underline{\rm n}; \bm{x}_1, \bm{x}_2 \rangle^{(0)}
= \psi^\dag (\bm{x}_1) \chi(\bm{x}_2) 
| n; \bm{x}_1, \bm{x}_2 \rangle^{(0)}, 
\end{equation}
where the states $| n; \bm{x}_1, \bm{x}_2 \rangle^{(0)}$ encode the gluonic
content of $| \underline{\rm n}; \bm{x}_1, \bm{x}_2 \rangle^{(0)}$. 
Then, the heavy quark and antiquark fields can be integrated out by using Wick theorem. 
Note that $| n; \bm{x}_1, \bm{x}_2 \rangle^{(0)}$ implicitly carries fundamental
SU(3) indices originating from the quark and antiquark fields. 
In the computation of contact term, we make use of the identities
($\bm{D}_c = \bm{\nabla}+ig\bm{A}^T$)
\begin{subequations}
\label{eq:Didentities}
\begin{eqnarray}
{}^{(0)} \langle n ; \bm{x}_1, \bm{x}_2| \bm{D}(\bm{x}_1) | n ; \bm{x}_1,
\bm{x}_2 \rangle ^{(0)} 
&=& \bm{\nabla}_1, 
\\
{}^{(0)} \langle n ; \bm{x}_1, \bm{x}_2| \bm{D}_c(\bm{x}_2) | n ; \bm{x}_1,
\bm{x}_2 \rangle ^{(0)} 
&=& \bm{\nabla}_2, 
\\
{}^{(0)} \langle n ; \bm{x}_1, \bm{x}_2| \bm{D}(\bm{x}_1) | k ; \bm{x}_1,
\bm{x}_2 \rangle ^{(0)} 
&=& 
\frac{{}^{(0)} \langle n ; \bm{x}_1, \bm{x}_2| g \bm{E}(\bm{x}_1) 
| k ; \bm{x}_1, \bm{x}_2 \rangle ^{(0)} }{E_n^{(0)} (\bm{x}_1,\bm{x}_2) -
E_k^{(0)} (\bm{x}_1,\bm{x}_2) } 
\\
{}^{(0)} \langle n ; \bm{x}_1, \bm{x}_2| \bm{D}_c(\bm{x}_2) | k ; \bm{x}_1,
\bm{x}_2 \rangle ^{(0)} 
&=& 
- \frac{{}^{(0)} \langle n ; \bm{x}_1, \bm{x}_2| g \bm{E}^T (\bm{x}_2) 
| k ; \bm{x}_1, \bm{x}_2 \rangle ^{(0)} }{E_n^{(0)} (\bm{x}_1,\bm{x}_2) -
E_k^{(0)} (\bm{x}_1,\bm{x}_2) }, 
\end{eqnarray}
\end{subequations}
which hold for $n \neq k$. 
We also use, for $N \geq 0$, 
\begin{eqnarray}
\label{eq:timeorderedidentity}
&& \hspace{-5ex} 
\frac{{}^{(0)} \langle n ; \bm{x}_1, \bm{x}_2| O(\bm{x}_1) 
| k ; \bm{x}_1, \bm{x}_2 \rangle ^{(0)} }{\left[ E_n^{(0)} (\bm{x}_1,\bm{x}_2) -
E_k^{(0)} (\bm{x}_1,\bm{x}_2) \right]^{N+1} } 
\nonumber \\ &=& 
\frac{(-i)^{N+1}}{N!}
\int_0^\infty dt \, t^{N} \, 
{}^{(0)} \langle n ; \bm{x}_1, \bm{x}_2| T 
\bigg\{
 O(t,\bm{x}_1) 
P \exp \left[ -i g \int_0^t dt' \, A_0(t',\bm{x}_1) \right] 
\nonumber\\ && \hspace{20ex} \times
\bar{P} \exp \left[ +i g \int_0^t dt' \, A_0(t',\bm{x}_2) \right] 
\bigg\}
| k ; \bm{x}_1, \bm{x}_2 \rangle ^{(0)},  
\end{eqnarray}
which holds for any gluonic operator $O(\bm{x}_1)$ acting on the heavy quark. 
Here, $A_0$ is the temporal gluon field in the fundamental representation, 
$P$ and $\bar{P}$ are path and anti path ordering for products of color
matrices, respectively, and $T$ is time ordering for products of field operators. 
An analogous identity holds for operators acting on the heavy antiquark. 
The expression for the case where the operators are anti time ordered
can be obtained by taking the complex conjugate.
Finally, the above equality can be extended to products of matrix elements of gluonic operators
  (see the following eq.~(\ref{eq:3S1EE})).

\subsection[$\langle {\cal O}^V (^3S_1^{[1]}) \rangle$] 
{\boldmath $\langle {\cal O}^V (^3S_1^{[1]}) \rangle$} 

We begin with the color-singlet LDME at leading order in $v$. The contact term
can be computed at leading order in the QMPT as 
\begin{eqnarray}
\label{eq:singletcomp}
&& \hspace{-8ex}
-V_{{\cal O}(^3S_1^{[1]})} (\bm{x}_1, \bm{x}_2; \bm{\nabla}_1, \bm{\nabla}_2)
\delta^{(3)} (\bm{x}_1-\bm{x}_1') \delta^{(3)} (\bm{x}_2-\bm{x}_2')
\nonumber \\ &=&
\sum_{n \in {\mathbb S}} \int d^3x \langle \Omega | \left( \chi^\dag \sigma^i
\psi \right) (\bm{x}) | \underline{\rm n}; \bm{x}_1, \bm{x}_2 \rangle^{(0)}\,
{}^{(0)}\langle \underline{\rm n}; \bm{x}_1', \bm{x}_2'| \left( \psi^\dag \sigma^i 
\chi \right) (\bm{x}) | \Omega \rangle
\nonumber \\ &=&
\sum_{n \in {\mathbb S}} \int d^3x \langle \Omega | \left( \chi^\dag \sigma^i
\psi \right) (\bm{x}) 
\psi^\dag(\bm{x}_1) 
\chi(\bm{x}_2) 
| n; \bm{x}_1, \bm{x}_2 \rangle^{(0)}
\nonumber \\ && \hspace{17ex} \times
{}^{(0)}\langle n; \bm{x}_1', \bm{x}_2'| 
\chi^\dag(\bm{x}_2') 
\psi(\bm{x}_1') 
\left( \psi^\dag \sigma^i 
\chi \right) (\bm{x}) | \Omega \rangle
\nonumber \\ &=&
\sum_{n} \langle \Omega | 
\sigma^i
| n; \bm{x}_1, \bm{x}_2 \rangle^{(0)}
{}^{(0)}\langle n; \bm{x}_1, \bm{x}_2| 
\sigma^i 
| \Omega \rangle
\delta^{(3)}(\bm{r}) 
\delta^{(3)}(\bm{x}_1-\bm{x}_1') 
\delta^{(3)}(\bm{x}_2-\bm{x}_2'), 
\end{eqnarray}
where $\bm{r} = \bm{x}_1-\bm{x}_2$. 
Here, we used eq.~(\ref{eq:QQexplicit}) in the second equality. 
We then used the Wick theorem to integrate out the heavy quark and antiquark
fields, integrated over $\bm{x}$, and 
lifted the restriction on the sum over $n$ because only the
states  $| n; \bm{x}_1, \bm{x}_2 \rangle^{(0)}$ 
with color-singlet SU(3) fundamental indices contribute to the sum. 
Then, by using the completeness relation for the 
$| n; \bm{x}_1, \bm{x}_2 \rangle^{(0)}$ states, 
we obtain the contact term 
for the color-singlet matrix element at leading order in the QMPT given by 
\begin{equation}
-V_{{\cal O}(^3S_1^{[1]})} 
= N_c \,\sigma^i \otimes \sigma^i \delta^{(3)}(\bm{r}). 
\end{equation}
The factor of $N_c$ comes from the trace over the  SU(3) fundamental
indices in the last line of eq.~(\ref{eq:singletcomp}).\footnote{
  Writing explicitly the color indices $i$, $j$ in the fundamental representation, the state
    $| n; \bm{x}_1, \bm{x}_2 \rangle^{(0)}$ in the last line of eq.~(\ref{eq:singletcomp}),
    and the following eqs.~(\ref{eq:pwave_calc3}), (\ref{eq:1s0calc3}) and (\ref{eq:3S1EE}),  
    has to be interpreted as
    $$
    | n; \bm{x}_1, \bm{x}_2;i, i \rangle^{(0)} = \delta_{ij}| n; \bm{x}_1, \bm{x}_2;i, j \rangle^{(0)}\,.
    $$
    The completeness relation reads
    $$
    \sum_n | n; \bm{x}_1, \bm{x}_2;i, j \rangle^{(0)} \,{}^{(0)}\langle n; \bm{x}_1, \bm{x}_2;i', j'| = \delta_{ii'}\delta_{jj'},
    $$
    and finally it holds that $\delta_{ij} \delta_{ii'}\delta_{jj'} \delta_{i'j'} = \delta_{ii} = N_c$.
}
The Pauli matrices on the left and right of the $\otimes$ symbol apply to the wavefunction on the left and right of the contact term in
eq.~(\ref{eq:masterformula}), respectively. 
Plugging the contact term into eq.~(\ref{eq:masterformula}) gives us the pNRQCD expression for the color-singlet LDME
\begin{equation}
\label{eq:singletresult}
\langle {\cal O}^V(^3S_1^{[1]}) \rangle  
= 3 \times 2 \, N_c |\phi_V^{(0)} (\bm{0})|^2, 
\end{equation}
where $\phi_V^{(0)} (\bm{x})$ is the wavefunction of the quarkonium $V$ at
leading order in $v$, and 
the factor $3$ comes from the sum over polarizations of $V$. 
Equation.~(\ref{eq:singletresult}) reproduces the known result obtained
in the vacuum-saturation approximation~\cite{Bodwin:1994jh}.

\subsection[$\langle {\cal O}^V (^3P_J^{[8]}) \rangle$] 
{\boldmath $\langle {\cal O}^V (^3P_J^{[8]}) \rangle$} 

We now proceed with computing the color-octet LDME $\langle {\cal O}^V
(^3P_J^{[8]}) \rangle$. 
At leading order in the QMPT, the contact term is given by 
\begin{eqnarray}
\label{eq:pwave_calc1}
&& \hspace{-8ex}
-V_{{\cal O}(^3P_J^{[8]})} (\bm{x}_1, \bm{x}_2; \bm{\nabla}_1, \bm{\nabla}_2)
\delta^{(3)} (\bm{x}_1-\bm{x}_1') \delta^{(3)} (\bm{x}_2-\bm{x}_2')
\nonumber \\&=& 
\sum_{n \in {\mathbb S}}
{\cal T}_{1J}^{ii'jj'}
\langle \Omega | \left[ \chi^\dag 
\left( - \frac{i}{2} \overleftrightarrow{\bm{D}}^i \sigma^{i'} \right) 
T^a \psi \right] (\bm{x})
\Phi_\ell^{\dag ab} (0, \bm{x})
| \underline{\rm n} ; \bm{x}_1, \bm{x}_2 \rangle^{(0)}
\nonumber \\ && \times 
{}^{(0)} \langle \underline{\rm n} ; \bm{x}_1', \bm{x}_2' |
\Phi_\ell^{bc} (0, \bm{x}) 
\left[ \psi^\dag 
\left( - \frac{i}{2} \overleftrightarrow{\bm{D}}^j \sigma^{j'} \right)
T^c \chi \right] (\bm{x}) | \Omega \rangle,
\end{eqnarray}
where 
\begin{eqnarray}
{\cal T}_{10}^{ii'jj'} &=& \frac{1}{3} \delta^{ii'} \delta^{jj'},
\\
{\cal T}_{11}^{ii'jj'} &=& \frac{1}{2} \epsilon_{kim} \epsilon_{kjn}
\delta^{mi'} \delta^{nj'},
\\
{\cal T}_{12}^{ii'jj'} &=& 
\left( \frac{\delta_{im} \delta^{ni'} + \delta_{in} \delta^{mi'}}{2}
- \frac{\delta_{mn}}{3} \delta^{ii'} \right) 
\left( \frac{\delta_{jm} \delta^{nj'} + \delta_{jn} \delta^{j'm}}{2}
- \frac{\delta_{mn}}{3} \delta^{jj'} \right). 
\end{eqnarray}
Note that $\sum_{J=0,1,2} T_{1J}^{ii'jj'} = \delta^{ij} \delta^{i'j'}$. 
The $\underline{\rm n}$-to-vacuum matrix element can be computed as 
\begin{eqnarray}
\label{eq:pwave_calc2}
&& \hspace{-5ex}
\langle \Omega | \left[ \chi^\dag
\left( - \frac{i}{2} \overleftrightarrow{\bm{D}}^i \sigma^{i'} \right)
T^a \psi \right] (\bm{x})
\Phi_\ell^{\dag ab} (0, \bm{x})
| \underline{\rm n} ; \bm{x}_1, \bm{x}_2 \rangle^{(0)}
\\ 
&=& 
- \frac{i}{2} \sum_{k \neq n}
\langle \Omega | 
T^a \Phi_\ell^{\dag ab} (0, \bm{x}_1) 
|k \rangle^{(0)} {}^{(0)} \langle k |  
\frac{(g \bm{E}_1 + g \bm{E}_{2}^T )^i \sigma^{i'} } {E_k^{(0)}-E_n^{(0)}} 
| n \rangle^{(0)}
\delta^{(3)} (\bm{x}-\bm{x}_1) \delta^{(3)} (\bm{x}-\bm{x}_2) ,
\nonumber 
\end{eqnarray}
where we used Wick contraction to integrate out the heavy quark fields, 
and used the identities in eqs.~(\ref{eq:Didentities}) to compute the 
matrix elements of $\bm{D}$ in terms of chromoelectric fields. 
Here $\bm{E}_1 = \bm{E}(\bm{x}_1)$ and $\bm{E}_2^T =\bm{E}^T(\bm{x}_2)$. 
We suppressed the quark and antiquark positions in 
$E_n^{(0)} (\bm{x}_1, \bm{x}_2)$, $E_k^{(0)} (\bm{x}_1, \bm{x}_2)$, 
and the states 
$|n; \bm{x}_1, \bm{x}_2 \rangle^{(0)}$ and 
$|k; \bm{x}_1, \bm{x}_2 \rangle^{(0)}$, because they are all computed at the
same positions. 
By using the complex conjugate of the identity given in
eq.~(\ref{eq:timeorderedidentity}), we can write the matrix element in
the last line of eq.~(\ref{eq:pwave_calc2}) as 
\begin{eqnarray}
\label{eq:pwave_calc3}
&& \hspace{-5ex}
 \sum_{k \neq n}
\langle \Omega |
T^a \Phi_\ell^{\dag ab} (0, \bm{x}_1)
|k \rangle^{(0)} {}^{(0)} \langle k |
\frac{g \bm{E}_1 } {E_k^{(0)}-E_n^{(0)}}
| n \rangle^{(0)}
\nonumber \\
&&= - \frac{i}{2 N_c}  \int_0^\infty dt \,
\langle \Omega | 
\Phi_\ell^{\dag ab} (0, \bm{x}_1) 
\Phi_0^{\dag ad} (0,\bm{x}_1;t,\bm{x}_1)
g \bm{E}^d (t,\bm{x}_1) 
| n \rangle^{(0)},
\end{eqnarray}
where 
$\Phi_0(t,\bm{x}_1;t',\bm{x}_1) = P \exp \left[ -i g \int_t^{t'} d \tau \,
A_0^{\rm adj} (\tau, \bm{x}_1) \right]$ is the Schwinger line in the adjoint representation. 
The Schwinger line in the adjoint representation is obtained by combining the path ordered and anti-path
ordered Wilson lines in the fundamental representation that appear in
eq.~(\ref{eq:timeorderedidentity}) with the color matrices on the left-hand
side of eq.~(\ref{eq:pwave_calc3}).
In the case of eq.~(\ref{eq:pwave_calc3}), the expression involving the
Schwinger line on the right-hand side can be easily verified by using the
temporal gauge ($A_0 = 0$), and requiring gauge invariance to obtain
a general expression. 
Note that the operators in eq.~(\ref{eq:pwave_calc3}) are anti time ordered. 
The $\bm{E}_2^T$ term yields the same result, with $\bm{x}_1$ replaced by $\bm{x}_2$. 

The vacuum-to-$\underline{\rm n}$ matrix element in 
eq.~(\ref{eq:pwave_calc1}) can be computed
in the same way. Plugging in the result in eq.~(\ref{eq:pwave_calc3}) to 
into eq.~(\ref{eq:pwave_calc1}) we find 
\begin{equation}
\label{eq:pwave_calc4}
-V_{{\cal O}(^3P_J^{[8]})} 
=
T_{1J}^{ij} 
\delta^{(3)} (\bm{r}) 
\frac{1}{4 N_c} {\cal E}_{00}^{ij}, 
\end{equation}
where $T_{1J}^{ij}={\cal T}_{1J}^{ii'jj'} \sigma^{i'} \otimes \sigma^{j'}$ and 
\begin{eqnarray}
\label{eq:E00tensordef}
{\cal E}_{00}^{ij}
&=&
\int_0^\infty dt \int_0^\infty dt'
\langle \Omega | \Phi_\ell^{\dag ab} (0)
\Phi_0^{\dag ad} (0;t) g E^{d,i} (t)
g E^{e,j} (t') \Phi_0^{ec} (0;t') \Phi_\ell^{bc} (0) | \Omega \rangle.
\end{eqnarray}
This result leads to the LDMEs 
\begin{equation}
\label{eq:3pj_result}
\langle {\cal O}^V(^3P_J^{[8]}) \rangle 
=
3 \times \frac{2 J+1}{18 N_c} \, {\cal E}_{00} \, |\phi_V^{(0)} (\bm{0})|^2 ,
\end{equation}
where ${\cal E}_{00} = {\cal E}_{00}^{ij} \delta^{ij}$. 
This result is valid at leading order in $v$, up to corrections of order
$1/N_c^2$.

\subsection[$\langle {\cal O}^V (^1S_0^{[8]}) \rangle$] 
{\boldmath $\langle {\cal O}^V (^1S_0^{[8]}) \rangle$} 

Now we consider the color-octet $^1S_0$ LDME. 
The contact term is given by 
\begin{eqnarray}
\label{eq:1s0calc1}
&&
\hspace{-18ex}
- V_{{\cal O}(^1S_0^{[8]})} (\bm{x}_1, \bm{x}_2; \bm{\nabla}_1, \bm{\nabla}_2)
\delta^{(3)} (\bm{x}_1-\bm{x}_1') \delta^{(3)} (\bm{x}_2-\bm{x}_2')
\nonumber \\&=& 
\int d^3x 
\sum_{n \in {\mathbb S}} 
\langle \Omega | \left( \chi^\dag T^a \psi \right) (\bm{x}) 
\Phi_\ell^{\dag ab} (0, \bm{x})
| \underline{\rm n} ; \bm{x}_1, \bm{x}_2 \rangle
\nonumber \\ && \hspace{10ex}  \times 
\langle \underline{\rm n} ; \bm{x}_1', \bm{x}_2' | 
\Phi_\ell^{bc} (0, \bm{x}) \left( \psi^\dag T^c \chi \right) (\bm{x}) 
| \Omega \rangle. 
\end{eqnarray}
The contribution at leading order in the QMPT is given by replacing 
the $| \underline{\rm n} ; \bm{x}_1, \bm{x}_2 \rangle$ and 
$\langle \underline{\rm n} ; \bm{x}_1', \bm{x}_2' |$ by 
$| \underline{\rm n} ; \bm{x}_1, \bm{x}_2 \rangle^{(0)}$ and 
${}^{(0)}\langle \underline{\rm n} ; \bm{x}_1', \bm{x}_2' |$, respectively.
This contribution vanishes because both the vacuum-to-$\underline{\rm n}$ and 
$\underline{\rm n}$-to-vacuum matrix elements are proportional to the trace of 
a color matrix. Hence, the nonvanishing contribution to the contact term comes
from the order-$1/m$ correction to the state 
$| \underline{\rm n} ; \bm{x}_1, \bm{x}_2 \rangle$.
Since the operator ${\cal O}^V (^1S_0^{[8]})$ does not contain Pauli matrices,
contributions that do not vanish when applied to the $^3S_1$ state 
can only come from the spin-flip term in  $| \underline{\rm n} ; \bm{x}_1, \bm{x}_2 \rangle^{(0)}$ given by
\begin{equation}
\label{eq:spinflip} 
|\underline{\rm n} \rangle^{(1)}_{\rm spin-flip} 
= \frac{1}{2} c_F \sum_{k \neq n} |\underline{\rm k} \rangle^{(0)}
\frac{{}^{(0)} \langle k | \bm{\sigma}_1 \cdot g \bm{B}_1
+ \bm{\sigma}_2^T \cdot g \bm{B}_2^T 
| n \rangle^{(0)} }
{E_k^{(0)} - E_n^{(0)} },
\end{equation}
where  $\bm{B}_1 = \bm{B}(\bm{x}_1)$ and $\bm{B}_2^T =\bm{B}^T(\bm{x}_2)$.
The Pauli matrix $\bm{\sigma}_2^T$ comes from $\bm{\sigma}$ acting on the $\chi$ field\footnote{
The sign of the $\bm{\sigma}_2$ term differs from
ref.~\cite{Brambilla:2002nu} because in ref.~\cite{Brambilla:2002nu},
$\bm{\sigma}_2$ acts on the charge conjugated field $\chi_c$,
and $\bm{\sigma}^T = - C \bm{\sigma} C^{-1}$.}, 
while $\bm{\sigma}_1$ comes from $\bm{\sigma}$ acting on the $\psi$ field. 
Plugging this into the $\underline{\rm n}$-to-vacuum matrix element in eq.~(\ref{eq:1s0calc1}) we find 
\begin{eqnarray}
\label{eq:1s0calc2}
&& \hspace{-5ex} 
\frac{1}{2} c_F \sum_{k \neq n} 
\langle \Omega | \left( \chi^\dag T^a \psi \right) (\bm{x})
\Phi_\ell^{\dag ab} (0, \bm{x}) |\underline{\rm k} \rangle^{(0)} 
\frac{{}^{(0)} \langle k | \bm{\sigma}_1 \cdot g \bm{B}_1 | n \rangle^{(0)} }
{E_k^{(0)} - E_n^{(0)} }
\nonumber \\
&=& 
\delta^{(3)} (\bm{x}-\bm{x}_1) 
\delta^{(3)} (\bm{x}-\bm{x}_2) 
\frac{\bm{\sigma}^i}{2} c_F \sum_{k \neq n} 
\langle \Omega | 
T^a 
\Phi_\ell^{\dag ab} (0, \bm{x}_1) 
|k \rangle^{(0)} 
\frac{{}^{(0)} \langle k | g B_1^i | n \rangle^{(0)} }
{E_k^{(0)} - E_n^{(0)} }. 
\end{eqnarray}
The Pauli matrix $\bm{\sigma}$ acts on the $Q \bar Q$ wavefunction. 
Similarly to the calculation of the color-octet $^3P_J$ LDME, we can rewrite this matrix element as 
\begin{eqnarray}
\label{eq:1s0calc3}
&& \hspace{-5ex}
\sum_{k \neq n} \langle \Omega | T^a \Phi_\ell^{\dag ab} (0, \bm{x}_1)
|k \rangle^{(0)} \frac{{}^{(0)} \langle k | g
\bm{B}_1 | n \rangle^{(0)} } {E_k^{(0)} - E_n^{(0)} }
\nonumber \\ &=& 
- \frac{i}{2 N_c}
\int_0^\infty dt \, \langle \Omega | \Phi_\ell^{\dag ab} (0, \bm{x}_1) 
\Phi_0^{\dag ad} (0,\bm{x}_1;t,\bm{x}_1) g \bm{B}^d (t,\bm{x}) 
| n \rangle^{(0)}. 
\end{eqnarray}
The $\bm{B}_2^T$ term yields the same result, with $\bm{x}_1$ replaced by $\bm{x}_2$.
From this we find the result for the contact term at leading nonvanishing order in QMPT given by 
\begin{equation}
\label{eq:1s0calc4}
- \left. V_{{\cal O}(^1S_0^{[8]})} \right|_{^3S_1} 
= \frac{ {\sigma}^i \otimes {\sigma}^j }{4 N_c m^2}
\delta^{(3)} (\bm{r}) c_F^2 {\cal B}_{00}^{ij}, 
\end{equation}
where we neglect any contribution to the contact term that vanishes when applied to the wavefunction in the $^3S_1$ state. 
The tensor ${\cal B}_{00}^{ij}$ is defined by 
\begin{equation}
\label{eq:B00tensordef}
{\cal B}_{00}^{ij} = 
\int_0^\infty dt \int_0^\infty dt' \langle \Omega | \Phi_\ell^{\dag ab} (0) 
\Phi_0^{\dag ad} (0;t) g B^{d,i} (t) 
g B^{e,j} (t') \Phi_0^{ec} (0;t') \Phi_\ell^{bc} (0) | \Omega \rangle. 
\end{equation}
This gives the following result for the color-octet $^1S_0$ LDME
\begin{equation}
\label{eq:1s0result}
\langle {\cal O}^V (^1S_0^{[8]}) \rangle
= 3 \times \frac{1}{6 N_c m^2} c_F^2 {\cal B}_{00} |\phi_V^{(0)}(\bm{0})|^2,
\end{equation}
where ${\cal B}_{00} = \delta^{ij} {\cal B}_{00}^{ij}$. 
This result is valid at leading order in $v$, up to corrections of order
$1/N_c^2$.

\subsection[$\langle {\cal O}^V (^3S_1^{[8]}) \rangle$] 
{\boldmath $\langle {\cal O}^V (^3S_1^{[8]}) \rangle$} 

The color-octet $^3S_1$ LDME is the last remaining one to compute at leading order.
The contact term is given by 
\begin{eqnarray}
\label{eq:3s1calc1}
&& \hspace{-8ex}
-V_{{\cal O}(^3S_1^{[8]})} (\bm{x}_1, \bm{x}_2; \bm{\nabla}_1, \bm{\nabla}_2)
\delta^{(3)} (\bm{x}_1-\bm{x}_1') \delta^{(3)} (\bm{x}_2-\bm{x}_2')
\nonumber \\&=&
\int d^3x
\sum_{n \in {\mathbb S}}
\langle \Omega | \left( \chi^\dag \sigma^i T^a \psi \right) (\bm{x})
\Phi_\ell^{\dag ab} (0, \bm{x})
| \underline{\rm n} ; \bm{x}_1, \bm{x}_2 \rangle
\nonumber \\ && \hspace{5ex}  \times
\langle \underline{\rm n} ; \bm{x}_1', \bm{x}_2' |
\Phi_\ell^{bc} (0, \bm{x}) \left( \psi^\dag \sigma^i T^c \chi \right) (\bm{x})
| \Omega \rangle.
\end{eqnarray}
Again, the contribution at leading order in the QMPT vanishes, 
because both the vacuum-to-$\underline{\rm n}$ and 
$\underline{\rm n}$-to-vacuum matrix elements are proportional to the trace of 
a color matrix. Hence, similarly to the color-octet $^1S_0$ LDME, the leading
nonvanishing contribution comes from the order-$1/m$ correction to the state 
$|\underline{\rm n}; \bm{x}_1, \bm{x}_2 \rangle$. 
Since the operator ${\cal O}^V (^3S_1^{[8]})$ already contains Pauli matrices,
only the spin-independent terms give nonvanishing contributions when applied to
the $^3S_1$ state. That is, we only keep the terms in 
$|\underline{\rm n}; \bm{x}_1, \bm{x}_2 \rangle^{(1)}$ given by 
\begin{eqnarray}
\label{eq:3s1calc2}
&& \left. | \underline{\rm n} \rangle^{(1)}\right|_{^3S_1} =
- \sum_{k \neq n} \left[ - \frac{1}{2} \frac{{}^{(0)} \langle k | [\bm{D}_1\cdot, g \bm{E}_1 ] | n \rangle^{(0)}}{(E_n^{(0)}-E_k^{(0)})^2}
+ \sum_{j \neq n}
\frac{
{}^{(0)}\langle k|g \bm{E}_1|j\rangle^{(0)} \cdot
{}^{(0)}\langle j|g \bm{E}_1|n\rangle^{(0)} }
{(E_n^{(0)}-E_k^{(0)})^2 (E_n^{(0)}-E_j^{(0)}) } \right.
\nonumber \\ && 
\left. + 2 (\bm{\nabla}_1 E_n^{(0)} ) \cdot
\frac{ {}^{(0)} \langle k | g \bm{E}_1| n \rangle^{(0)}}
{(E_n^{(0)}-E_k^{(0)})^3}
\right] |\underline{\rm k} \rangle^{(0)}
                + \left[ \bm{\nabla}_1 \to \bm{\nabla}_2,\,
                \bm{D} \to \bm{D}_{2\,c},\, g\bm{E}_1 \to - g \bm{E}_2^T \right]. 
\end{eqnarray}
In eq.~(\ref{eq:3s1calc2}) we also neglected terms that give rise to a derivative acting on the wavefunction,
because the first derivative of an $S$-wave wavefunction vanishes at the origin. 
Since we are computing local operator matrix elements, we only need to compute $\bm{\nabla}_1 E_n^{(0)}$ at $\bm{x}_1 = \bm{x}_2$, 
which is scaleless and vanishes in dimensional regularization. 
Finally, the contribution from $[\bm{D}_1 \cdot, g \bm{E}_1 ] = (\bm{D}_1 \cdot g \bm{E}_1)^a T^a$ can be eliminated by using the Gauss law,
which states that on physical states we can replace at leading order in QMPT $(\bm{D} \cdot \bm{E})^a$  with
  $g \left(\psi^\dag T^a \psi+\chi^\dag T^a \chi + \sum_j \bar{q}_j \gamma^0 T^a q_j \right)$.
  The heavy quark terms $g\psi^\dag T^a \psi$ and $g\chi^\dag T^a \chi$ lead to matrix elements
  proportional to ${}^{(0)} \langle k | T^aT^a | n \rangle^{(0)} = C_F N_c {}^{(0)} \langle k | n \rangle^{(0)}$
  when evaluated at  $\bm{x}_1 = \bm{x}_2$. They vanish since the states $|k \rangle^{(0)}$ is orthogonal to $|n\rangle^{(0)}$ by definition.
  The light quark term  $\sum_j \bar{q}_j \gamma^0 T^a q_j$ originating from $[\bm{D}_1 \cdot, g \bm{E}_1]$ cancels against the one
  originating from  $[\bm{D}_{2\,c} \cdot, g \bm{E}_2]$. 

The non-vanishing terms in eq.~(\ref{eq:3s1calc2}) give the following contribution to the 
$\underline{\rm n}$-to-vacuum matrix element 
\begin{eqnarray}
&& \hspace{-5ex} 
- \bm{\sigma} \sum_{k \neq n} 
\sum_{j \neq n}
\langle \Omega | T^a 
\Phi_\ell^{\dag ad} (0,\bm{x}_1) | k \rangle^{(0)} 
\frac{
{}^{(0)}\langle k| g \bm{E}_1 |j\rangle^{(0)} \cdot
{}^{(0)}\langle j|g \bm{E}_1|n\rangle^{(0)} }
{(E_n^{(0)}-E_k^{(0)})^2 (E_n^{(0)}-E_j^{(0)}) }
+ \left[ g \bm{E}_1 \to - g \bm{E}_2^T \right]_{\bm{x}_2 = \bm{x}_1}
\nonumber \\ &=&
\bm{\sigma} \frac{i}{2 N_c} d^{a'bc'}
\int_0^\infty dt_1\, t_1  \int_{t_1}^\infty dt_2
\langle \Omega |
\Phi_\ell^{\dag ad} (0,\bm{x}_1)
\Phi_0^{\dag a'a} (0,\bm{x}_1;t_1,\bm{x}_1) 
g E^{b,i} (t_1,\bm{x}_1)
\nonumber \\ && \hspace{30ex} \times 
\Phi_0^{\dag cc'} (t_1,\bm{x}_1;t_2,\bm{x}_1) 
g E^{c,i} (t_2,\bm{x}_1)
|n\rangle^{(0)},
\label{eq:3S1EE}                
\end{eqnarray}
where $d^{a'bc'} = 2 \, {\rm tr} ( \{ T^{a'}, T^b \} T^{c'} )$ 
comes from the trace of three color matrices; the contribution proportional to 
$f^{abc}$ cancels between the $g\bm{E}_1$ and $-g\bm{E}_2^T$ terms. From this we obtain 
\begin{eqnarray}
\left. -V_{{\cal O} (^3S_1^{[8]})} \right|_{^3S_1}
&=&
\sigma^i \otimes \sigma^i\frac{1}{4 N_c m^2} \delta^{(3)}(\bm{r}) {\cal E}_{10;10}, 
\end{eqnarray}
where we have neglected the contributions that vanish when applied to
wavefunctions in the $^3S_1$ state. The ${\cal E}_{10;10}$ is defined by 
\begin{eqnarray}
\label{eq:E1010def}
{\cal E}_{10;10} 
&=& d^{a' b c'} d^{e' x y'}
\int_0^\infty dt_1 \, t_1 \int_{t_1}^\infty dt_2
\langle \Omega |
\Phi_\ell^{\dag ad} (0) \Phi_0^{a'a \dag} (0;t_1)
g E^{b,i} (t_1) \Phi_0^{c c' \dag} (t_1;t_2) g E^{c,i} (t_2)
\nonumber \\ && 
\times
\int_0^\infty dt'_1 \, t'_1 \int_{t'_1}^\infty dt'_2  \,
g E^{y,j} (t_2') \Phi_0^{y y'} (t'_1;t_2') g E^{x,j} (t_1')
\Phi_0^{e' e} (0;t_1') \Phi_\ell^{de} (0)
|\Omega\rangle.  
\end{eqnarray}
This leads to the following result for the color-octet $^3S_1$ LDME given by 
\begin{equation}
\label{eq:3s1result}
\langle {\cal O}^V (^3S_1^{[8]}) \rangle 
=
3\times \frac{1}{2 N_c m^2} \, {\cal E}_{10;10} \, |\phi_V^{(0)} (\bm{0})|^2, 
\end{equation}
which is valid at leading order in $v$, up to corrections of order
$1/N_c^2$.

\subsection{Heavy quark spin symmetry} 
\label{sec:HQSS}

Since our calculations of the LDMEs are valid at leading nonvanishing orders 
in $v$, they follow the heavy-quark spin symmetry relations, which are valid 
up to corrections of order $v^2$. 
As we have already seen in the calculation of the color-octet $^3P_J$ LDMEs, 
our results reproduce the relations 
$\langle {\cal O}^V(^3P_1^{[8]}) \rangle
= 3 \times \langle {\cal O}^V(^3P_0^{[8]}) \rangle$
and $\langle {\cal O}^V(^3P_2^{[8]}) \rangle
= 5 \times \langle {\cal O}^V(^3P_0^{[8]}) \rangle$.

Heavy quark spin symmetry also gives rise to relations between LDMEs for the
$^3S_1$ state and LDMEs for the $^1S_0$ state. 
For example, the color-singlet LDME 
\begin{equation}
\langle {\cal O}^{P} (^1S_0^{[1]}) \rangle = 
\langle \Omega|\chi^\dag \psi 
{\cal P}_{P (\bm{P}=\bm{0})}
\psi^\dag \chi|\Omega \rangle
\end{equation}
for a $^1S_0$ quarkonium $P$ can be computed in the same way as 
$\langle {\cal O}^{V} (^3S_1^{[1]}) \rangle$. 
The contact term for this LDME is 
\begin{equation}
-V_{{\cal O}(^1S_0^{[1]})}
= N_c \delta^{(3)} (\bm{r}), 
\end{equation}
which gives the LDME 
\begin{equation}
\langle {\cal O}^{P} (^1S_0^{[1]}) \rangle = 
2 N_c |\phi_P^{(0)} (\bm{0})|^2. 
\end{equation}
Since $\phi_P^{(0)}(\bm{r}) = \phi_V^{(0)}(\bm{r})$ at leading order in $v$, 
this result reproduces the heavy-quark spin symmetry relation 
$ \langle {\cal O}^{P} (^1S_0^{[1]}) \rangle
= 1/3 \times \langle {\cal O}^{V} (^3S_1^{[1]}) \rangle$. 

Similarly, the color-octet LDME 
\begin{equation}
\langle {\cal O}^{P} (^1P_1^{[8]}) \rangle =
\langle \Omega|\chi^\dag
\left( -\frac{i}{2} \overleftrightarrow{\bm{D}}^i \right)
T^a \psi \Phi^{\dag ab}_\ell (0)
{\cal P}_{P (\bm{P}=\bm{0})}
\Phi^{bc}_\ell (0) \psi^\dag
\left( -\frac{i}{2} \overleftrightarrow{\bm{D}}^i \right)
T^c \chi|\Omega \rangle
\end{equation}
can be computed in the same way as 
$\langle {\cal O}^{V} (^3P_J^{[8]}) \rangle$. The contact term for this
LDME is 
\begin{equation}
-V_{{\cal O}(^1P_1^{[8]})}
= \delta^{(3)} (\bm{r}) \frac{1}{4 N_c} {\cal E}_{00},
\end{equation}
which gives the following result for the LDME 
\begin{equation}
\langle {\cal O}^{P} (^1P_1^{[8]}) \rangle = 
\frac{1}{2N_c} {\cal E}_{00} |\phi_P^{(0)} (\bm{0})|^2. 
\end{equation}
This reproduces the heavy-quark spin symmetry relation 
$\langle {\cal O}^{P} (^1P_1^{[8]}) \rangle = 3 \times 
\langle {\cal O}^{V} (^3P_0^{[8]}) \rangle$. 
We note that $\langle {\cal O}^{P} (^3P_J^{[8]}) \rangle$ vanish for all $J$ 
at leading order in $v$, because the contact terms 
$-V_{{\cal O}(^3P_J^{[8]})}$ at leading order in the QMPT vanish when 
applied to the $^1S_0$ state. 
Likewise, $\langle {\cal O}^{V} (^1P_1^{[8]}) \rangle$ vanishes at leading
order in $v$, and hence does not appear in the NRQCD factorization formula
in eq.~(\ref{eq:fac_jpsi}).

We can also compute the color-octet LDMEs 
$\langle {\cal O}^{P} (^3S_1^{[8]}) \rangle$ and 
$\langle {\cal O}^{P} (^1S_0^{[8]}) \rangle$ for the $^1S_0$ state. 
We note that the $^3S_1$ contributions to the contact terms for the 
LDMEs $\langle {\cal O}^{V} (^3S_1^{[8]}) \rangle$ and 
$\langle {\cal O}^{V} (^1S_0^{[8]}) \rangle$ 
that we found vanish when applied to the $^1S_0$ state. 
For the contact term $-V_{{\cal O}(^3S_1^{[8]})}$, the contribution
nonvanishing for the $^1S_0$ state comes from the spin-flip interaction: 
\begin{equation}
\left. -V_{{\cal O}(^3S_1^{[8]})} \right|_{^1S_0}
= 
\{\sigma^k, \sigma^i\} \otimes \{ \sigma^j, \sigma^k \} 
\frac{c_F^2 }{16 N_c m^2} \, \delta^{(3)} (\bm{r}) \, {\cal B}_{00}^{ij}
=
\frac{c_F^2 }{4 N_c m^2} \, \delta^{(3)} (\bm{r}) \, {\cal B}_{00}, 
\end{equation}
which gives $\langle {\cal O}^{P} (^3S_1^{[8]}) \rangle = \langle {\cal O}^{V}
(^1S_0^{[8]}) \rangle$. 
Similarly, the contribution to the contact term $-V_{{\cal O}(^1S_0^{[8]})}$
that is nonvanishing for the $^1S_0$ state comes from the spin-independent 
terms:
\begin{equation}
\left. -V_{{\cal O}(^1S_0^{[8]})} \right|_{^1S_0}
=
\frac{1}{4 N_c m^2} \, \delta^{(3)} (\bm{r}) \, {\cal E}_{10;10}, 
\end{equation}
so that 
$\langle {\cal O}^{P} (^1S_0^{[8]}) \rangle = 1/3 \times \langle {\cal O}^{V} (^3S_1^{[8]}) \rangle$. 

\subsection{Evolution equations} 

The NRQCD LDMEs contain ultraviolet divergences, which must be renormalized. 
Since we employ dimensional regularization, power divergences are automatically
discarded, while logarithmic divergences lead to logarithmic dependences on the
scale at which the LDMEs are renormalized. The evolution equations for the
LDMEs at one loop have been computed in refs.~\cite{Bodwin:1994jh,Bodwin:2012xc}. 
Since $S$-wave wavefunctions at the origin first develop logarithmic ultraviolet divergences 
from two loops~\cite{Czarnecki:1997vz, Beneke:1997jm, Chung:2020zqc},
the scale dependence in the LDMEs must come from the gluonic correlators. 

\begin{figure}[tbp]
\centering
\includegraphics[width=.95\textwidth]{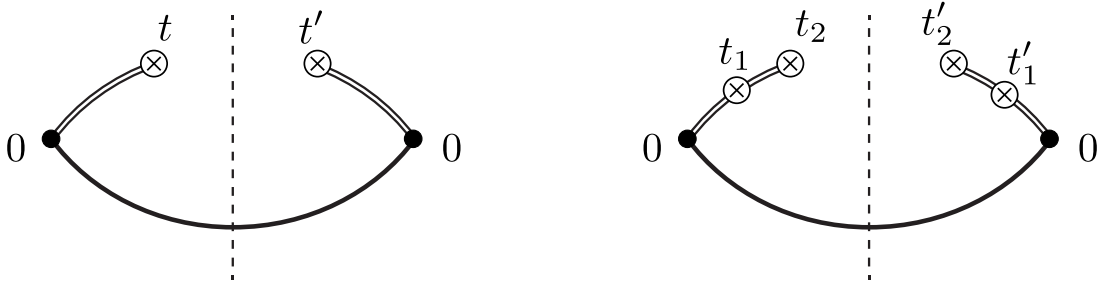}
\caption{\label{fig:figw}
Left: graphical representation of the gluon field strengths and Wilson lines of
the integrand of eqs.~(\ref{eq:B00tensordef}) and (\ref{eq:E00tensordef}). 
The symbols $\otimes$ represent insertions of gluon field strengths at the times $t$ and $t'$. 
Right: graphical representation of the field strengths and Wilson lines of the integrand of eq.~(\ref{eq:E1010def}). 
The symbols $\otimes$ represent insertions of chromoelectric fields at the times $t_1$, $t_1'$, $t_2$, and $t_2'$. 
In both diagrams, filled circles represent the spacetime origin, double lines
are Schwinger lines, solid lines are gauge-completion Wilson lines in the $\ell$ direction, and the dashed line is the cut. 
}
\end{figure}

The gluonic correlators  ${\cal B}_{00}$ and  ${\cal E}_{00}$ are defined
through the relations ${\cal B}_{00}= \delta^{ij} {\cal B}_{00}^{ij}$ 
and ${\cal E}_{00}= \delta^{ij} {\cal E}_{00}^{ij}$, where the tensors 
${\cal B}_{00}^{ij}$ and ${\cal E}_{00}^{ij}$ are defined in eqs.~(\ref{eq:B00tensordef}) and (\ref{eq:E00tensordef}), respectively. 
The correlator ${\cal E}_{10;10}$ is defined in eq.~(\ref{eq:E1010def}). 
These quantities take the form of time moments of gluon field strengths 
attached to Schwinger lines, with gauge completion Wilson lines in the $\ell$ direction. 
We show graphical representations of the configurations of Wilson lines and
insertions of gluon field strengths in figure~\ref{fig:figw}.
Note that all three correlators have mass dimension 2, so that if we compute
them in perturbation theory, they will contain quadratic power divergences. 

\begin{figure}[tbp]
\centering
\includegraphics[width=.8\textwidth]{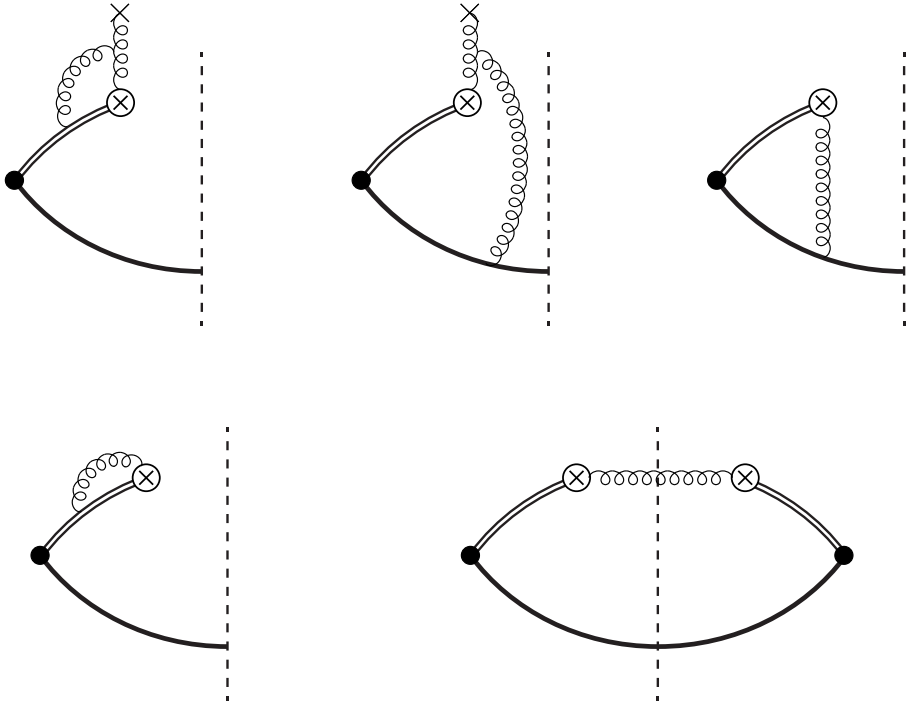}
\caption{\label{fig:B00diags}
Representative one-loop Feynman diagrams for ${\cal B}_{00}$. 
The $\otimes$ symbol is the chromomagnetic field, and 
the symbol $\times$ represent contributions from nonperturbative/external gluon fields. 
}
\end{figure}

We first examine the correlator ${\cal B}_{00}$. Representative Feynman diagrams that 
contribute to the correlator at one loop are shown in figure~\ref{fig:B00diags}.
The last three diagrams, which involve only perturbative gluons, 
diverge quadratically, and hence do not contain logarithmic
divergences in dimensional regularization. On the other hand, the first two 
diagrams in figure~\ref{fig:B00diags} involve nonperturbative gluon fields,
which we represent through external gluon lines.
These diagrams can give rise to logarithmic divergences with nonperturbative coefficients. 
The second diagram may be discarded, however, since momentum conservation either requires
all gluons to be nonperturbative, hence not giving rise to any ultraviolet divergence,
or all loop gluons to be perturbative, hence giving rise to a scaleless integral that vanishes in dimensional regularization.
The first diagram on the other hand may give rise to a logarithmic divergence with a nonperturbative coefficient.
It is similar to the one-loop correction to the operators $\psi^\dag \bm{\sigma} \cdot g \bm{B} \psi$ and 
$\chi^\dag \bm{\sigma} \cdot g \bm{B} \chi$ in the NRQCD Lagrangian at leading
power in $1/m$, except that the gluon fields are in the adjoint representation. 
That is, the scale dependence of ${\cal B}_{00}$ at one-loop level is equal to a color factor times the anomalous dimension of the operator 
$\psi^\dag \bm{\sigma} \cdot g \bm{B} \psi$ or $\chi^\dag \bm{\sigma} \cdot g \bm{B} \chi$. 
By explicit calculation, we find that the scale dependence of ${\cal B}_{00}$ is given by 
\begin{equation}
\frac{d}{d \log \Lambda} {\cal B}_{00}
= - \frac{\alpha_s C_A}{\pi} {\cal B}_{00} 
+ O(\alpha_s^2), 
\end{equation}
where $\Lambda$ is the renormalization scale for ${\cal B}_{00}$. 
We note that the renormalization of the $\psi^\dag \bm{\sigma} \cdot g \bm{B} \psi$ term in the NRQCD Lagrangian requires
\begin{equation}
\label{eq:cFevolution}
\frac{d}{d \log \Lambda} c_F (m;\Lambda) = 
\frac{\alpha_s C_A}{2 \pi} + O(\alpha_s^2),
\end{equation}
so that $c_F^2 {\cal B}_{00}$ is scale invariant at one-loop level. 
This implies that the $^1S_0^{[8]}$ LDME does not evolve at one loop, which
agrees with the known result obtained in perturbative calculations in NRQCD. 

It is straightforward to compute the same diagrams in figure~\ref{fig:B00diags} with the
chromomagnetic fields replaced by chromoelectric fields and find that they
vanish at one loop. Hence, the ${\cal E}_{00}$ does not involve logarithmic UV
divergences at one loop. Similarly to the $^1S_0^{[8]}$  case, 
this implies that the $^3P_J^{[8]}$ LDMEs do not evolve at one loop, which
agrees with the known result obtained in perturbative calculations in NRQCD.

\begin{figure}[tbp]
\centering
\includegraphics[width=.5\textwidth]{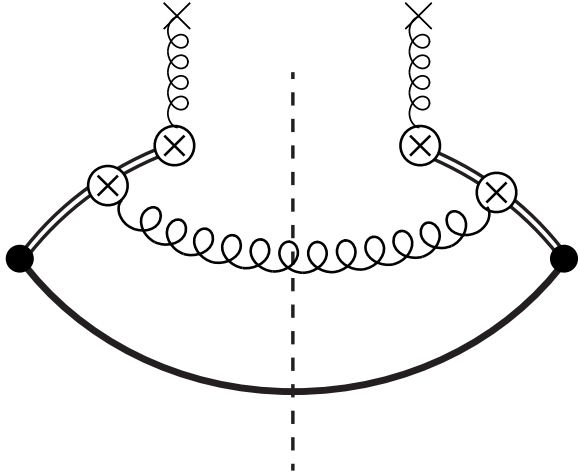}
\caption{\label{fig:E1010diag}
One-loop Feynman diagram for the logarithmically divergent contribution to 
${\cal E}_{10;10}$. 
The symbol $\otimes$ is the chromoelectric field, and 
symbol $\times$ represent contributions from nonperturbative/external gluon fields.
}
\end{figure}

We now turn to the computation of the logarithmic divergence in 
${\cal E}_{10;10}$. Similarly to the ${\cal B}_{00}$, direct evaluation of 
${\cal E}_{10;10}$ in perturbative QCD can only produce scaleless power
divergences. 
By dimensional analysis, we see that the
logarithmically divergent contribution can only arise from perturbatively 
integrating out the 
chromoelectric fields at times $t_1$ and $t_1'$ in eq.~(\ref{eq:E1010def}), 
because this is the only dimensionless integral. 
The Feynman diagram for this contribution is shown in
figure~\ref{fig:E1010diag}.
By computing the correlator ${\cal E}_{10;10}$ through order $\alpha_s$, we find
\begin{eqnarray}
{\cal E}_{10;10} |_{\textrm{1-loop log UV}} 
&=& 
\frac{d^{a b c} d^{a b c}}{N_c^2-1}
{\cal E}_{00} 
\frac{g^2}{6 \pi^2}
\int_0^\infty dt_1 \, t_1 
\int_0^\infty dt'_1 \, t'_1 
\int_0^\infty dk \, k^{3-2 \epsilon}
e^{-i k (t_1-t_1')}
\nonumber \\ &=& 
\frac{1}{2 \epsilon_{\rm UV}} 
\frac{2 \alpha_s}{3 \pi} \frac{N_c^2 -4}{N_c} {\cal E}_{00}, 
\end{eqnarray}
where we identified ${\cal E}_{00}$ from the low-energy mode
contributions to the chromoelectric fields at the times $t_2$ and $t_2'$, 
and we discarded any contribution that does not produce a logarithmic ultraviolet divergence. 
This result gives the following evolution equation
\begin{equation}
\label{eq:RG_E1010}
\frac{d}{d \log \Lambda} {\cal E}_{10;10}
= \frac{2 \alpha_s}{3 \pi} \frac{N_c^2 -4}{N_c} {\cal E}_{00} + O(\alpha_s^2), 
\end{equation}
where $\Lambda$ is the renormalization scale for ${\cal E}_{10;10}$. 
This result implies that $\langle {\cal O}^V(^3S_1^{[8]})\rangle$ satisfies the following evolution equation 
\begin{equation}
\label{eq:RG_LDME}
\frac{d}{d \log \Lambda} \langle {\cal O}^V (^3S_1^{[8]})\rangle
= \frac{6 (N_c^2-4)}{N_c m^2} 
\frac{\alpha_s}{\pi}
\langle {\cal O}^V (^3P_0^{[8]})\rangle, 
\end{equation}
which agrees with ref.~\cite{Bodwin:2012xc}, 
after using the heavy-quark spin symmetry relation 
$\sum_J \langle {\cal O}^V (^3P_J^{[8]})\rangle$ 
$=$ $9 \times \langle {\cal O}^V (^3P_0^{[8]})\rangle$. 
We note that eq.~(\ref{eq:RG_LDME}) can also be obtained from the evolution
equations for decay LDMEs derived in ref.~\cite{Bodwin:1994jh}, 
by using the fact that at one-loop
level the perturbative NRQCD calculations of the decay and production LDMEs 
involve the same Feynman diagrams. 

In calculations of short-distance coefficients, it is customary to choose the
NRQCD factorization scale $\Lambda$ to be the heavy quark mass $m$. In this
case, the correlators ${\cal E}_{10;10}$ and ${\cal B}_{00}$ must be evaluated 
at different scales in computations of charmonium and bottomonium LDMEs. 
We compute ${\cal E}_{10;10}$ and ${\cal B}_{00}$ at different scales by using the one-loop
renormalization group improved formulae 
\begin{subequations}
\label{eq:RGsolution}
\begin{eqnarray}
{\cal E}_{10;10} (\Lambda) &=& 
{\cal E}_{10;10} (\Lambda_0) + \frac{4 (N_c^2-4)}{3 N_c\, \beta_0}
{\cal E}_{00} \log \frac{\alpha_s(\Lambda_0)}{\alpha_s(\Lambda)},
\\
{\cal B}_{00} (\Lambda) &=& 
{\cal B}_{00} (\Lambda_0)  \times \left(
\frac{\alpha_s(\Lambda)}{\alpha_s(\Lambda_0)}\right)^{2 C_A/\beta_0}, 
\end{eqnarray}
\end{subequations}
where $\beta_0 = 11 N_c/3-2 n_f/3$. 

\subsection{Summary of the LDMEs} 
\label{sec:LDMEsummary}

The pNRQCD results for the polarization-summed LDMEs that appear in the NRQCD 
factorization formula in
eq.~(\ref{eq:fac_jpsi}) for production of a $^3S_1$ quarkonium $V$ are given in
eqs.~(\ref{eq:singletresult}), (\ref{eq:3pj_result}), (\ref{eq:1s0result}), and 
(\ref{eq:3s1result}) at leading nonvanishing orders in $v$. The pNRQCD
expressions for the color-octet LDMEs are valid up to corrections of order
$1/N_c^2$.  
The LDMEs can be written in terms of the radial wavefunction $R_V^{(0)}(r)$, 
defined through the relation $\phi_V^{(0)} (r) = R_V^{(0)}(r)/(4 \pi)$ for
$S$-wave states, as
\begin{subequations}
\label{eq:LDMEs_summary}
\begin{eqnarray}
\langle {\cal O}^V(^3S_1^{[1]}) \rangle
&=& \frac{3 N_c}{2 \pi} |R_V^{(0)}(0)|^2,
\\
\langle {\cal O}^V(^3P_J^{[8]}) \rangle
&=& \frac{2 J+1}{18 N_c} {\cal E}_{00} \frac{3 |R_V^{(0)}(0)|^2}{4 \pi},
\\
\langle {\cal O}^V (^1S_0^{[8]}) \rangle
&=& \frac{1}{6 N_c m^2} \frac{3 |R_V(0)|^2}{4 \pi} c_F^2(m;\Lambda) {\cal
B}_{00}(\Lambda),
\\
\langle {\cal O}^V (^3S_1^{[8]}) \rangle (\Lambda) 
&=& \frac{1}{2 N_c m^2} \frac{3 |R_V^{(0)} (0)|^2 }{4 \pi} {\cal E}_{10;10}(\Lambda),
\end{eqnarray}
\end{subequations}
where we have made explicit the scale dependence of the gluonic correlators and of the $^3S_1^{[8]}$ LDME. 
The expression for $\langle {\cal O}^V (^3S_1^{[8]}) \rangle$ is valid
when the LDME and the correlator ${\cal E}_{10;10}$ are
regularized dimensionally and renormalized in
the same scheme and at the same scale.
These expressions have first been reported in ref.~\cite{Brambilla:2022rjd}.

While the color-singlet LDME $\langle {\cal O}^V(^3S_1^{[1]}) \rangle$
can be determined from the quarkonium wavefunction at the origin, 
the expressions for the color-octet LDMEs also involve the gluonic correlators 
${\cal E}_{00}$, ${\cal B}_{00}$, and ${\cal E}_{10;10}$. 
The correlators ${\cal E}_{00}$ and ${\cal B}_{00}$ are defined through the relations 
${\cal E}_{00} = \delta^{ij} {\cal E}_{00}^{ij}$ and 
${\cal B}_{00} = \delta^{ij} {\cal B}_{00}^{ij}$, 
where the tensors ${\cal E}_{00}^{ij}$ and ${\cal B}_{00}^{ij}$ are defined in
eqs.~(\ref{eq:E00tensordef}) and (\ref{eq:B00tensordef}), respectively. 
The correlator ${\cal E}_{10;10}$ is defined in eq.~(\ref{eq:E1010def}). 
Since the quarkonium wavefunctions can be computed by solving the Schr\"odinger
equation from the known QCD potential, or extracted from the leptonic width, 
and the gluonic correlators are universal quantities that do not depend on the quarkonium state, 
the determination of the three gluonic correlators ${\cal E}_{00}$, ${\cal B}_{00}$, and ${\cal E}_{10;10}$ fixes the three color-octet LDMEs, 
and the inclusive production cross section for all strongly coupled $^3S_1$ heavy quarkonia.
That is, the pNRQCD results for the LDMEs greatly reduce the number of
independent color-octet LDMEs. As the strongly coupled pNRQCD formalism is
expected to be valid for $J/\psi$, $\psi(2S)$, $\Upsilon(2S)$, and
$\Upsilon(3S)$ states, the pNRQCD results reduce the number of independent color-octet
LDMEs from $4 \times 3 = 12$ to 3. We note that the pNRQCD results 
for the LDMEs (\ref{eq:LDMEs_summary}) imply 
at leading order in $v$ the universal relations between two strongly coupled
$^3S_1$ quarkonia $V$ and $V'$ given by 
\begin{subequations}
\label{eq:LDMEs_universal_relations}
\begin{eqnarray}
\frac{\langle {\cal O}^{V'} (^3S_1^{[1]}) \rangle}
{\langle {\cal O}^V (^3S_1^{[1]}) \rangle}
&=& \frac{|R_{V'}^{(0)} (0)|^2}{|R_V^{(0)} (0)|^2},
\\
\frac{\langle {\cal O}^{V'} (^3P_J^{[8]}) \rangle}
{\langle {\cal O}^V (^3P_J^{[8]}) \rangle}
&=& 
\frac{|R_{V'}^{(0)} (0)|^2}{|R_V^{(0)} (0)|^2},
\\
\frac{\langle {\cal O}^{V'} (^3S_1^{[8]}) \rangle(\Lambda)}
{\langle {\cal O}^V (^3S_1^{[8]}) \rangle(\Lambda)}
&=& 
\frac{m_Q^2}{m_{Q'}^2} 
\frac{|R_{V'}^{(0)} (0)|^2}{|R_V^{(0)} (0)|^2},
\\
\frac{\langle {\cal O}^{V'} (^1S_0^{[8]}) \rangle}
{\langle {\cal O}^V (^1S_0^{[8]}) \rangle}
&=& 
\frac{m_Q^2}{m_{Q'}^2} \frac{c_F^2 (m_{Q'};\Lambda)}{c_F^2 (m_Q;\Lambda)}
\frac{|R_{V'}^{(0)} (0)|^2}{|R_V^{(0)} (0)|^2},
\end{eqnarray}
\end{subequations}
which we obtain by taking ratios of the right-hand sides of 
eq.~(\ref{eq:LDMEs_summary}). Here, 
$V$ and $V'$ are bound states of $Q \bar Q$ and $Q' \bar{Q'}$,
respectively, and $\Lambda$ is the NRQCD scale. 
Hence, once the LDMEs are determined for one $^3S_1$ quarkonium state, 
the pNRQCD results fix the LDMEs for all other $^3S_1$ charmonium and bottomonium states. 
Note that if $Q = Q'$, the heavy quark masses and the short-distance
coefficient $c_F$ cancel in the ratios, so that the ratios of the LDMEs are
given simply by $|R_{V'}^{(0)} (0)|^2/|R_V^{(0)} (0)|^2$ for all four of the LDMEs. 
These relations take the form given in eqs.~(\ref{eq:LDMEs_universal_relations}) when the $^3S_1^{[8]}$ LDMEs 
for the $V$ and $V'$ states are computed at the same scale; 
the LDMEs at different scales can be obtained by solving the evolution
equations. At one-loop level, the relation for the 
$^1S_0^{[8]}$ LDME can be written as 
\begin{equation}
\label{eq:LDMEs_universal_relation_1S0}
\frac{\langle {\cal O}^{V'} (^1S_0^{[8]}) \rangle}
{\langle {\cal O}^V (^1S_0^{[8]}) \rangle}
=
\frac{m_Q^2}{m_{Q'}^2} 
\left( \frac{\alpha_s(m_{Q'})}{\alpha_s(m_Q)}\right)^{2 C_A/\beta_0}
\frac{|R_{V'}^{(0)} (0)|^2}{|R_V^{(0)} (0)|^2},
\end{equation}
which is obtained by using the one loop renormalization group improved
expression for the solution of eq.~(\ref{eq:cFevolution}).
The relations (\ref{eq:LDMEs_universal_relations}) are satisfied by 
LDMEs computed from
eqs.~(\ref{eq:LDMEs_summary}), regardless of the specific values of the
gluonic correlators.
Hence,
these universal relations are expected to hold for strongly coupled $S$-wave
quarkonia that may include $J/\psi$, $\psi(2S)$, and excited $\Upsilon$ states.
Remarkably, these relations imply that ratios of production rates of strongly coupled
spin-1 quarkonia with same heavy quark flavor are simply given by ratios of squares of
quarkonium wavefunctions at the origin, up to corrections of higher orders in
$v$. In the next section, we make predictions of the ratio of $J/\psi$ and 
$\psi(2S)$ production rates, as well as the $\Upsilon(3S)$ and $\Upsilon(2S)$ cross
section ratio by using these universal relations and compare with data.

We note that the gluonic correlators ${\cal E}_{00}$, ${\cal B}_{00}$, and ${\cal E}_{10;10}$ take the form 
$\langle {\cal O}^\dag {\cal O} \rangle = \lVert {\cal O} | \Omega \rangle \lVert^2$, 
where ${\cal O}$ is a time-ordered product of gluonic operators. 
That is, we can write ${\cal E}_{00}$, ${\cal B}_{00}$, and ${\cal E}_{10;10}$
as norms of states obtained from applying time-ordered gluonic operators to the vacuum:
\begin{subequations}
\label{eq:correlators_norm}
\begin{eqnarray}
{\cal E}_{00} &=& 
\left \lVert 
\int_0^\infty dt \, 
g \bm{E}^{a} (t) \Phi_0^{ac} (0;t) \Phi_\ell^{bc} (0) | \Omega \rangle
\right \lVert^2, 
\\
{\cal B}_{00} &=& 
\left \lVert 
\int_0^\infty dt \, 
g \bm{B}^{a} (t) \Phi_0^{ac} (0;t) \Phi_\ell^{bc} (0) | \Omega \rangle
\right \lVert^2, 
\\
{\cal E}_{10;10} &=& 
\left \lVert 
d^{dac} 
\int_0^\infty dt_1 \, t_1 \int_{t_1}^\infty dt_2  \,
g E^{i,b} (t_2) \Phi_0^{bc} (t_1;t_2) g E^{i,a} (t_1)
\Phi_0^{df} (0;t_1) \Phi_\ell^{ef} (0)
|\Omega\rangle
\right \lVert^2. 
\hspace{6ex} 
\end{eqnarray}
\end{subequations}
As we have mentioned in the previous section, these correlators contain
quadratic power divergences when computed in perturbative QCD. 
Furthermore, as we have shown, 
${\cal B}_{00}$ and ${\cal E}_{10;10}$ develop logarithmic divergences at one
loop, which must be removed through renormalization. 
We recall that the pNRQCD results for the LDMEs are valid only in dimensional
regularization, because we have discarded scaleless integrals in deriving the
expressions for the LDMEs. Since in dimensional regularization, 
power and logarithmic divergences are removed through subtraction, 
the values of the correlators are not necessarily positive definite, even though they can be
written as norms of states as shown in eqs.~(\ref{eq:correlators_norm}). 
Hence, in this paper, we do not make any assumptions on the signs of 
${\cal E}_{10;10}$, ${\cal B}_{00}$, and ${\cal E}_{00}$.

\section{\boldmath Phenomenology of inclusive production of $S$-wave quarkonia}
\label{sec:pheno} 

We now use our results for the color-singlet and color-octet LDMEs for
$S$-wave spin-triplet quarkonia to compute inclusive cross sections of 
$J/\psi$, $\psi(2S)$, $\Upsilon(2S)$, and $\Upsilon(3S)$. 
For our phenomenological results, we compute the $p_T$-differential 
short-distance coefficients from $pp$ collisions at
next-to-leading order (NLO) in $\alpha_s$ by using the \textsc{FDCHQHP}
package~\cite{Wan:2014vka}. 
We take the heavy quark masses $m_c = 1.5$~GeV and $m_b = 4.75$~GeV, and take
the NRQCD factorization scale to be $\Lambda = m_c$ for charmonium, and 
$\Lambda = m_b$ for bottomonium\footnote{
Although from an effective field theory perspective the scale $\Lambda$ should be
taken close to the soft scale $mv$, the specific choice of it 
is without phenomenological consequences, for the scale dependence
cancels in the cross sections at the given accuracy.
Our choice of $\Lambda$, of about the heavy quark mass,
provides a better convergence of the perturbative series in the short distance
coefficients at the price of possibly affecting the natural power counting  
of the low energy correlators.
}. 
We use CTEQ6M parton distribution functions and compute $\alpha_s$ at two loops
with $n_f = 5$ light quark flavors and $\Lambda_{\rm QCD}^{(5)} = 226$~MeV. 
The scale at which the parton distribution
functions and $\alpha_s$ are computed are taken to be 
$\sqrt{p_T^2 + 4 m_Q^2}$, where $Q = c$ for charmonium and $Q = b$ for
bottomonium. 
When computing the scale dependences of the gluonic correlators using
eqs.~(\ref{eq:RGsolution}), we use 
$\alpha_s (m_c) = 0.30$ and $\alpha_s (m_b) = 0.21$.
When we take into account the effect of feeddowns, we compute the contribution
from the decay of $n'S$ quarkonium into $nS$ quarkonium by the product of the 
branching fraction $B_{n'S \to nS + X}$ and the direct production rate 
$\sigma_{n'S}$. In the case of feeddowns from $P$-wave quarkonia, we employ the
measured $p_T$-dependent feeddown fractions in refs.~\cite{ATLAS:2014ala, 
LHCb:2014ngh}.  Although it is possible to compute the production rates of
$P$-wave quarkonia in NRQCD, for example by using the results for the $P$-wave
LDMEs in refs.~\cite{Brambilla:2020ojz, Brambilla:2021abf}, the measured
feeddown fractions are generally more accurate than NRQCD calculations. 

The determinations of the LDMEs from their pNRQCD expressions require the
gluonic correlators ${\cal E}_{10;10}$, ${\cal B}_{00}$, and ${\cal E}_{00}$, 
as well as the wavefunctions at the origin $|R_V^{(0)} (0)|^2$. 
Since lattice QCD calculations of the gluonic correlators are not available
yet, we determine the correlators by comparing with measured cross section data.
In contrast, the wavefunctions at the origin could be 
  computed by solving a Schr\"odinger equation based on the lattice QCD determination
  of the quarkonium potential, which is known. 
  However, since accurate measurements of 
the leptonic decay rates of $^3S_1$ heavy quarkonia are available,
it is more straightforward to 
determine $|R_V^{(0)}(0)|^2$ by using 
\begin{equation}
\Gamma(V \to \ell^+ \ell^-) 
= \frac{4 N_c}{3 m_V^2} \alpha^2 e_Q^2 \left( 1- \frac{2 \alpha_s C_F }{\pi} 
\right)^2 |R_V^{(0)}(0)|^2, 
\end{equation}
where $e_Q = 2/3$ for $Q =c$ and $e_Q = -1/3$ for $Q = b$, and $\alpha$ is the 
fine structure constant.  This expression is valid at leading order in $v$ and
through NLO in $\alpha_s$ to determine $|R_V^{(0)}(0)|^2$. 
Here, $m_V$ is the mass of the quarkonium $V$. By using the measured 
decay rates into $e^+ e^-$ from ref.~\cite{ParticleDataGroup:2018ovx}, 
we obtain the central values 
$|R_{J/\psi}^{(0)} (0)|^2 = 0.825\textrm{~GeV}^3$, 
$|R_{\psi(2S)}^{(0)} (0)|^2 = 0.492\textrm{~GeV}^3$, 
$|R_{\Upsilon(2S)}^{(0)} (0)|^2 = 3.46\textrm{~GeV}^3$ and 
$|R_{\Upsilon(3S)}^{(0)} (0)|^2 = 2.67\textrm{~GeV}^3$. 
Here we used $\alpha_s = 0.25$ for charmonium and $\alpha_s = 0.21$ for
bottomonium, which are computed at the scale of the quarkonium mass. 
The color-singlet LDME can already be computed by using these values of $|R_V^{(0)}(0)|^2$. We obtain 
\begin{subequations}
\label{eq:CS_ldmes}
\begin{eqnarray}
\langle {\cal O}^{J/\psi} (^3S_1^{[1]}) \rangle 
&=& 1.18 \pm 0.35\textrm{~GeV}^3, \\
\langle {\cal O}^{\psi(2S)} (^3S_1^{[1]}) \rangle 
&=& 0.71 \pm 0.21\textrm{~GeV}^3, \\
\langle {\cal O}^{\Upsilon(2S)} (^3S_1^{[1]}) \rangle 
&=& 4.96 \pm 0.50\textrm{~GeV}^3, \\
\langle {\cal O}^{\Upsilon(3S)} (^3S_1^{[1]}) \rangle 
&=& 3.83 \pm 0.38\textrm{~GeV}^3,
\end{eqnarray}
\end{subequations}
where the uncertainties come from the fact that the pNRQCD expression for the
color-singlet LDME is valid up to corrections of relative order $v^2$, which
are estimated to be 30\% and 10\% of the central values for charmonium and
bottomonium, respectively, based on the typical sizes $v^2 \approx 0.3$ for
charmonium and $v^2 \approx 0.1$ for bottomonium. 
These values are compatible within uncertainties
with the potential-model calculations from refs.~\cite{Eichten:1995ch, 
Bodwin:2007fz, Chung:2010vz}
that are widely adopted in quarkonium phenomenology. 

Because the short-distance coefficients are computed at the $\overline{\rm MS}$
scale $\Lambda = m$, the gluonic correlators ${\cal B}_{00} (\Lambda)$ and 
${\cal E}_{10;10} (\Lambda)$ are evaluated at different scales for charmonium
and bottomonium. We take into account the difference in the scale by using the
one-loop renormalization group improved formulae in eqs.~(\ref{eq:RGsolution}). 
The effect of this running is numerically small for ${\cal B}_{00}(\Lambda)$;
${\cal B}_{00} (m_b)$ is smaller than ${\cal B}_{00} (m_c)$ 
by a factor of about $0.8$. 
On the other hand, the evolution of ${\cal E}_{10;10} (\Lambda)$ depends on 
the value of ${\cal E}_{00}$. For example, if ${\cal E}_{00}$ is positive, then 
${\cal E}_{10;10} (\Lambda)$ takes a larger value at the scale of the bottom
quark mass compared to its value at the scale of the charm quark mass. 
As we will see later, this point will play an important r\^{o}le in the
phenomenological determinations of the correlators.

\subsection{Cross section ratios}
\label{sec:ratios}

We begin with the ratios of cross sections $\sigma_{\psi(2S)}/\sigma_{J/\psi}$ 
and $\sigma_{\Upsilon(3S)}/\sigma_{\Upsilon(2S)}$. 
Because the LDMEs in the factorization formula in eq.~(\ref{eq:fac_jpsi})
satisfy the universal relations in eqs.~(\ref{eq:LDMEs_universal_relations}), 
the ratios do not depend on the values of the gluonic correlators. 
That is, the ratios of direct cross sections also satisfy 
\begin{subequations}
\label{eq:ratios}
\begin{eqnarray}
\frac{\sigma_{\psi(2S)}^{\rm direct}}{\sigma_{J/\psi}^{\rm direct}} 
&=& \frac{|R_{\psi(2S)}^{(0)} (0)|^2}{|R_{J/\psi}^{(0)} (0)|^2},
\\
\frac{\sigma_{\Upsilon(3S)}^{\rm direct}}{\sigma_{\Upsilon(2S)}^{\rm direct}} 
&=& \frac{|R_{\Upsilon(3S)}^{(0)} (0)|^2}{|R_{\Upsilon(2S)}^{(0)} (0)|^2}.
\end{eqnarray}
\end{subequations}
We expect these relations to hold at large $p_T$. 

\begin{figure}[tbp]
\centering
\includegraphics[width=.49\textwidth]{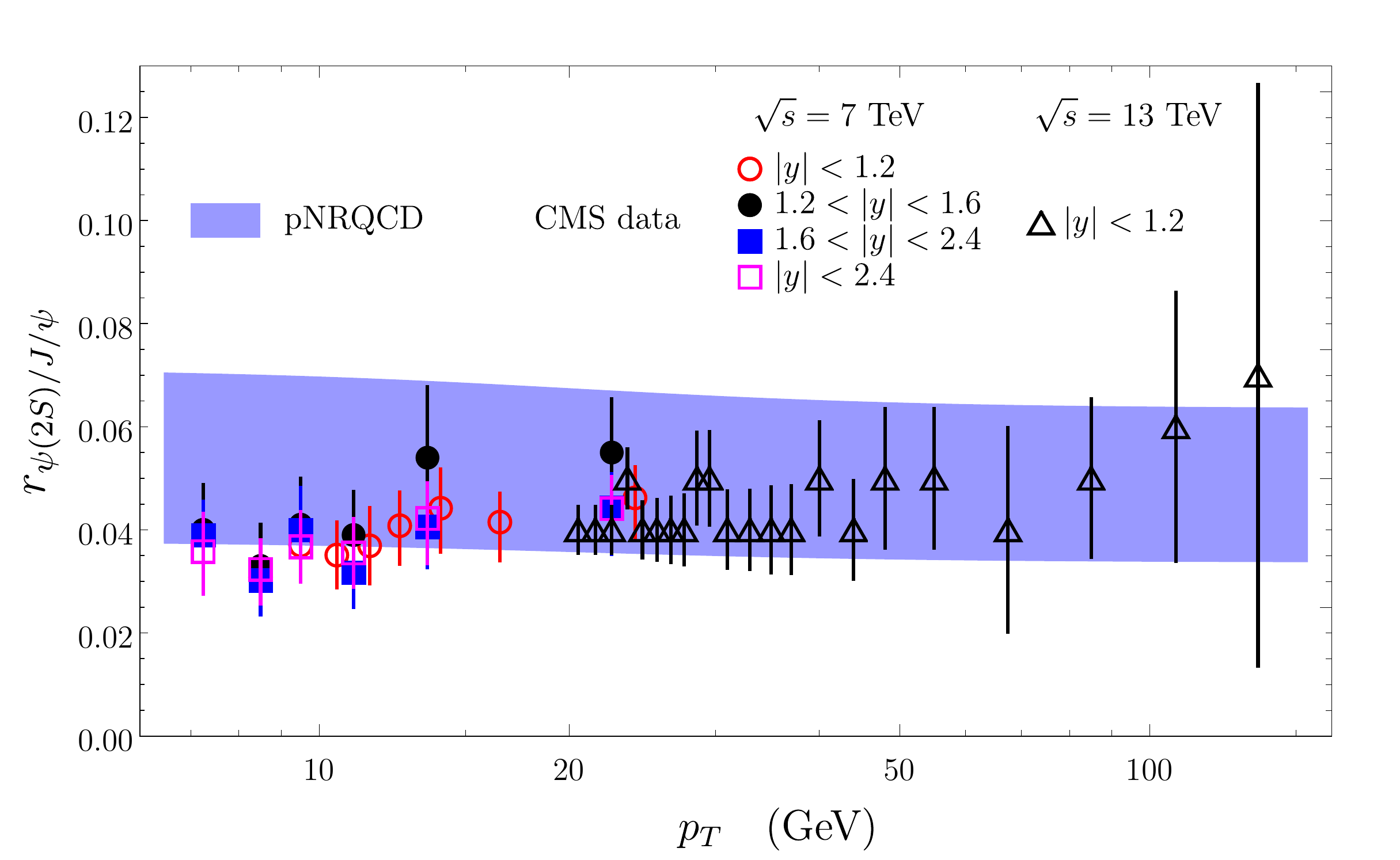}
\includegraphics[width=.49\textwidth]{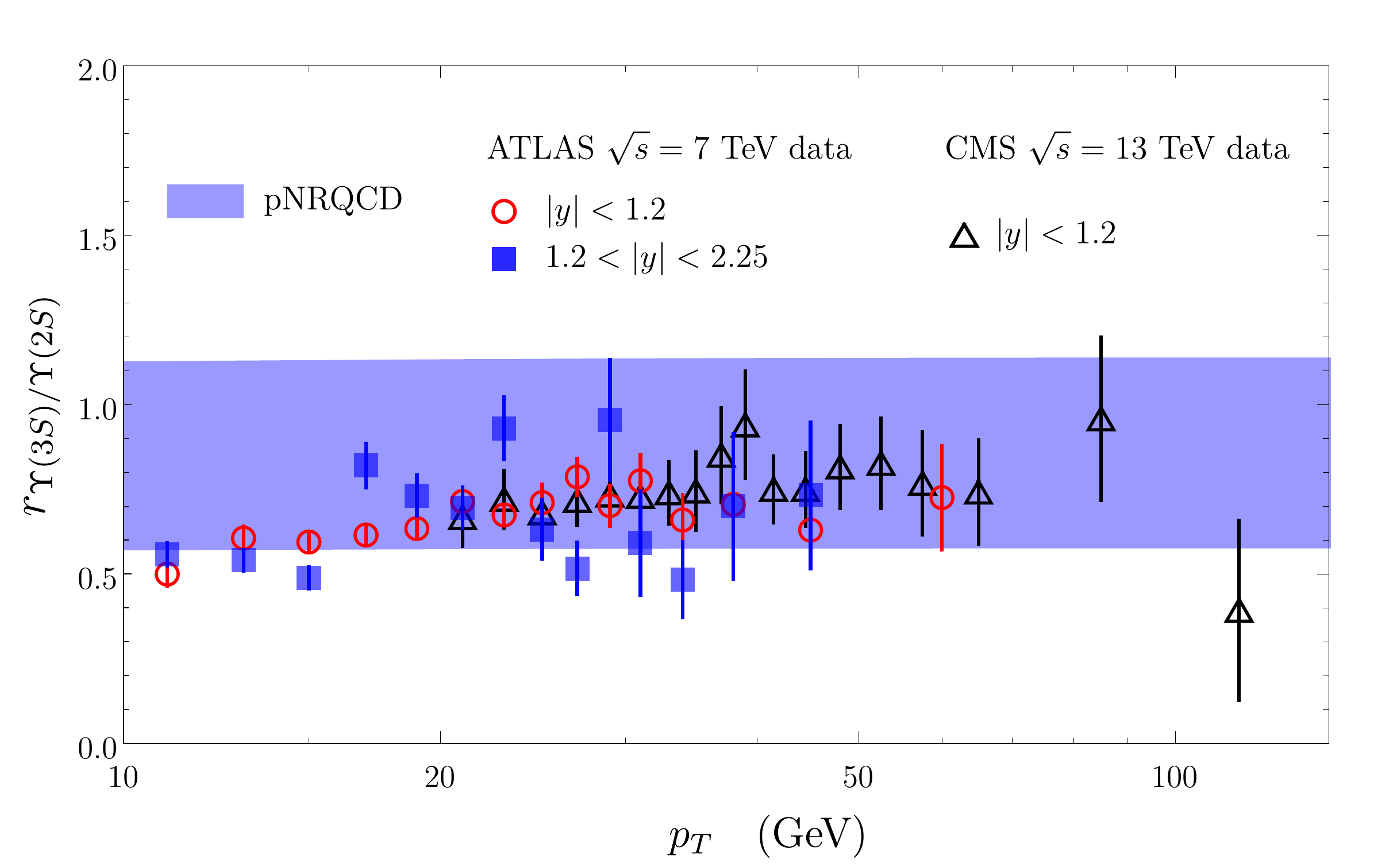}
\caption{\label{fig:ratios}
Left: pNRQCD result for the ratio $r_{\psi(2S)/J/\psi}$ defined in
eq.~(\ref{eq:charmratio}) compared to CMS data at center of mass energies 
$\sqrt{s}=7$~TeV~\cite{CMS:2011rxs}
and $\sqrt{s} = 13$~TeV~\cite{CMS:2017dju}.
Right: pNRQCD result for the ratio 
${r}_{\Upsilon(3S)/\Upsilon(2S)}$ defined in 
eq.~(\ref{eq:upsratio}) compared to the experimental values obtained from 
measurements of 
${r}_{\Upsilon(3S)/\Upsilon(1S)}$ and 
${r}_{\Upsilon(2S)/\Upsilon(1S)}$ from ATLAS at
$\sqrt{s}=7$~TeV~\cite{ATLAS:2012lmu}
and from CMS at $\sqrt{s} = 13$~TeV~\cite{CMS:2017dju}. 
}
\end{figure}

In order to compare with measured cross section ratios, we must take into
account the feeddown contributions. 
While $\sigma_{\psi(2S)}^{\rm prompt} = \sigma_{\psi(2S)}^{\rm direct}$, 
$\sigma_{J/\psi}^{\rm prompt}$ includes feeddowns from decays of $\psi(2S)$ and $\chi_{c}$. 
That is, 
\begin{align}
\sigma_{J/\psi}^{\rm prompt} &= 
\sigma_{J/\psi}^{\rm direct} + 
B_{\psi(2S) \to J/\psi+X} \times \sigma_{\psi(2S)}^{\rm prompt}
+ R_{J/\psi}^{\chi_c} \times \sigma_{J/\psi}^{\rm prompt}
\nonumber\\
&= \sigma_{J/\psi}^{\rm direct} +
B_{\psi(2S) \to J/\psi+X} \, \sigma_{\psi(2S)}^{\rm direct}
+ \frac{R_{J/\psi}^{\chi_c}
\left( \sigma_{J/\psi}^{\rm direct} + B_{\psi(2S) \to J/\psi+X}\,\sigma_{\psi(2S)}^{\rm direct} \right)
}{1-R_{J/\psi}^{\chi_c}} ,
\label{eq:jpsiprompt}
\end{align}
where $B_{\psi(2S) \to J/\psi+X}$ is the branching fraction of $\psi(2S)$ into
$J/\psi + X$, and $R_{J/\psi}^{\chi_c}$ is the feeddown fraction of prompt $J/\psi$ from decays 
of $\chi_{c}$ into $J/\psi + X$.
From eqs.~(\ref{eq:ratios}) and (\ref{eq:jpsiprompt}) we can compute the ratio 
\begin{equation}
\label{eq:charmratio}
r_{\psi(2S)/J/\psi}
= 
\frac{B_{\psi(2S) \to \mu^+ \mu^-} \times \sigma_{\psi(2S)}^{\rm prompt}}
{B_{J/\psi \to \mu^+ \mu^-} \times \sigma_{J/\psi}^{\rm prompt}} 
\end{equation}
by using the measured branching fractions from ref.~\cite{ParticleDataGroup:2018ovx}, 
$R_{J/\psi}^{\chi_c}$ from ref.~\cite{ATLAS:2014ala}, and the ratio of wavefunctions at the origin 
$|R_{\psi(2S)}^{(0)} (0)|^2/|R_{J/\psi}^{(0)} (0)|^2$.
The ratio $r_{\psi(2S)/J/\psi}$ is a function of $p_T$, where the $p_T$ in the
numerator and the denominator are the transverse momenta of the $\psi(2S)$ and
$J/\psi$, respectively. Note that in the feeddown contribution from decays of
$\psi(2S)$ into $J/\psi$, the $p_T$ of the $\psi(2S)$ is larger than the $p_T$
of the $J/\psi$ by approximately a factor of $m_{\psi(2S)}/m_{J/\psi}$. 
Because the measured $p_T$-differential cross section falls off like $1/p_T^n$ as $p_T$ increases 
where $n \approx$ 5--6, we can take this effect into 
account by multiplying $\sigma_{\psi(2S)}^{\rm direct}$ in the denominator 
of eq.~(\ref{eq:charmratio}) by $(m_{J/\psi}/m_{\psi(2S)})^{n}$ and fix 
$n=5.5$. 
We estimate the uncertainties in $r_{\psi(2S)/J/\psi}$ from unaccounted
corrections of higher orders in $v$ by 30\% of the central value, based on the
typical size $v^2 \approx 0.3$ for charmonia. We also take into account the
uncertainty in the measured values of $R_{J/\psi}^{\chi_c}$. 
Since the effect of the difference in $p_T$ of the $\psi(2S)$ and $J/\psi$ 
in the feeddown contribution is about 15\% of the central value of
$r_{\psi(2S)/J/\psi}$, and changes mildly under variations of the power $n$ in
the factor $(m_{J/\psi}/m_{\psi(2S)})^{n}$, we do not consider varying 
$n$. We add the uncertainties in quadrature. 
We compare our calculation of $r_{\psi(2S)/J/\psi}$ with CMS measurements
at center of mass energies
$\sqrt{s}=7$~TeV~\cite{CMS:2011rxs}
and $\sqrt{s} = 13$~TeV~\cite{CMS:2017dju}
in figure~\ref{fig:ratios}. 
We see that the pNRQCD result for $r_{\psi(2S)/J/\psi}$ is in fair agreement
with CMS data, and the agreement improves with increasing $p_T$. 
We note that the pNRQCD result implies that  $r_{\psi(2S)/J/\psi}$ is
independent of the center of mass energy or
the rapidity of the produced quarkonia, which is also supported by experiment. 

We can also compute ratios of inclusive cross sections of $\Upsilon(2S)$ and
$\Upsilon(3S)$ in a similar way. The inclusive cross section of $\Upsilon(3S)$
includes feeddowns from $\chi_b(3P)$, so that 
\begin{equation}
\sigma_{\Upsilon(3S)}^{\rm inclusive} =
\sigma_{\Upsilon(3S)}^{\rm direct}
+ R_{\Upsilon(3S)}^{\chi_b(3P)} \times 
\sigma_{\Upsilon(3S)}^{\rm inclusive} 
= 
\sigma_{\Upsilon(3S)}^{\rm direct}
+ \frac{R_{\Upsilon(3S)}^{\chi_b(3P)}}{1-R_{\Upsilon(3S)}^{\chi_b(3P)}} \times 
\sigma_{\Upsilon(3S)}^{\rm direct}. 
\end{equation}
Similarly, the inclusive cross section of $\Upsilon(2S)$ including feeddowns
from $\Upsilon(3S)$, $\chi_b(2P)$, and $\chi_b(3P)$ is given by 
\begin{eqnarray}
\sigma_{\Upsilon(2S)}^{\rm inclusive}
&=&
\sigma_{\Upsilon(2S)}^{\rm direct}
+ 
B_{\Upsilon(3S) \to \Upsilon(2S)+X} \, 
\frac{ \sigma_{\Upsilon(3S)}^{\rm direct} }{1-R_{\Upsilon(3S)}^{\chi_b(3P)}} 
\nonumber \\ && 
+ \frac{R_{\Upsilon(2S)}^{\chi_b}}{1-R_{\Upsilon(2S)}^{\chi_b}} 
\left( \sigma_{\Upsilon(2S)}^{\rm direct} + 
\frac{B_{\Upsilon(3S) \to \Upsilon(2S)+X} \, \sigma_{\Upsilon(3S)}^{\rm direct} 
}{1-R_{\Upsilon(3S)}^{\chi_b(3P)}} \right), 
\end{eqnarray}
where $R_{\Upsilon(2S)}^{\chi_b} = R_{\Upsilon(2S)}^{\chi_b(3P)} +
R_{\Upsilon(2S)}^{\chi_b(2P)}$. 
By using these expressions for $\sigma_{\Upsilon(3S)}^{\rm inclusive}$ and 
$\sigma_{\Upsilon(2S)}^{\rm inclusive}$, we can compute the ratio
\begin{equation}
\label{eq:upsratio}
{r}_{\Upsilon(3S)/\Upsilon(2S)}
=
\frac{B_{\Upsilon(3S) \to \mu^+ \mu^-} \times 
\sigma_{\Upsilon(3S)}^{\rm inclusive}}
{B_{\Upsilon(2S) \to \mu^+ \mu^-} \times \sigma_{\Upsilon(2S)}^{\rm inclusive}}
\end{equation}
just from the measured branching fractions, $R_{\Upsilon(n'S)}^{\chi_b(nP)}$, 
and the ratios 
$|R_{\Upsilon(3S)}^{(0)} (0)|^2/|R_{\Upsilon(2S)}^{(0)} (0)|^2$.
Similarly to the charmonium case, we also compute
${r}_{\Upsilon(3S)/\Upsilon(2S)}$ as a function of $p_T$, where the $p_T$ in
the numerator and the denominator are the transverse momenta of the
$\Upsilon(3S)$ and $\Upsilon(2S)$, respectively. 
We use the measured feeddown fractions $R_{\Upsilon(n'S)}^{\chi_b(nP)}$ 
from ref.~\cite{LHCb:2014ngh}, and the branching fractions in
ref.~\cite{ParticleDataGroup:2018ovx}.
We also take into account the difference in $p_T$ in the feeddown from decays
of $\Upsilon(3S)$ into $\Upsilon(2S)$ by multiplying  $\sigma^{\rm
direct}_{\Upsilon(3S)}$ in the denominator by a factor of 
$(m_{\Upsilon(2S)}/m_{\Upsilon(3S)})^{n}$ with $n=5.5$. 
We estimate the uncertainty in ${r}_{\Upsilon(3S)/\Upsilon(2S)}$ from
uncalculated corrections of order $v^2$ by 10\% of the central value, based on
the typical size $v^2 \approx 0.1$ for bottomonium. We also take into account
the uncertainty in the measured values of $R_{\Upsilon(nS)}^{\chi_b(n'P)}$.
In the case of bottomonium, the effect of the difference in the $p_T$ of 
$\Upsilon(3S)$ and $\Upsilon(2S)$ increase ${r}_{\Upsilon(3S)/\Upsilon(2S)}$
by less than 2\%, so we do not consider varying the power $n$ in the ratio 
$(m_{\Upsilon(2S)}/m_{\Upsilon(3S)})^{n}$. 
We add the uncertainties in quadrature. 
We compare our calculation of ${r}_{\Upsilon(3S)/\Upsilon(2S)}$ with
experiments in figure~\ref{fig:ratios}.
The experimental values in figure~\ref{fig:ratios} are computed from 
measurements of the
ratios $r_{\Upsilon(3S)/\Upsilon(1S)}$ and $r_{\Upsilon(2S)/\Upsilon(1S)}$ 
at $\sqrt{s}=7$~TeV by ATLAS in ref.~\cite{ATLAS:2012lmu}
and at $\sqrt{s}=13$~TeV by CMS in ref.~\cite{CMS:2017dju}. 
Similarly to the charmonium case, the pNRQCD result is in fair agreement with
experiment for values of $p_T$ larger than the quarkonium mass, and is
independent of the rapidity or the center of mass energy. 
We note that the theoretical uncertainty in ${r}_{\Upsilon(3S)/\Upsilon(2S)}$ 
is dominated by the uncertainties in $R_{\Upsilon(n'S)}^{\chi_b(nP)}$.

\subsection
[Phenomenological determination of ${\cal E}_{10;10}$, ${\cal E}_{00}$, and
$c_F^2 {\cal B}_{00}$]
{\boldmath Phenomenological determination of ${\cal E}_{10;10}$, ${\cal
E}_{00}$, and $c_F^2 {\cal B}_{00}$}
\label{sec:LDMEdeterminations}

We now determine the gluonic correlators ${\cal E}_{10;10}$,
${\cal E}_{00}$, and ${\cal B}_{00}$ by comparing the NRQCD factorization
formula in eq.~(\ref{eq:fac_jpsi}) with measured cross section data. 
We consider the $p_T$-differential cross section measurements of $J/\psi$ and
$\psi(2S)$ from CMS in refs.~\cite{CMS:2011rxs, CMS:2015lbl}, and the
$p_T$-differential cross section measurements of $\Upsilon(2S)$ and
$\Upsilon(3S)$ from ATLAS in ref.~\cite{ATLAS:2012lmu}, which provide
data from a wide range of $p_T$. 
We use the $p_T$-differential short-distance coefficients that we compute at NLO in
$\alpha_s$ using the {\textsc FDCHQHP} package~\cite{Wan:2014vka}. 
As we have mentioned, we take into account the effect of feeddowns from
decays of $P$-wave quarkonia by using the measured feeddown fractions in
refs.~\cite{ATLAS:2014ala, LHCb:2014ngh}
and compute the feeddown contributions from decays of $S$-wave quarkonia by
using the measured branching fractions in
ref.~\cite{ParticleDataGroup:2018ovx}. In the case of feeddowns from 
decays of $n'S$ into $nS$ quarkonium, 
we take into account the difference in the $p_T$ of the $n'S$ and $nS$
quarkonium by setting $p_T^{nS} = (m_{nS}/m_{n'S}) p_T^{n'S}$. 
We consider the theoretical uncertainty from uncalculated relativistic 
corrections to be 30\% and 10\% of the central values for charmonium and 
bottomonium, and the experimental uncertainties in the measured values of 
cross sections and feeddown fractions. We add the uncertainties in quadrature. 
We neglect the uncertainty from corrections of order $1/N_c^2$, because it is 
smaller than the uncertainties that we consider. 

\begin{figure}[tbp]
\centering
\includegraphics[width=.95\textwidth]{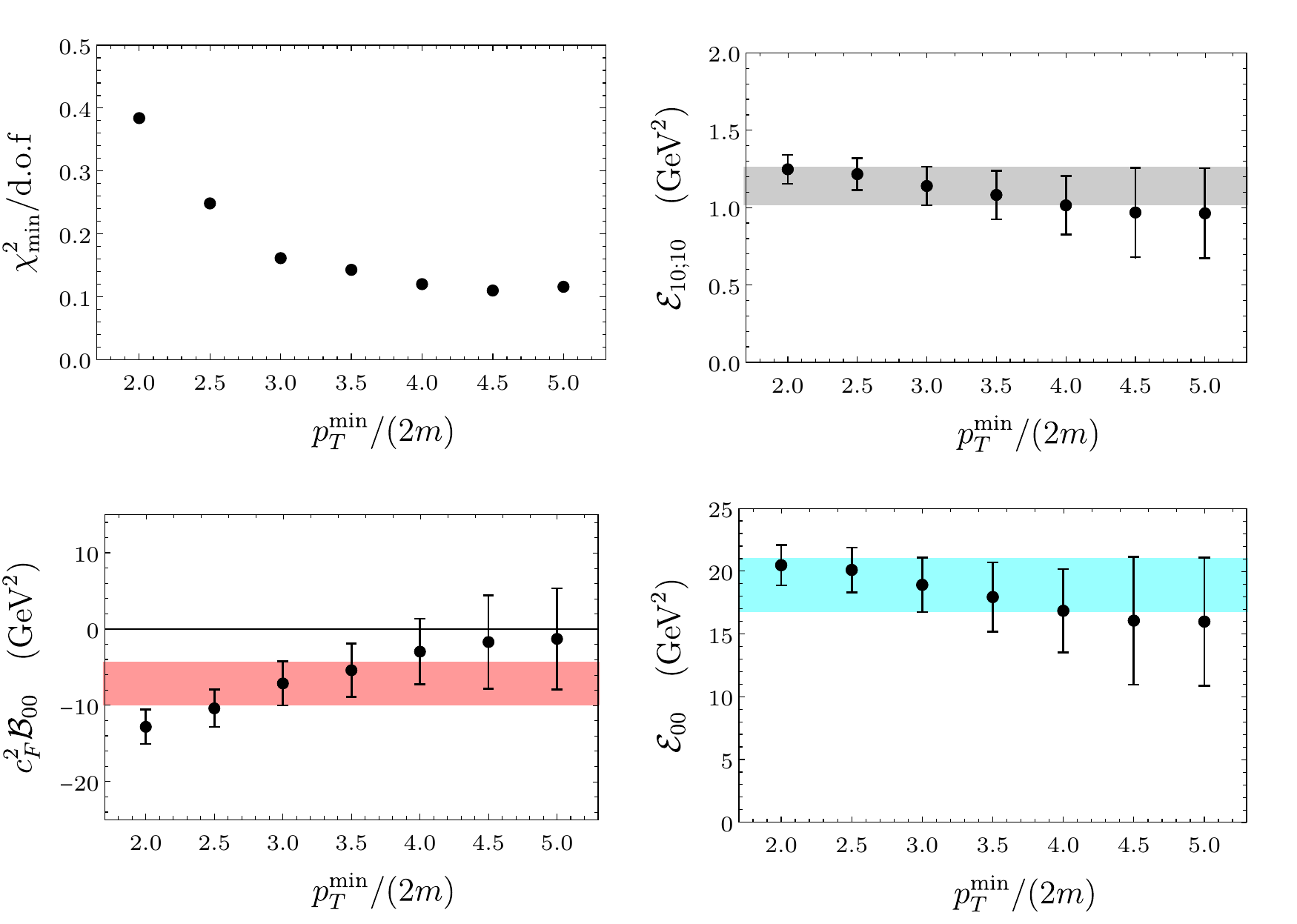}
\caption{\label{fig:ptmindep}
Dependence on the lower $p_T$ cut 
$p_T^{\rm min}$ of the $\chi^2_{\rm min}/{\rm d.o.f}$, 
and the values of ${\cal E}_{10;10}$, $c_F^2 {\cal B}_{00}$, and 
${\cal E}_{00}$ determined from fits to cross section data. 
The ${\cal E}_{10;10}$ and ${\cal B}_{00}$ are renormalized in the
$\overline{\rm MS}$ scheme at the scale $\Lambda = 1.5$~GeV, 
and $c_F$ is computed at the same scale with the charm quark mass $m_c
=1.5$~GeV. 
The bands represent the results of the fit for $p_T^{\rm min}/(2 m)=3$. 
}
\end{figure}

Because the NRQCD factorization formula in eq.~(\ref{eq:fac_jpsi}) 
is expected to hold as an expansion in powers
of $m/p_T$, we exclude measurements with $p_T < p_T^{\rm min}$ from the
fit, and vary $p_T^{\rm min}/(2 m)$ between 2 and 5. 
We perform least-square fits to the cross section data in refs.~\cite{CMS:2011rxs,
CMS:2015lbl, ATLAS:2012lmu}. 
The dependence on $p_T^{\rm min}$ of the values of 
$\chi^2_{\rm min}/{\rm d.o.f}$, as well as the fit values of 
${\cal E}_{10;10}$, $c_F^2 {\cal B}_{00}$, and ${\cal E}_{00}$ are shown in
fig.~\ref{fig:ptmindep}. 
We see that the quality of the fit improves with increasing $p_T^{\rm min}$, 
although $\chi^2_{\rm min}/{\rm d.o.f}$ is less than one for the whole range of
$p_T^{\rm min}$ that we consider. The individual values of the gluonic
correlators vary mildly as $p_T^{\rm min}$ increases, and are consistent within
uncertainties for $3<p_T^{\rm min}/(2 m)<5$. 

\begin{figure}[tbp]
\centering
\includegraphics[width=.95\textwidth]{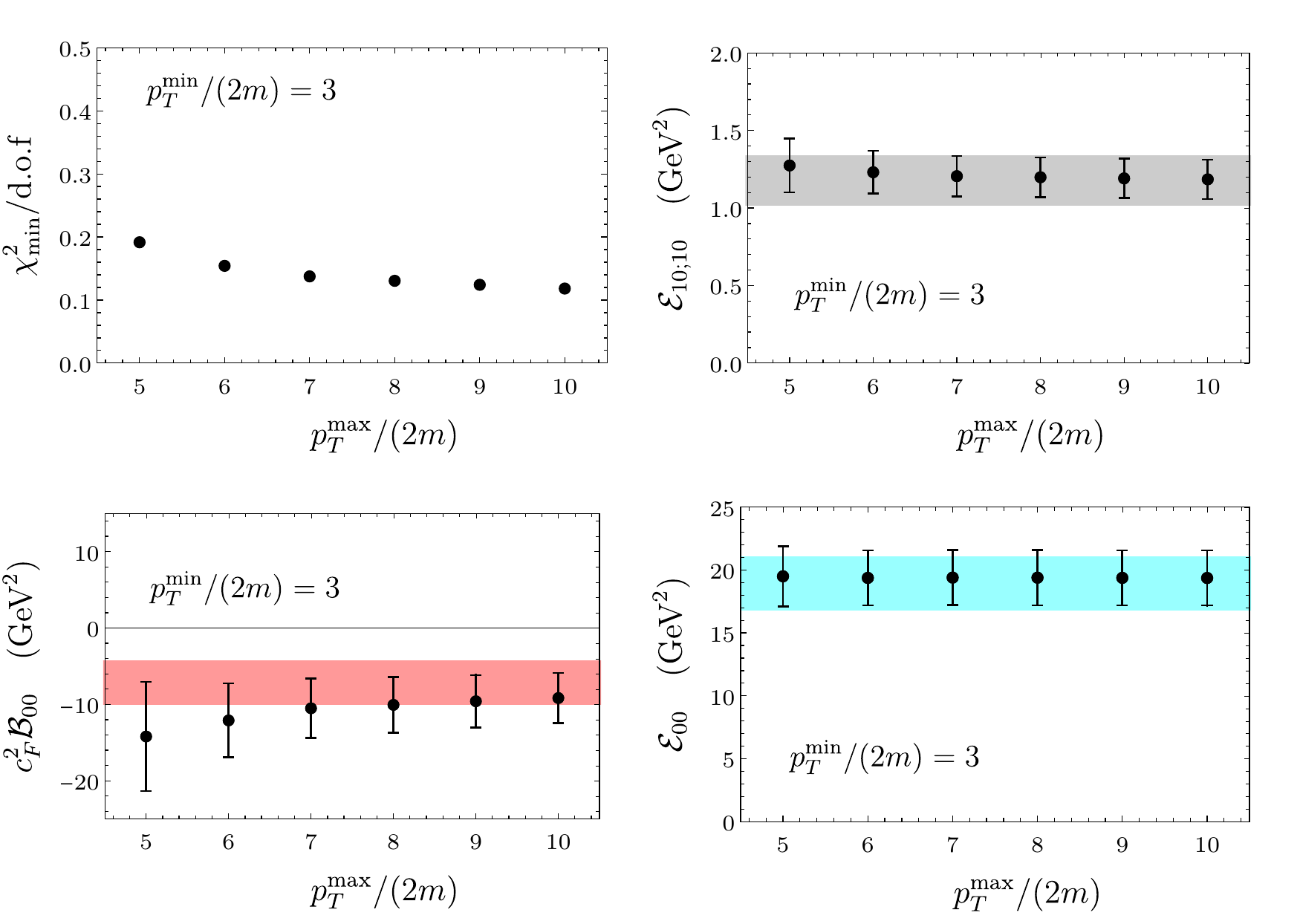}
\caption{\label{fig:ptmaxdep3}
Dependence on the upper $p_T$ cut 
$p_T^{\rm max}$ of the $\chi^2_{\rm min}/{\rm d.o.f}$,
and the values of ${\cal E}_{10;10}$, $c_F^2 {\cal B}_{00}$, and
${\cal E}_{00}$ determined from fits to cross section data with fixed
$p_T^{\rm min}/(2 m) = 3$.
The ${\cal E}_{10;10}$ and ${\cal B}_{00}$ are renormalized in the
$\overline{\rm MS}$ scheme at the scale $\Lambda = 1.5$~GeV,
and $c_F$ is computed at the same scale with the charm quark mass $m_c
=1.5$~GeV.
The bands represent the results of the fit for $p_T^{\rm min}/(2 m)=3$ and 
$p_T^{\rm max} = \infty$.
}
\end{figure}

\begin{figure}[tbp]
\centering
\includegraphics[width=.95\textwidth]{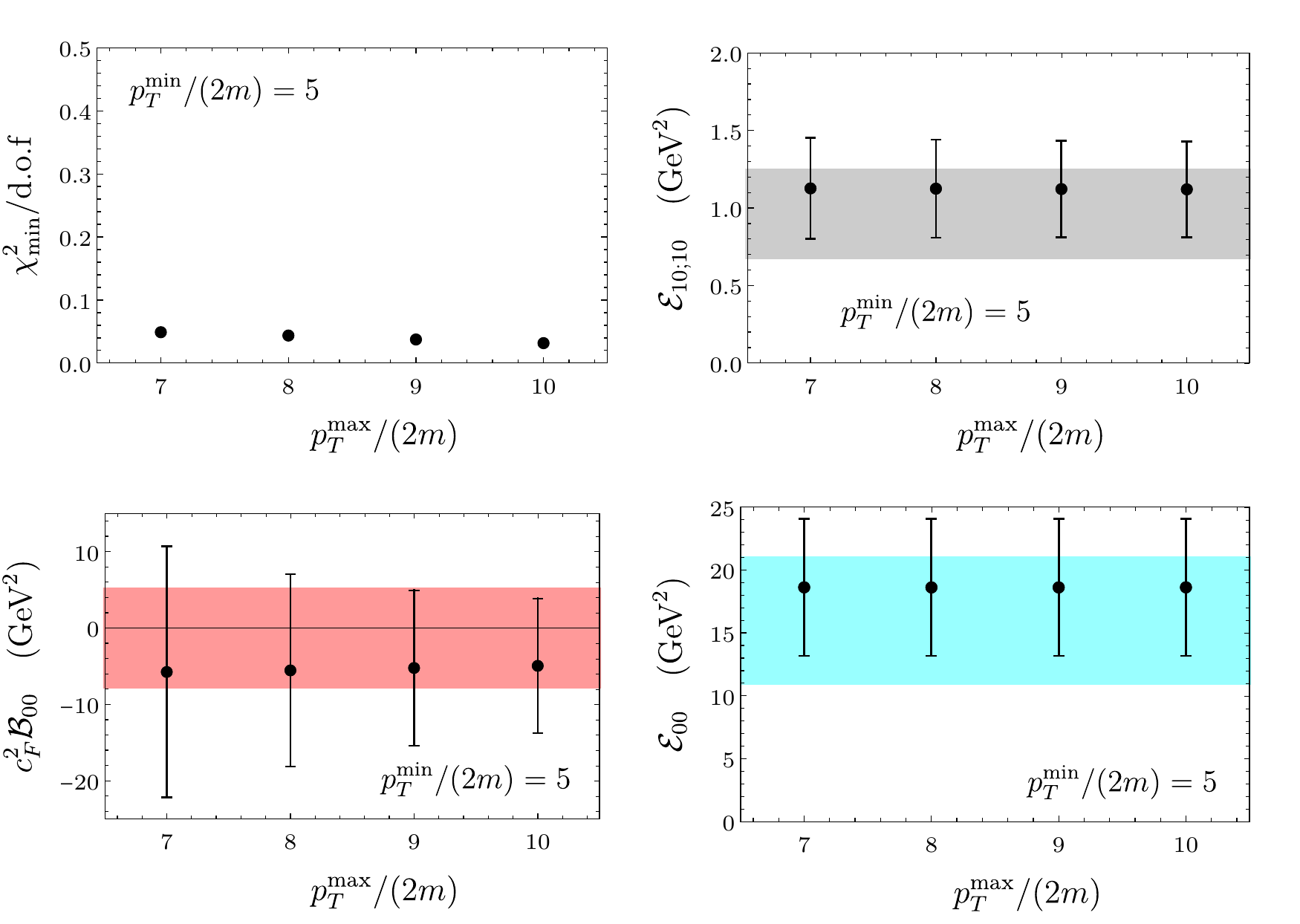}
\caption{\label{fig:ptmaxdep5}
Dependence on the upper $p_T$ cut
$p_T^{\rm max}$ of the $\chi^2_{\rm min}/{\rm d.o.f}$,
and the values of ${\cal E}_{10;10}$, $c_F^2 {\cal B}_{00}$, and
${\cal E}_{00}$ determined from fits to cross section data with fixed
$p_T^{\rm min}/(2 m) = 5$.
The ${\cal E}_{10;10}$ and ${\cal B}_{00}$ are renormalized in the
$\overline{\rm MS}$ scheme at the scale $\Lambda = 1.5$~GeV,
and $c_F$ is computed at the same scale with the charm quark mass $m_c
=1.5$~GeV.
The bands represent the results of the fit for $p_T^{\rm min}/(2 m)=5$ and
$p_T^{\rm max} = \infty$.
}
\end{figure}

We also consider the effect of a high $p_T$ cut, $p_T < p_T^{\rm max}$, 
because the fit may be affected by radiative corrections associated with
logarithms of $p_T/m$, which can become significant for large $p_T$. 
For this we fix $p_T^{\rm min}/(2 m) = 3$ and vary 
$p_T^{\rm max}/(2 m)$ between 5 and 10. The results of the fits with both low and
high $p_T$ cuts are shown in figure~\ref{fig:ptmaxdep3}. 
In all cases, the results with a high $p_T$ cut are consistent with what we
obtain with $p_T^{\rm max} = \infty$.  
We also show results of fits with fixed $p_T^{\rm min}/(2 m) = 5$
and $p_T^{\rm max}/(2 m)$ between 7 and 10 in figure~\ref{fig:ptmaxdep5}.
Similarly to the $p_T^{\rm min}/(2 m) = 3$ case, the results are consistent
with what we obtain with $p_T^{\rm max} = \infty$.

\begin{table}[tbp]
\centering
\begin{tabular}{|c||c|c|c|}
\hline
$p_T$ cut & ${\cal E}_{10;10}$ & $c_F^2 {\cal B}_{00}$ & ${\cal E}_{00}$ \\ 
\hline
\hline
$p_T/(2 m) > 3$ & $1.14 \pm 0.12$
& $-7.13 \pm 2.89$ & $18.9 \pm 2.16$
\\
\hline
$p_T/(2 m) > 5$ & $0.96 \pm 0.29$
& $-1.29 \pm 6.63$& $16.0 \pm 5.11$
\\
\hline
\end{tabular}
\caption{\label{tab:corrfitresults}
Fit results for the gluonic correlators 
${\cal E}_{10;10}$, $c_F^2 {\cal B}_{00}$, and 
${\cal E}_{00}$ in units of GeV$^2$ 
for $p_T$ cuts $p_T/(2 m) >3$ and $p_T/(2 m)>5$. 
The ${\cal B}_{00}$ and ${\cal E}_{00}$ are renormalized in the $\overline{\rm
MS}$ scheme at the scale $\Lambda = 1.5$~GeV, 
and $c_F$ is computed at the heavy quark mass $m = 1.5$~GeV and at the 
$\overline{\rm MS}$ scale $\Lambda = 1.5$~GeV. 
}
\end{table}

\begin{table}[tbp]
\centering
\begin{tabular}{|c|c||>{\centering}p{0.2\textwidth}|>{\centering}p{0.2\textwidth}|>{\centering\arraybackslash}p{0.2\textwidth}|}
\hline
$V$ & $p_T$ cut & $\langle {\cal O}^V(^3S_1^{[8]})\rangle$ & 
$\langle {\cal O}^V(^1S_0^{[8]})\rangle$ 
& $\langle {\cal O}^V(^3P_0^{[8]})\rangle/m^2$ 
\\
\hline
\hline
\multirow{2}{*}{$J/\psi$} 
& $p_T/(2 m) > 3$ & $1.66 \pm 0.18$ & $-3.47 \pm 1.41$ & $3.07 \pm 0.35$
\\
\cline{2-5}
& $p_T/(2 m) > 5$ & $1.40 \pm 0.42$ & $-0.63 \pm 3.22$ & $2.59 \pm 0.83$
\\
\hline
\hline
\multirow{2}{*}{$\psi(2S)$} 
& $p_T/(2 m) > 3$ & $0.99 \pm 0.11$ & $-2.07 \pm 0.84$ & $1.83 \pm 0.21$
\\
\cline{2-5}
& $p_T/(2 m) > 5$ & $0.84 \pm 0.25$ & $-0.37 \pm 1.92$ & $1.55 \pm 0.49$  
\\
\hline
\hline
\multirow{2}{*}{$\Upsilon(2S)$} 
& $p_T/(2 m) > 3$ & $1.79 \pm 0.20$ & $-1.12 \pm 0.46$ & $1.28 \pm 0.15$
\\
\cline{2-5}
& $p_T/(2 m) > 5$ & $1.52 \pm 0.47$ & $-0.20 \pm 1.04$ & $1.08 \pm 0.35$
\\
\hline
\hline
\multirow{2}{*}{$\Upsilon(3S)$} 
& $p_T/(2 m) > 3$ & $1.39 \pm 0.16$ & $-0.87 \pm 0.35$ & $0.99 \pm 0.11$
\\
\cline{2-5}
& $p_T/(2 m) > 5$ & $1.17 \pm 0.37$ & $-0.16 \pm 0.81$ & $0.84 \pm 0.27$
\\
\hline
\end{tabular}
\caption{\label{tab:LDMEfitresults}
Results for the color-octet LDMEs for $V=J/\psi$, $\psi(2S)$, 
$\Upsilon(2S)$, and $\Upsilon(3S)$ states 
computed from eqs.~(\ref{eq:LDMEs_summary}) and the fit results for the 
gluonic correlators in table~\ref{tab:corrfitresults},
in units of $10^{-2}$~GeV$^3$
for $p_T$ cuts $p_T/(2 m) >3$ and $p_T/(2 m)>5$. 
The $^3S_1^{[8]}$ LDME is renormalized in the $\overline{\rm MS}$ scheme at 
scale $\Lambda = m$. 
}
\end{table}

\begin{table}[tbp]
\centering
\begin{tabular}{|c|c||>{\centering}p{0.2\textwidth}|>{\centering}p{0.2\textwidth}|>{\centering\arraybackslash}p{0.2\textwidth}|}
\hline
$V$ & $p_T$ cut & $\langle {\cal O}^V(^3S_1^{[8]})\rangle$ & 
$\langle {\cal O}^V(^1S_0^{[8]})\rangle$ & $\langle {\cal
O}^V(^3P_0^{[8]})\rangle/m^2$ 
\\
\hline
\hline
\multirow{2}{*}{$J/\psi$} 
& $p_T/(2 m) > 3$ & $1.72 \pm 0.18$ & $-4.70 \pm 1.55$ & $3.14 \pm 0.35$
\\
\cline{2-5}
& $p_T/(2 m) > 5$ & $1.57 \pm 0.45$ & $-2.73 \pm 3.64$ & $2.89 \pm 0.87$
\\
\hline
\hline
\multirow{2}{*}{$\psi(2S)$} 
& $p_T/(2 m) > 3$ & $0.96 \pm 0.11$ & $-0.52 \pm 1.17$ & $1.80 \pm 0.21$
\\
\cline{2-5}
& $p_T/(2 m) > 5$ & $0.85 \pm 0.26$ & \phantom{$-$}$0.54 \pm 2.40$ & $1.58 \pm 0.50$  
\\
\hline
\hline
\multirow{2}{*}{$\Upsilon(2S)$} 
& $p_T/(2 m) > 3$ & $1.46 \pm 0.30$ & $-0.53 \pm 0.61$ & $1.04 \pm 0.22$
\\
\cline{2-5}
& $p_T/(2 m) > 5$ & $1.09 \pm 0.69$ & \phantom{$-$}$0.59 \pm 1.39$ & $0.77 \pm 0.50$
\\
\hline
\hline
\multirow{2}{*}{$\Upsilon(3S)$} 
& $p_T/(2 m) > 3$ & $1.52 \pm 0.20$ & $-1.11 \pm 0.42$ & $1.09 \pm 0.15$
\\
\cline{2-5}
& $p_T/(2 m) > 5$ & $1.15 \pm 0.45$ & $-0.13 \pm 0.95$ & $0.83 \pm 0.33$
\\
\hline
\end{tabular}
\caption{\label{tab:LDMEpredictions}
The color-octet LDMEs for $V=J/\psi$, $\psi(2S)$, $\Upsilon(2S)$, and $\Upsilon(3S)$ in units of $10^{-2}$~GeV$^3$ 
obtained by excluding cross section measurements of $V$ from fits, for $p_T$ cuts $p_T/(2 m) >3$ and $p_T/(2 m)>5$. 
The $^3S_1^{[8]}$ LDME is renormalized in the $\overline{\rm MS}$ scheme at scale $\Lambda = m$. 
}
\end{table}

In the phenomenological analysis in the following sections, 
we take the results of the fit from the ranges $p_T/(2 m) > 3$ and 
$p_T/(2 m) > 5$. The results for the gluonic correlators obtained 
from the fits are listed in table~\ref{tab:corrfitresults}.
The color-octet LDMEs for $J/\psi$, $\psi(2S)$, $\Upsilon(2S)$, and 
$\Upsilon(3S)$ states computed from eqs.~(\ref{eq:LDMEs_summary}) and
the results for the gluonic
correlators in table~\ref{tab:corrfitresults}
are shown in table~\ref{tab:LDMEfitresults}. 
These results differ slightly from a previous analysis in
ref.~\cite{Brambilla:2022rjd}, because we have improved the numerical accuracy 
of our calculation of the short-distance coefficients. 
We note that the results for the LDMEs in table~\ref{tab:LDMEfitresults}
satisfy the universal relations in eq.~(\ref{eq:LDMEs_universal_relations}) 
exactly once the evolution of the $^3S_1^{[8]}$ LDMEs is taken into account, 
because the relations (\ref{eq:LDMEs_universal_relations}) follow from 
the pNRQCD expressions for the LDMEs in eqs.~(\ref{eq:LDMEs_summary}).

The uncertainties in the gluonic correlators are correlated. The correlation
matrix of the uncertainties in ${\cal E}_{10;10}$,  ${\cal B}_{00}$, 
and  ${\cal E}_{00}$ are given by 
\begin{subequations}
\begin{eqnarray}
C_{p_T/(2 m) > 3} &=& 
\begin{pmatrix}
0.0153 & -0.308 & 0.267 \\
-0.308 & 8.35 & -5.17 \\
0.267 & -5.17 & 4.68
\end{pmatrix}
\textrm{~GeV$^4$}
,
\\
C_{p_T/(2 m) > 5} &=&  
\begin{pmatrix}
0.0846 & -1.68 & 1.48 \\
-1.68 & 44.0 & -28.6 \\
1.48 & -28.6 & 26.1
\end{pmatrix} 
\textrm{~GeV$^4$}. 
\end{eqnarray}
\end{subequations}
The normalized eigenvectors $v_n$ and eigenvalues $\lambda_n$ of the 
correlation matrices are given by (from the full precision correlation
matrix)
\begin{subequations}
\begin{eqnarray}
v_1 &=& \begin{pmatrix} 0.0338 \\ -0.816  \\ 0.577 \end{pmatrix}, \quad
v_2  =  \begin{pmatrix} 0.0387 \\ 0.578   \\ 0.815 \end{pmatrix}, \quad
v_3  =  \begin{pmatrix} 0.999  \\ 0.00520 \\ -0.0511 \end{pmatrix}, \quad
\\
\lambda_1 &=& 12.0~\textrm{~GeV$^4$}, \quad
\lambda_2 = 1.03~\textrm{~GeV$^4$}, \quad
\lambda_3 = 4.88 \times 10^{-5}~\textrm{~GeV$^4$},
\end{eqnarray}
\end{subequations}
for $p_T/(2 m) > 3$, and 
\begin{subequations}
\begin{eqnarray}
v_1 &=& \begin{pmatrix} 0.0343 \\ -0.805 \\ 0.592 \end{pmatrix}, \quad 
v_2 = \begin{pmatrix} 0.0400 \\ 0.593 \\ 0.804 \end{pmatrix}, \quad 
v_3 = \begin{pmatrix} 0.999 \\ 0.00392 \\ -0.0525 \end{pmatrix}, \quad 
\\
\lambda_1 &=& 65.1~\textrm{~GeV$^4$}, \quad 
\lambda_2 = 5.04~\textrm{~GeV$^4$}, \quad
\lambda_3 = 5.50 \times 10^{-5}~\textrm{~GeV$^4$}, 
\end{eqnarray}
\end{subequations}
for $p_T/(2 m) > 5$.
We note that the eigenvectors are almost insensitive to $p_T^{\rm min}$, 
while the eigenvalues depend on $p_T^{\rm min}$. 
The eigenvector $v_3$, which is the most strongly constrained, is almost purely 
the correlator ${\cal E}_{10;10}$. The eigenvectors $v_1$ and $v_2$ are 
mainly admixtures of ${\cal B}_{00}$ and ${\cal E}_{00}$, so that while 
the combination given by $v_2$ has a smaller uncertainty than $v_1$, 
the absolute uncertainties in 
${\cal B}_{00}$ and ${\cal E}_{00}$ are comparable in size. 

Thanks to the universal nature of the gluonic correlators, it is even possible
to predict the LDMEs for a specific $^3S_1$ quarkonium state from 
production rates of other quarkonia, without knowledge of the cross section 
data of that specific quarkonium. For example, predictions for $J/\psi$ color-octet 
LDMEs can be obtained from fits including only the $\psi(2S)$, $\Upsilon(2S)$, 
and $\Upsilon(3S)$ data, without using $J/\psi$ cross section data. 
We show the predictions for the 
$J/\psi$, $\psi(2S)$, $\Upsilon(2S)$, and $\Upsilon(3S)$ LDMEs obtained by
excluding that specific quarkonium state from the fits in
table~\ref{tab:LDMEpredictions}.
These results are consistent with the full fits in table~\ref{tab:LDMEfitresults} within uncertainties.

We note that our fits lead to stronger constraints for the LDMEs compared to
existing approaches based on hadroproduction data. 
Especially, both correlators ${\cal E}_{10;10}$ and ${\cal E}_{00}$ are
constrained to be positive, which leads to positive values of LDMEs 
$\langle {\cal O}^V (^3S_1^{[8]}) \rangle$ and 
$\langle {\cal O}^V (^3P_0^{[8]}) \rangle$. 
As we have stated
previously, because the large-$p_T$ cross section is in general given by a
linear combination of LP and NLP contributions, which behave like 
$d \sigma^{\textrm{LP}}/dp_T^2 \sim 1/p_T^4$ and 
$d \sigma^{\textrm{NLP}}/dp_T^2 \sim 1/p_T^6$, respectively, 
a fit from hadroproduction data of a single quarkonium can only strongly 
constrain two
linear combinations of LDMEs, and the remaining degree of freedom is poorly
determined. The fact that the NLO short-distance coefficients for the 
color-octet channels have an approximate degeneracy in their $p_T$ shapes has  
been pointed out in refs.~\cite{Ma:2010jj, Han:2014kxa}, where 
only two linear combinations of the color-octet LDMEs
were constrained, and the individual LDMEs left unconstrained\footnote{
In ref.~\cite{Shao:2014yta}, the authors determined upper and lower bounds for 
$\langle {\cal O}^V (^1S_0^{[8]}) \rangle$ by requiring the LDMEs 
$\langle {\cal O}^V (^3P_0^{[8]}) \rangle$ and
$\langle {\cal O}^V (^1S_0^{[8]}) \rangle$ to be both positive definite.}.
Similarly, the hadroproduction-based determination of $J/\psi$
LDMEs in ref.~\cite{Bodwin:2014gia} resulted in near 100\% uncertainties for 
$\langle {\cal O}^{J/\psi} (^3S_1^{[8]}) \rangle$ and
$\langle {\cal O}^{J/\psi} (^3P_0^{[8]}) \rangle$, which are strongly correlated. 
In contrast, in the pNRQCD case, the universality of the
gluonic correlators lets us employ both the charmonium and bottomonium data in
the fit, leading to stronger constraints. 
This happens because, while the $S$-wave charmonium cross section can be
described by different sets of LDMEs with different values of ${\cal E}_{00}$, 
different sets of charmonium LDMEs will lead to different predictions for 
the $\Upsilon$ cross sections, since the value of ${\cal E}_{10;10}$ for the
bottomonium case will depend through the running on the value of ${\cal E}_{00}$.

\begin{figure}[tbp]
\centering
\includegraphics[width=.48\textwidth]{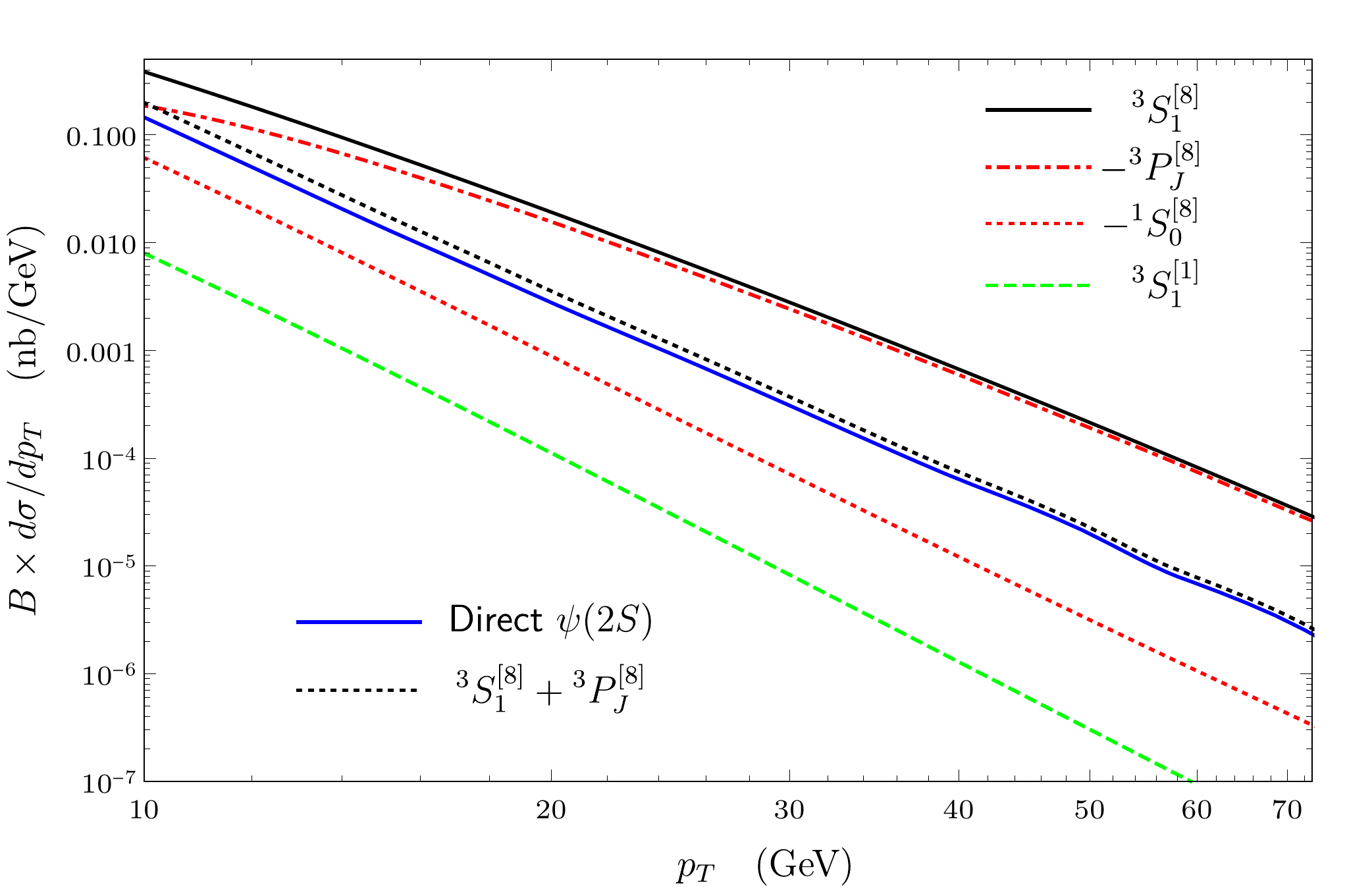}
\includegraphics[width=.48\textwidth]{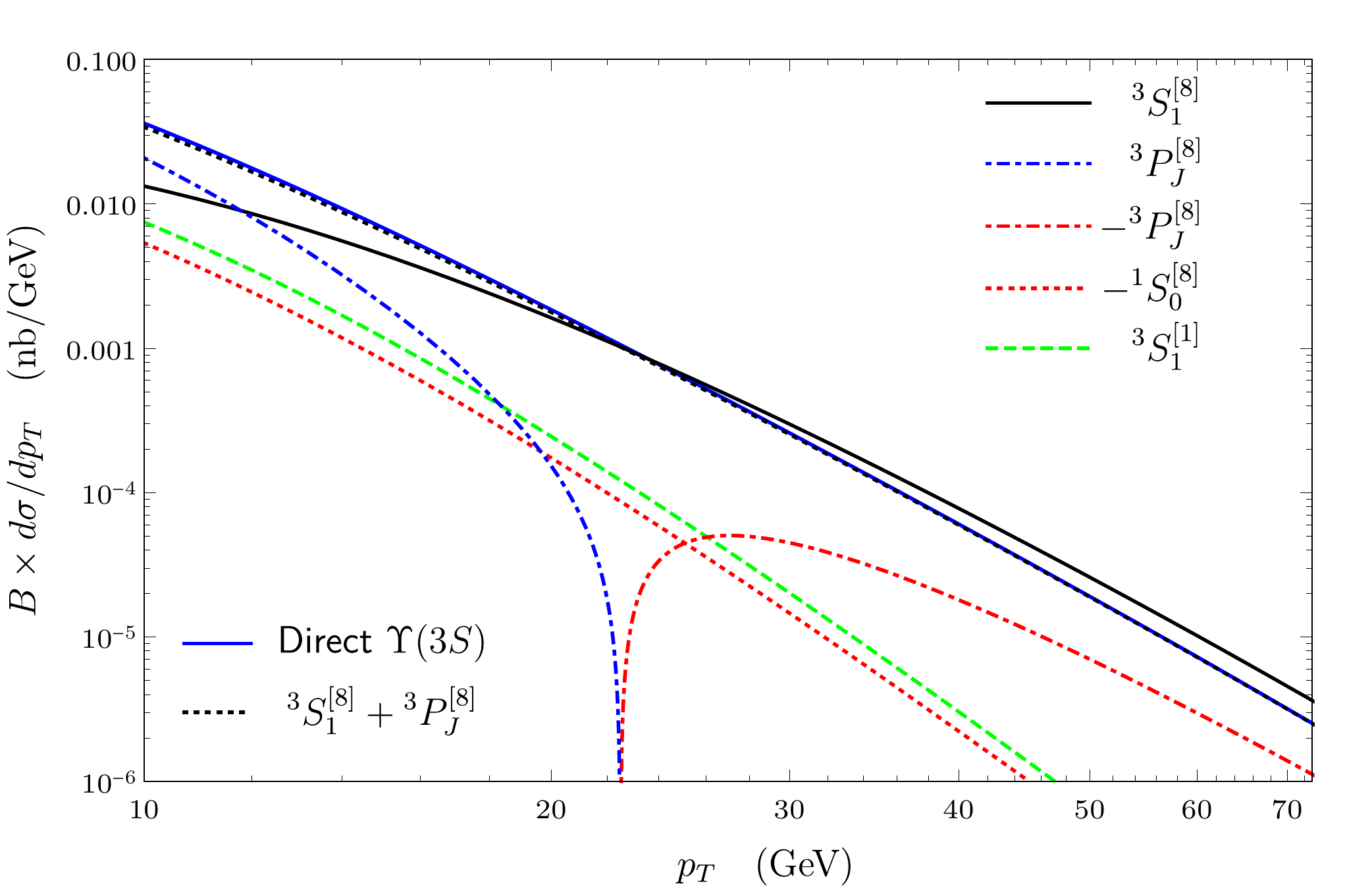}
\caption{\label{fig:channels}
Contributions from individual channels to the direct production rate of
$^3S_1$-wave charmonium (left) and bottomonium (right) at the LHC center of mass energy $\sqrt{s}=7$~TeV
integrated over the rapidity range $|y|<1.2$, computed with the LDMEs
determined from the fit with $p_T/(2 m) >3$. 
Here, $B$ is the branching fraction into a muon pair. 
Absolute values of negative contributions are shown in red. 
We also show the sum of $^3S_1^{[8]}$ and $^3P_J^{[8]}$ contributions
(black dotted lines), 
which make up for the bulk of the direct cross section (blue solid lines). 
}
\end{figure}

We show the contributions from each channel to direct charmonium and
bottomonium production cross sections in figure~\ref{fig:channels}. 
We see that at large $p_T$, the bulk of the direct cross section comes from the sum
of the $^3S_1^{[8]}$ and $^3P_J^{[8]}$ contributions, while the $^1S_0^{[8]}$
contribution is small. The color-singlet contribution is tiny\footnote{For the 
color singlet channel to contribute appreciably at large $p_T$, the gluon
fragmentation contribution must be included~\cite{Braaten:1993rw,
Braaten:1994xb}. In a fixed-order calculation, however, this occurs from
next-to-next-to-leading order, and is usually not included in NLO calculations.
Nonetheless, even after including gluon
fragmentation contributions, the color singlet contribution amounts to only
about 1\% of the large $p_T$ cross section at the LHC~\cite{Bodwin:2015iua}, 
and has negligible effects to our results.}. Because the
$^3S_1^{[8]}$ contribution is large and positive, while the $^3P_J^{[8]}$
contribution is large and negative, large cancellations occur in the sum of the
two channels. We note that while the LDME $\langle {\cal O}^V (^3S_1^{[8]})
\rangle$ and the short-distance coefficient $\hat{\sigma}_{Q
\bar{Q}(^3P_J^{[8]})}$ contain logarithms of the NRQCD factorization scale 
$\Lambda$ at one loop, the sum of the $^3S_1^{[8]}$ and $^3P_J^{[8]}$ 
contributions is independent of $\Lambda$. 
Since our fits strongly constrain ${\cal E}_{00}$ to be positive, 
${\cal E}_{10;10}$ takes a larger value at the scale of $m_b$ than its value at
the scale of $m_c$. Because of this, the cancellation between the 
$^3S_1^{[8]}$ and $^3P_J^{[8]}$ contributions is weaker in the bottomonium
case compared to the charmonium one.

We can compare our LDME determinations with existing results in the literature. 
Our results for charmonium and bottomonium are compatible with the partial 
determinations of two linear combinations of the three color-octet LDMEs in
ref.~\cite{Ma:2010jj, Han:2014kxa}. 
Interestingly, our charmonium results are similar to what was obtained
in refs.~\cite{Zhang:2014ybe, Han:2014jya} by using both $J/\psi$ and $\eta_c$ 
hadroproduction data based on heavy quark spin symmetry, 
as we also obtain small values of 
$\langle {\cal O}^{J/\psi} (^1S_0^{[8]}) \rangle$ and 
positive values for $\langle {\cal O}^{J/\psi} (^3S_1^{[8]}) \rangle$ and 
$\langle {\cal O}^{J/\psi} (^3P_0^{[8]}) \rangle$. 
However, the approach taken in 
refs.~\cite{Zhang:2014ybe, Han:2014jya} is very different from this work: 
in refs.~\cite{Zhang:2014ybe, Han:2014jya},
the upper and lower limits on $\langle {\cal O}^{J/\psi} (^1S_0^{[8]}) \rangle$ 
were obtained by applying eq.~(\ref{eq:fac_jpsi}) to 
$\eta_c$ production via heavy-quark spin symmetry and by making the assumption 
that $\langle {\cal O}^{J/\psi} (^1S_0^{[8]}) \rangle$ is 
positive definite, respectively, while our determination is based on the
universality of the gluonic correlators, and we do not rely on
the assumption of positivity. 
The charmonium results in refs.~\cite{Gong:2012ug, Bodwin:2014gia,
Bodwin:2015iua, Feng:2018ukp}, which are also based on $J/\psi$ and 
$\psi(2S)$ hadroproduction data, have same signs for the LDMEs 
$\langle {\cal O}^{J/\psi} (^3S_1^{[8]}) \rangle$ and
$\langle {\cal O}^{J/\psi} (^3P_0^{[8]}) \rangle$, which also leads to cancellations
between $^3S_1^{[8]}$ and $^3P_J^{[8]}$ contributions. 
However, the results in refs.~\cite{Gong:2012ug, Bodwin:2014gia, 
Bodwin:2015iua} involve large values of $\langle {\cal O}^{J/\psi} (^1S_0^{[8]})
\rangle$, so that the
direct cross section is dominated by the $^1S_0^{[8]}$ contribution. 
This is in contrast with our results, as we find the direct cross section to be
dominated by the sum of the $^3S_1^{[8]}$ and $^3P_J^{[8]}$ contributions. 
The global fit approach in refs.~\cite{Butenschoen:2011yh}, 
based on $J/\psi$ inclusive production data from $pp$, $p \bar p$, $ep$, and $e^+ e^-$ collider 
experiments, leads to a set of LDMEs where 
$\langle {\cal O}^{J/\psi} (^3P_0^{[8]}) \rangle$ is negative, while 
$\langle {\cal O}^{J/\psi} (^3S_1^{[8]}) \rangle$ and 
$\langle {\cal O}^{J/\psi} (^1S_0^{[8]}) \rangle$ 
are positive, so that every color-octet channel has a positive
contribution to the direct $J/\psi$ hadroproduction cross section at large $p_T$.
This leads to $p_T$-differential hadroproduction rates of $J/\psi$ that
are incompatible with LHC measurements at large $p_T$: 
the global fit in ref.~\cite{Butenschoen:2011yh} gives direct $J/\psi$ cross sections at the LHC 
that exceed the measured prompt cross sections in ref.~\cite{CMS:2015lbl} 
by more than a factor of 2 at $p_T = 30$~GeV, 
and by more than a factor of 3 at $p_T = 60$~GeV. 
The global fit of $\psi(2S)$ LDMEs in ref.~\cite{Butenschoen:2022orc} presented 
analyses with and without a lower $p_T$ cut given by $p_T > 7$~GeV. 
Unlike the global fit analysis of $J/\psi$ LDMEs, the available data for
$\psi(2S)$ employed in ref.~\cite{Butenschoen:2022orc} 
come only from hadroproduction in $pp$ and $p \bar p$ colliders. 
In their analysis without the $p_T$ cut, the $^3P_J^{[8]}$ LDME is negative,
similarly to the global fit of $J/\psi$ LDMEs, but once the data with 
$p_T < 7$~GeV are excluded from the fit, the $^3P_J^{[8]}$ LDME
turns positive. 
The quality of the fit also improves when the low $p_T$ data are excluded. 
The $\psi(2S)$ LDMEs in ref.~\cite{Butenschoen:2022orc} with the $p_T$ cut 
agree with our results for $p_T^{\rm min}/(2 m)> 5$ within uncertainties.

\subsection
[Production of $J/\psi$, $\psi(2S)$, and $\Upsilon$ at the LHC]
{\boldmath Production of $J/\psi$, $\psi(2S)$, and $\Upsilon$ at the LHC}
\label{sec:crosssection}

\begin{figure}[tbp]
\centering
\includegraphics[width=.48\textwidth]{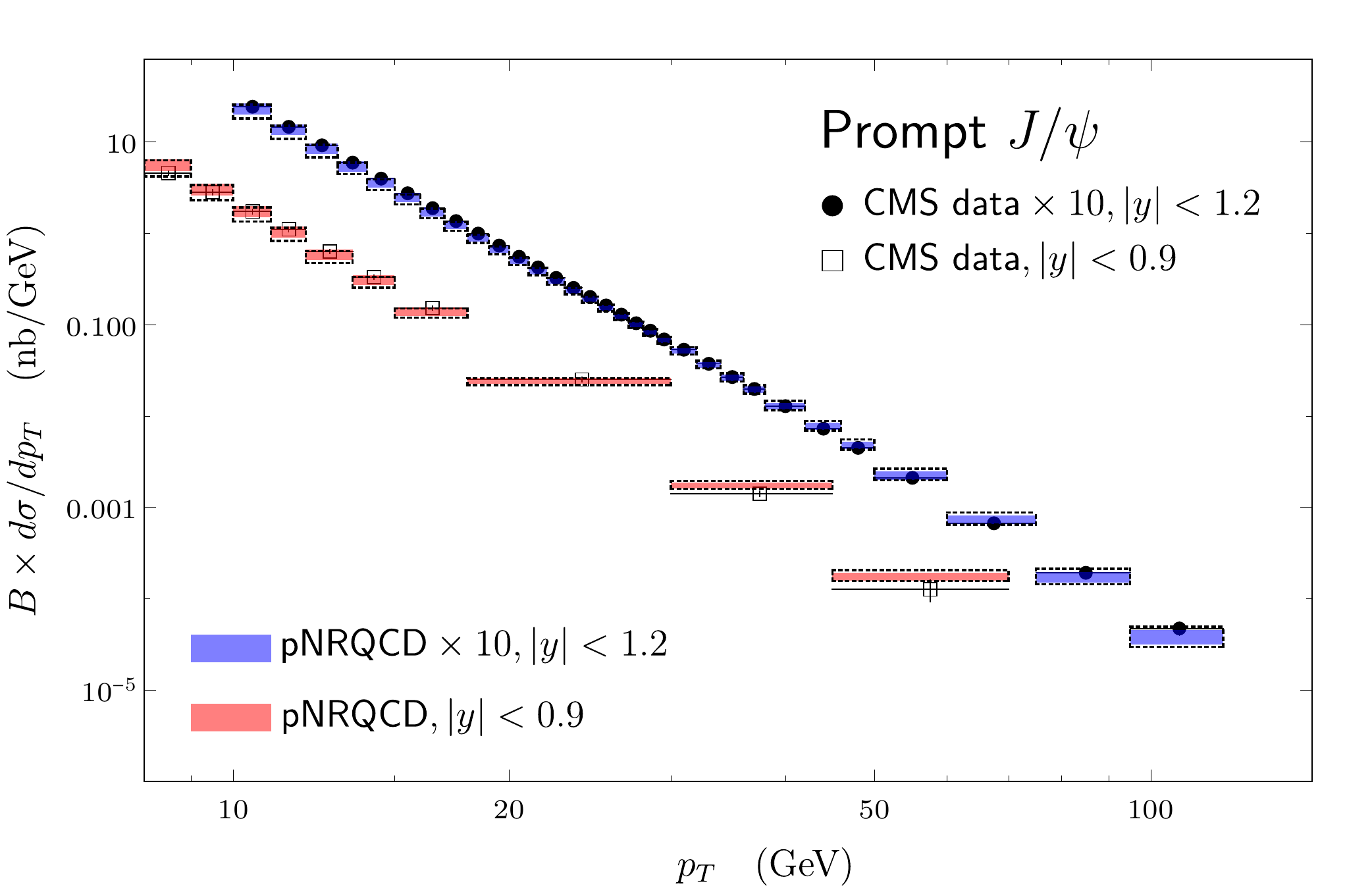}
\includegraphics[width=.48\textwidth]{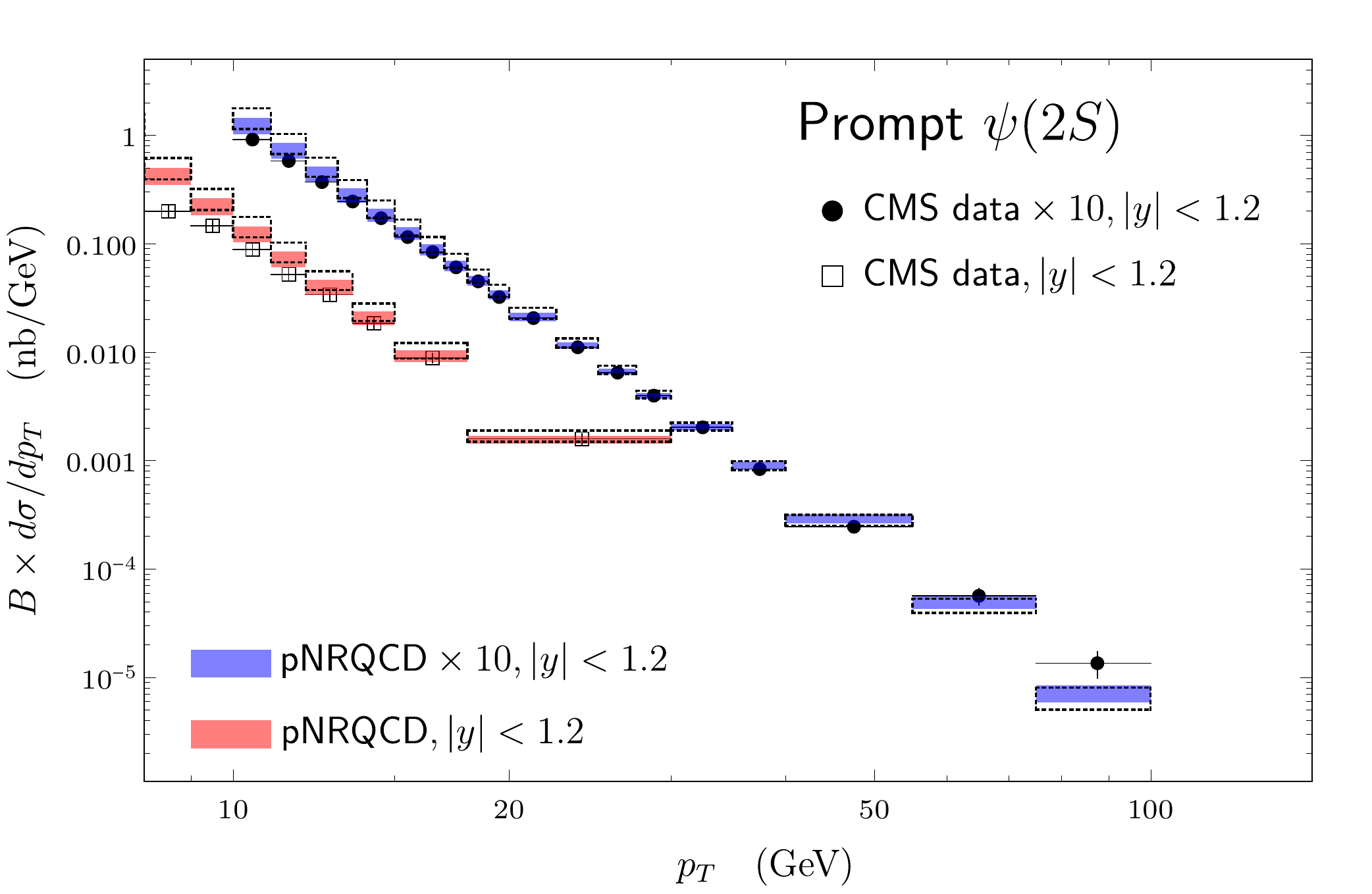}
\caption{\label{fig:psirates}
Production cross section of prompt $J/\psi$ and $\psi(2S)$ at the LHC center of mass energy $\sqrt{s}=7$~TeV
compared to CMS data~\cite{CMS:2011rxs, CMS:2015lbl}; $B$ is the dimuon branching fraction. 
Results from the LDMEs given in table~\ref{tab:LDMEpredictions} are shown as dotted outlined bands.
}
\end{figure}

\begin{figure}[tbp]
\centering
\includegraphics[width=.48\textwidth]{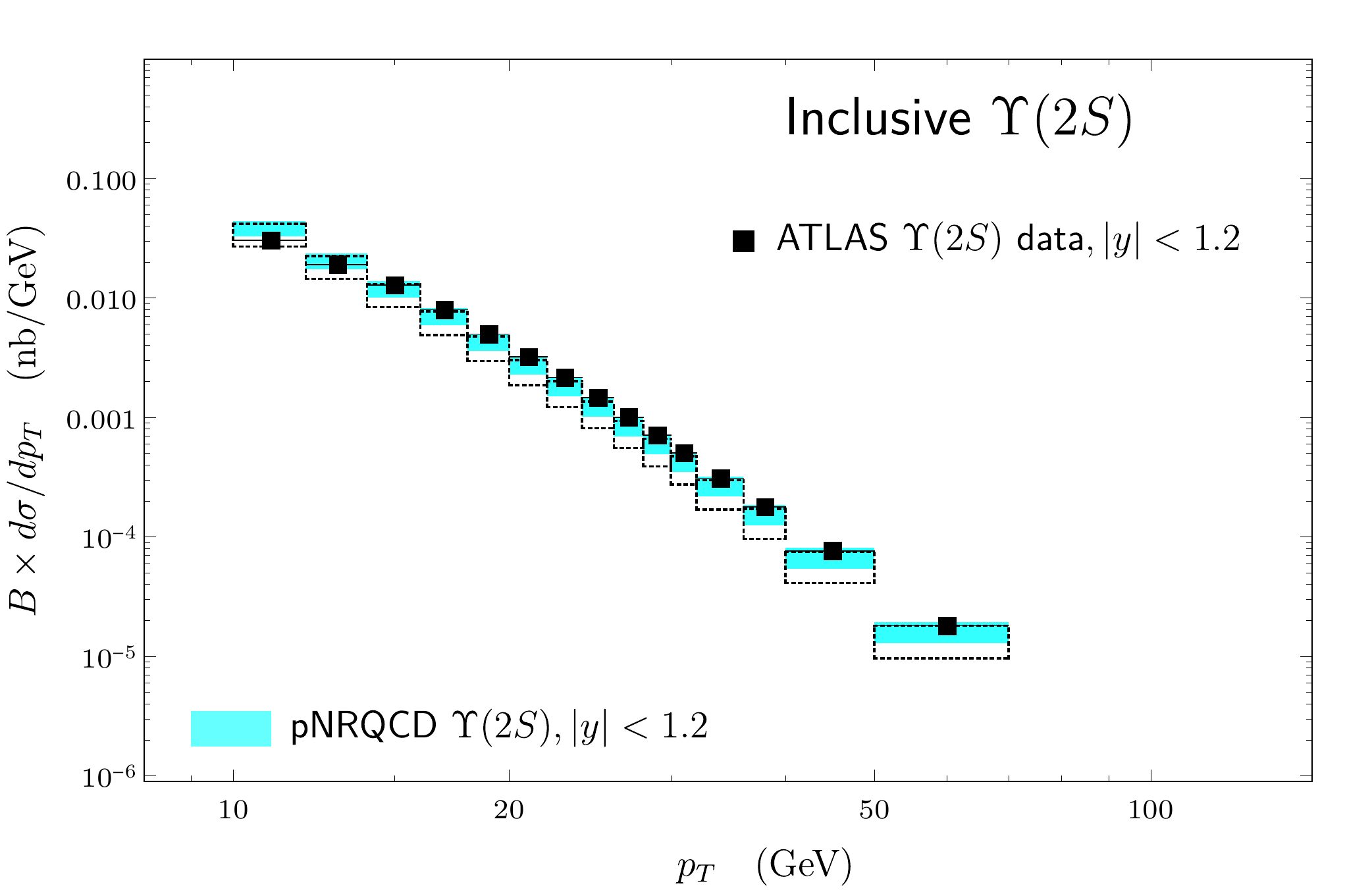}
\includegraphics[width=.48\textwidth]{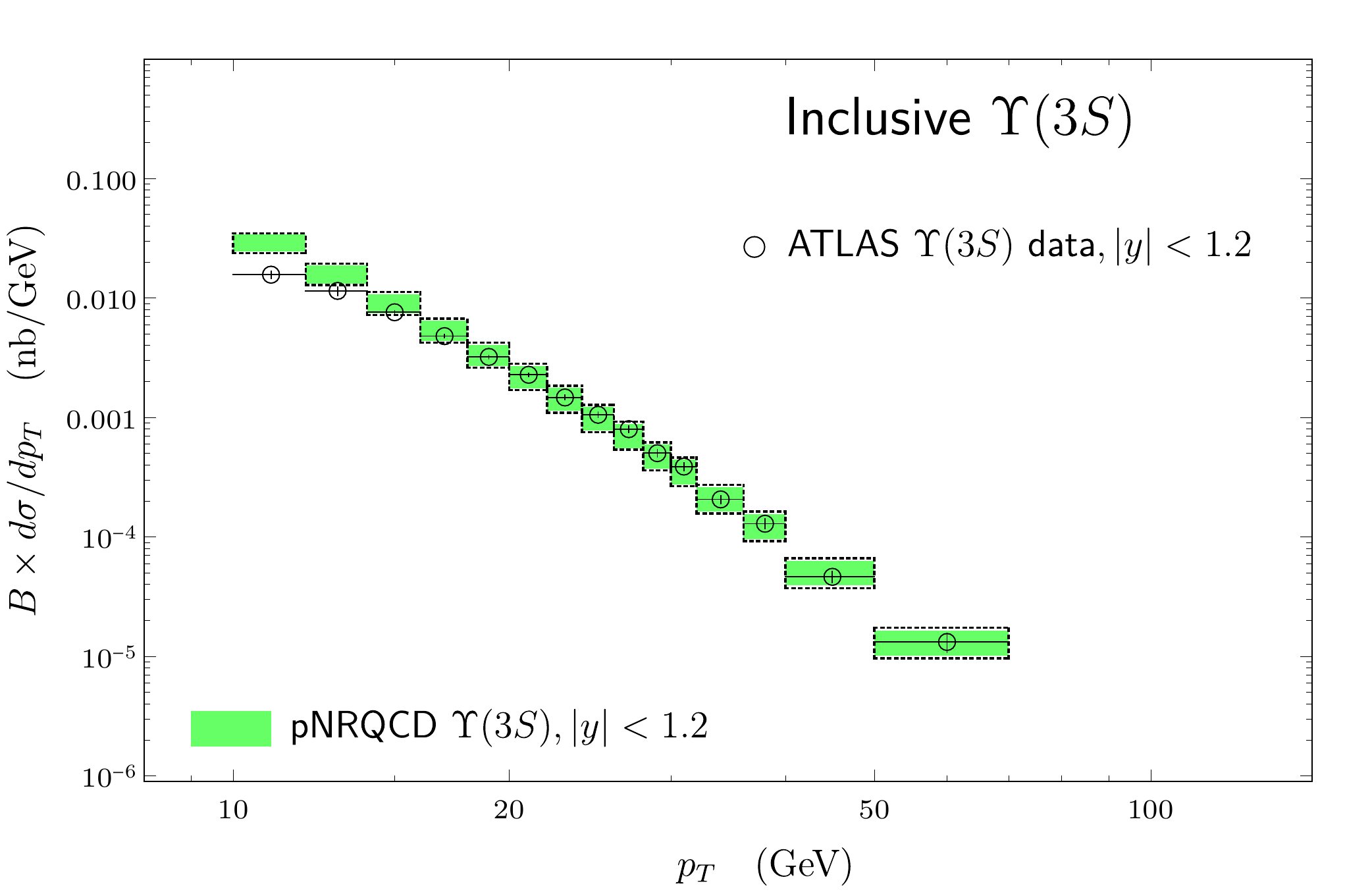}
\caption{\label{fig:upsrates}
Production cross section of inclusive $\Upsilon(2S)$ and $\Upsilon(3S)$ at the LHC center of mass energy $\sqrt{s}=7$~TeV
compared to ATLAS data~\cite{ATLAS:2012lmu}; $B$ is the dimuon branching fraction. 
Results from the LDMEs given in table~\ref{tab:LDMEpredictions} are shown as dotted outlined bands.
}
\end{figure}

We now show our results for the production cross sections of $J/\psi$, $\psi(2S)$, and
$\Upsilon$ at the LHC, based on the LDMEs determined in the previous section. 
Our results for the prompt $J/\psi$ and $\psi(2S)$ cross sections
are shown in figure~\ref{fig:psirates}, 
and the inclusive $\Upsilon(2S)$ and $\Upsilon(3S)$ cross sections are shown in
figure~\ref{fig:upsrates}, compared to CMS~\cite{CMS:2011rxs, CMS:2015lbl} 
and ATLAS measurements~\cite{ATLAS:2012lmu}. 
The theoretical uncertainties encompass the uncertainties in the LDMEs 
from the $p_T$ cuts $p_T/(2 m)>3$ and $p_T/(2 m)>5$. 
The pNRQCD results agree well with experiment, although there is some tension
in the highest and lowest $p_T$ bins. In the $\Upsilon(3S)$ case, the pNRQCD
results deviate from measurements at values of $p_T$ close to the $\Upsilon$
mass, which may signal a breakdown of the NRQCD factorization formalism given 
in the form of eq.~(\ref{eq:fac_jpsi}) at values of $p_T$ comparable to the
quarkonium mass. For the $\psi(2S)$, this already happens for $p_T \approx 10$~GeV. 
In figure~\ref{fig:psirates} and~\ref{fig:upsrates}, 
we also show results for the cross sections computed from the LDME determinations in 
table~\ref{tab:LDMEpredictions} as dotted outlined bands; the obtained cross
sections are consistent with the results of the full fit, which is a
strong indication that the pNRQCD approach is valid. 

\begin{figure}[tbp]
\centering
\includegraphics[width=.8\textwidth]{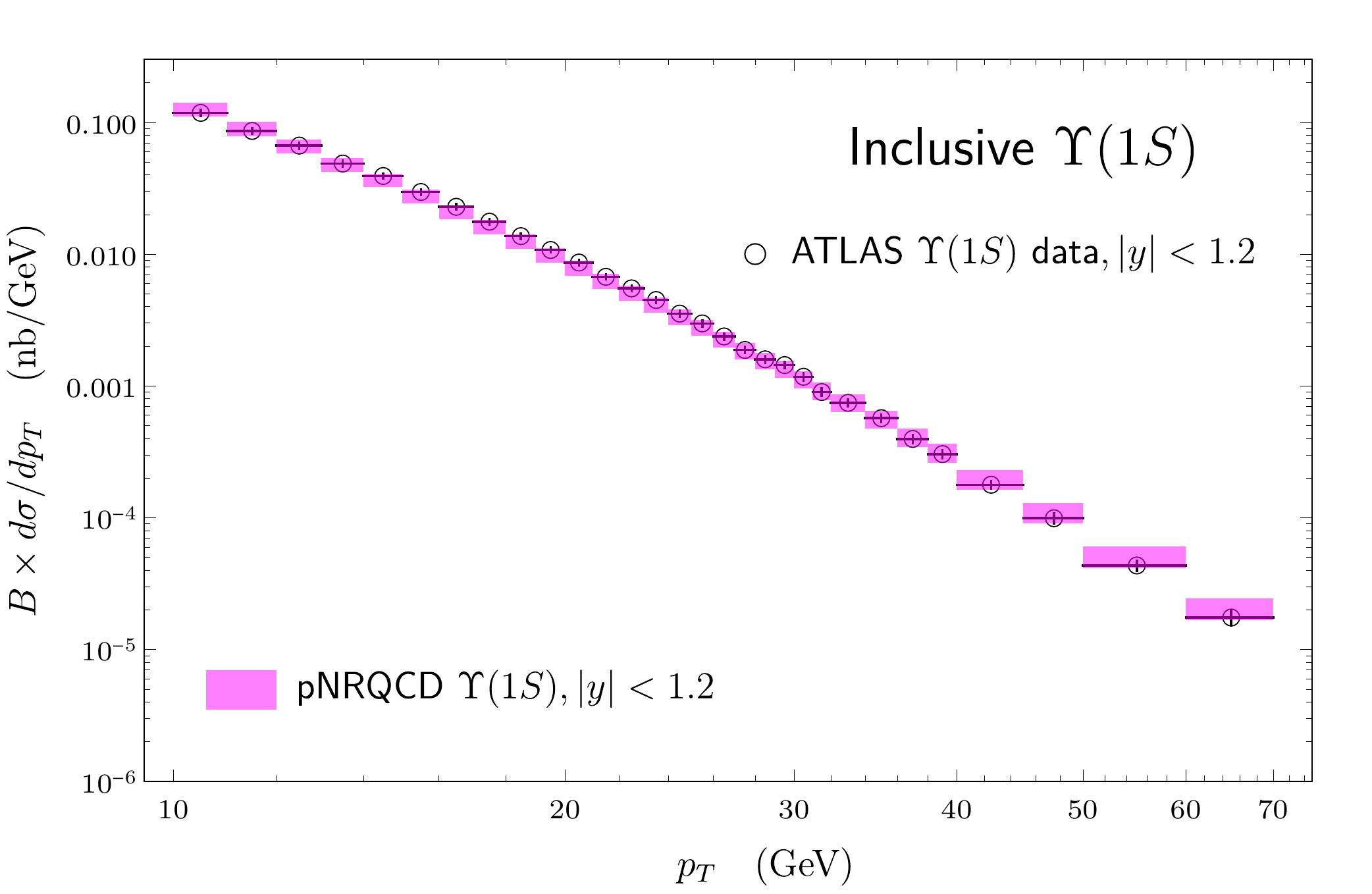}
\caption{\label{fig:ups1Srate}
Production cross section of inclusive $\Upsilon(1S)$ at the LHC center of mass energy $\sqrt{s}=7$~TeV
compared to ATLAS data~\cite{ATLAS:2012lmu}; $B$ is the dimuon branching fraction. 
The $\Upsilon(1S)$ LDMEs are computed from the $\Upsilon(2S)$ and $\Upsilon(3S)$ LDMEs through 
the universality relations in eqs.~(\ref{eq:LDMEs_universal_relations}). 
}
\end{figure}

Even though the $\Upsilon(1S)$ is likely to be a weakly coupled system, rather
than a strongly coupled one, it is still an interesting question whether the 
pNRQCD approach could explain the $\Upsilon(1S)$ production rate. 
We can compute the direct $\Upsilon(1S)$ cross sections under the assumption
that our calculations of the color-octet LDMEs is valid for the $1S$ state by
rescaling the direct $\Upsilon(nS)$ cross sections by a factor of 
$|R_{\Upsilon(1S)}^{(0)} (0)|^2/|R_{\Upsilon(nS)}^{(0)} (0)|^2$, 
where $n=2$ or 3. Then, we obtain the inclusive $\Upsilon(1S)$ cross section by
adding the feeddown contributions from $\Upsilon(2S)$ and $\Upsilon(3S)$ decays
into $\Upsilon(1S)$, and considering the feeddowns from $\chi_b(n'P)$ by using
the measured feeddown fractions $R_{\Upsilon(1S)}^{\chi_b(n'P)}$ with $n' = 1$,
2, and 3 from ref.~\cite{LHCb:2014ngh}. 
We use $|R_{\Upsilon(1S)}^{(0)} (0)|^2 = 6.75$~GeV$^3$, which we obtain from the measured decay rate into $e^+ e^-$. 
The pNRQCD results for the inclusive $\Upsilon(1S)$ cross section are 
shown in figure~\ref{fig:ups1Srate} compared to ATLAS data~\cite{ATLAS:2012lmu}.
We see that the pNRQCD prediction gives an excellent description of the inclusive $\Upsilon(1S)$ 
production rate at the LHC for a wide range of $p_T$, although our results may
not be reliable for values of $p_T$ comparable to the $\Upsilon(1S)$ mass, 
since the pNRQCD results overestimate the
$\Upsilon(3S)$ cross section for $p_T \approx m_{\Upsilon}$.

\subsection
[Polarization of $J/\psi$, $\psi(2S)$, and $\Upsilon$ at the LHC]
{\boldmath Polarization of $J/\psi$, $\psi(2S)$, and $\Upsilon$ at the LHC}
\label{sec:polarization}

In this section, we compute the polarization of $J/\psi$, $\psi(2S)$, and
$\Upsilon$ at the LHC. The polarization parameter $\lambda_\theta$ is defined by 
\begin{equation}
\label{eq:lamtheta_def}
\lambda_\theta = \frac{\sigma - 3 \sigma_L}{\sigma + \sigma_L}, 
\end{equation}
where $\sigma_L$ is the cross section for longitudinally produced quarkonium,
and $\sigma$ is the polariza\-tion-summed cross section. 
We can compute $\sigma_L$ by replacing the short-distance coefficients and
LDMEs in eq.~(\ref{eq:fac_jpsi}) by longitudinally polarized ones. 
If the produced quarkonium is totally transversely (longitudinally) polarized,
then $\lambda_\theta$ takes the value $+1$ ($-1$). The positivity of the
polarized cross sections gives the physical bounds $-1 < \lambda_\theta < 1$. 

While the polarized short-distance coefficients can be computed in perturbation
theory, the polarized LDMEs are a priori unknown, except for the 
polarized $^3S_1^{[1]}$ and $^3S_1^{[8]}$ LDMEs, which are given by 
$\langle {\cal O}^{V(\lambda)} (^3S_1^{[1]}) \rangle
= \frac{1}{3} \times \langle {\cal O}^{V} (^3S_1^{[1]}) \rangle$ and 
$\langle {\cal O}^{V(\lambda)} (^3S_1^{[8]}) \rangle
= \frac{1}{3} \times \langle {\cal O}^{V} (^3S_1^{[8]}) \rangle$,
because the contact terms $-V_{{\cal O} (^3S_1^{[1]})}$ and 
$\left. -V_{{\cal O} (^3S_1^{[8]})}\right|_{^3S_1}$ are isotropic.
On the other hand, the contact terms 
$-V_{{\cal O} (^3P_J^{[8]})}$ and $\left. -V_{{\cal O} (^1S_0^{[8]})} \right|_{^3S_1}$ 
depend on the tensors ${\cal E}_{00}^{ij}$ and ${\cal B}_{00}^{ij}$,
respectively, which contain gauge-completion Wilson lines in the $\ell$
direction. If the tensors ${\cal E}_{00}^{ij}$ and ${\cal B}_{00}^{ij}$ are not
isotropic, and instead develop a dependence on the direction $\ell$, 
then the polarized LDMEs $\langle {\cal O}^{V(\lambda)} (^3P_J^{[8]}) \rangle$ 
and $\langle {\cal O}^{V(\lambda)} (^1S_0^{[8]}) \rangle$ will also depend on 
the direction $\ell$. Since in the definitions of color-octet LDMEs the 
direction $\ell$ is arbitrary, it is in general not possible to obtain
polarization predictions if the polarized LDMEs are $\ell$ dependent. 
That is, for the NRQCD factorization formula to hold for polarized cross
sections, the LDMEs must be independent of the direction $\ell$ of the
gauge-completion Wilson line. In order to be able to make predictions for
quarkonium polarizations, we assume that the LDMEs are independent of $\ell$,
and take the polarized LDMEs to be 
$\langle {\cal O}^{V(\lambda)} (N) \rangle 
= \frac{1}{3} \times \langle {\cal O}^{V} (N) \rangle$
for all LDMEs appearing in eq.~(\ref{eq:fac_jpsi}). 
We note that this assumption has been taken implicitly in existing studies of
quarkonium polarizations based on NRQCD. 

We compute the polarized short-distance coefficients by using the FDCHQHP
package~\cite{Wan:2014vka}. In order to include feeddown effects, we also
compute the short-distance coefficients for the $P$-wave color singlet
channels. We note that, the short-distance coefficient for the $^3S_1^{[8]}$
channel is strongly transversely polarized, and has a small positive longitudinal
contribution, while the $^3P_J^{[8]}$ channel has a large negative transverse
contribution and a small positive longitudinal contribution. 
The short-distance coefficient for the $^1S_0^{[8]}$ channel is unpolarized.

The pNRQCD calculations of the LDMEs lead to two robust predictions for
polarizations of $^3S_1$ heavy quarkonia. First, thanks to the universal
relations in eqs.~(\ref{eq:LDMEs_universal_relations}), the polarization of
directly produced $^3S_1$ quarkonium is independent of the radial excitation,
because the wavefunction at the origin cancels in the definition of
$\lambda_\theta$ in eq.~(\ref{eq:lamtheta_def}), independently of the values of the gluonic correlators. 
Second, because the correlator ${\cal E}_{10;10}$ takes a larger value at the
scale of the bottom quark mass compared to the charmonium case due to its
running~[eq.~(\ref{eq:RG_E1010})], 
the directly produced $\Upsilon$ is more transverse than $J/\psi$ or
$\psi(2S)$ at comparable values of $p_T/m$.

\begin{figure}[ht]
\centering
\includegraphics[width=.95\textwidth]{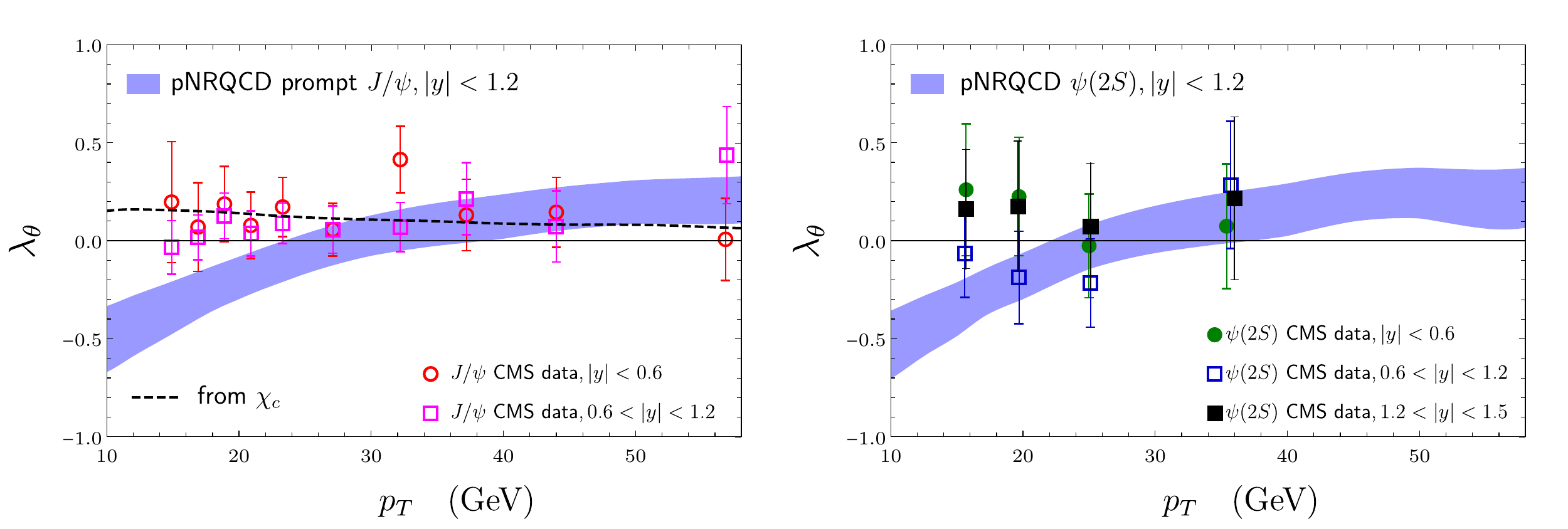}
\caption{\label{fig:psipol}
pNRQCD results for the polarization parameter $\lambda_\theta$ for prompt 
$J/\psi$ (left) and $\psi(2S)$ (right), compared to CMS
data~\cite{CMS:2013gbz}. 
The polarization of $J/\psi$ from $\chi_c$ decays is 
shown as a black dashed line. 
}
\end{figure}

We show the direct polarization of $\psi(2S)$ compared with 
CMS data from ref.~\cite{CMS:2013gbz} in figure~\ref{fig:psipol}. 
The theoretical uncertainties come from the uncertainties in the LDMEs, 
and encompass the two $p_T$ regions. 
The polarization parameter $\lambda_\theta$ of directly produced 
$\psi(2S)$ is negative at small $p_T$, and slowly rises with
increasing $p_T$. Since we neglect feeddown effects on $\psi(2S)$ production,
its direct polarization can be compared directly with measurements, which agree
with the pNRQCD result. The result for direct $\psi(2S)$ polarization is
slightly changed from the previous pNRQCD analysis in
ref.~\cite{Brambilla:2020ojz}, due to improved calculations of polarized
short-distance coefficients and small changes in the LDMEs. 
In the $J/\psi$ case, we consider the feeddowns
from $\psi(2S)$ and $\chi_c$. The feeddown from $\psi(2S)$ has little effect
on $J/\psi$ polarization, because the direct polarizations are same for
both states. We take the pNRQCD determinations of $\chi_c$ LDMEs in
ref.~\cite{Brambilla:2021abf} to
compute the polarization of $J/\psi$ produced in $\chi_c$ decays. The
polarization of prompt $J/\psi$, including effects of feeddowns from $\psi(2S)$
and $\chi_c$, is shown in figure~\ref{fig:psipol}, compared to CMS
data~\cite{CMS:2013gbz}. Our results are in fair agreement with measurements,
except for the smallest $p_T$ bins. 
The feeddowns from $\chi_c$ have little
effect on prompt $J/\psi$ polarization, because $J/\psi$ from $\chi_c$ decays 
is similarly polarized as directly produced $J/\psi$. 

As shown in figure~\ref{fig:psipol}, the pNRQCD results give values of
$\lambda_\theta$ for $J/\psi$ and $\psi(2S)$ that are positive but small at
large $p_T$, meaning that the transverse cross section is almost the same size
as the longitudinal cross section. In our case, this happens because the
large positive transverse cross section from the $^3S_1^{[8]}$ channel is
largely 
cancelled by the large negative transverse cross section from the $^3P_J^{[8]}$
channel; such cancellation does not occur in the 
longitudinal cross sections, because both channels have positive longitudinal
cross section contributions. We note that a similar mechanism for small
$\lambda_\theta$ has been suggested in refs.~\cite{Han:2014jya, Zhang:2014ybe}
based on hadroproduction data for $J/\psi$ and $\eta_c$ by using 
heavy quark spin symmetry. 
As it has been suggested in refs.~\cite{Bodwin:2014gia,
Faccioli:2014cqa, Bodwin:2015iua}, it is also possible to obtain small values of
$\lambda_\theta$ if the cross section is dominated by the $^1S_0^{[8]}$
channel, because the short-distance coefficient for this
channel is unpolarized.
The pNRQCD analysis disfavors this scenario.
In the case of the global fit of $J/\psi$ LDMEs in
ref.~\cite{Butenschoen:2011yh}, both the $^3S_1^{[8]}$ and $^3P_J^{[8]}$
channels have large positive transverse cross section contributions, 
because the $^3P_0^{[8]}$ LDME is negative, which results in values of 
$\lambda_\theta$ that are close to 1 at large $p_T$, 
which disagree with measurements.

\begin{figure}[tbp]
\centering
\includegraphics[width=.95\textwidth]{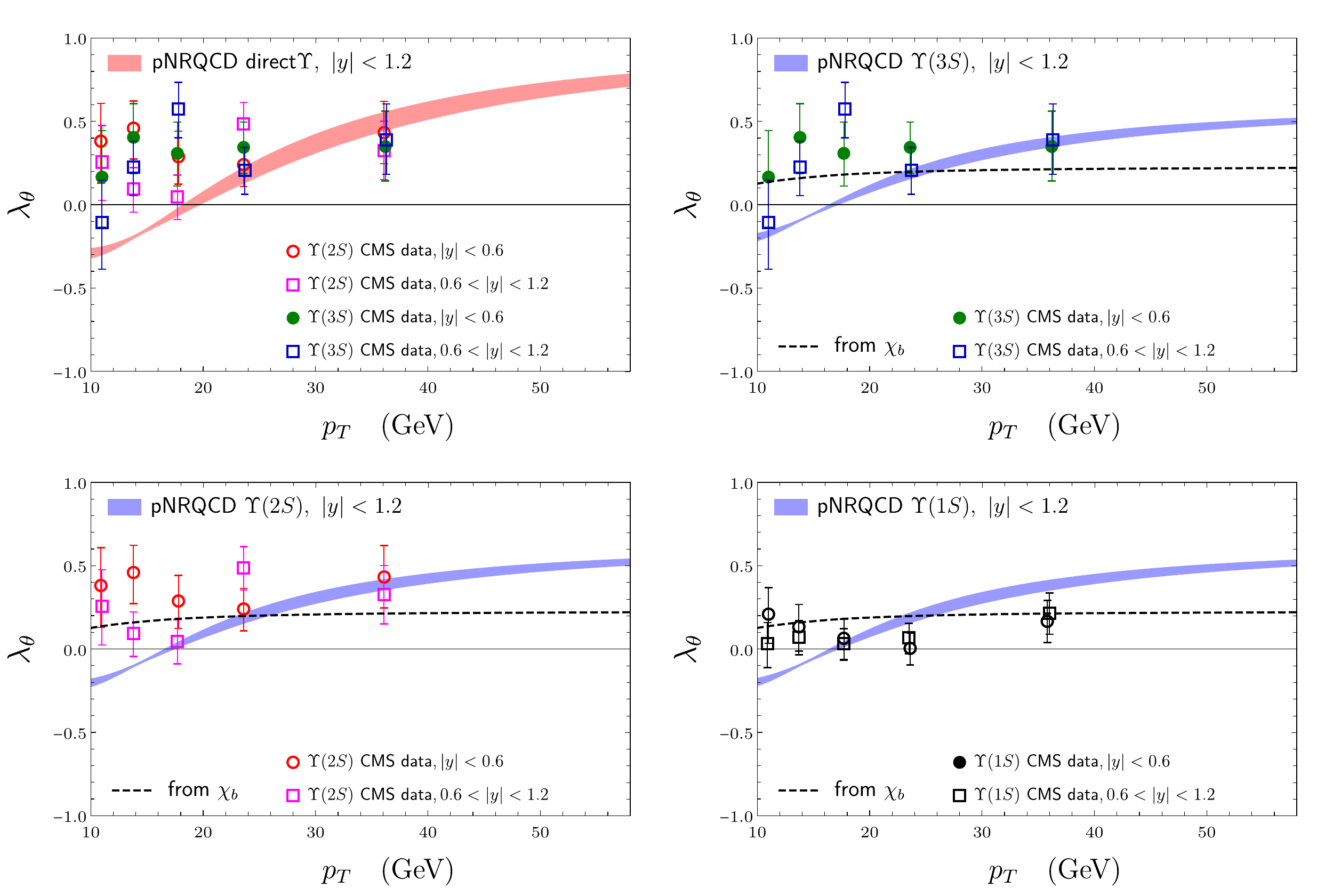}
\caption{\label{fig:upspol}
pNRQCD results for the polarization parameter $\lambda_\theta$ for 
directly produced $\Upsilon$ states (top left), 
inclusive $\Upsilon(3S)$ (top right), 
inclusive $\Upsilon(2S)$ (bottom left), and
inclusive $\Upsilon(1S)$ (bottom right), 
compared to CMS data~\cite{CMS:2012bpf}. 
The polarizations of $\Upsilon$ from $\chi_b$ decays are 
shown as black dashed lines. 
}
\end{figure}

We also show the direct polarization of $\Upsilon$ in figure~\ref{fig:upspol},
compared to the CMS measurements of $\Upsilon(2S)$ and $\Upsilon(3S)$ in ref.~\cite{CMS:2012bpf}. 
As we have done for the charmonium case, the theoretical uncertainties 
encompass the uncertainties in the LDMEs from the two $p_T$ regions. 
The result for direct $\Upsilon$ polarization is
slightly changed from the previous pNRQCD analysis in
ref.~\cite{Brambilla:2020ojz}, due to improved calculations of polarized
short-distance coefficients and small changes in the LDMEs.
As we have done for the cross sections, we take into account the effect of
feeddowns from $\chi_b$, as well as $\Upsilon(3S)$ decay into $\Upsilon(2S)$, 
by using the pNRQCD results for the $\chi_b$ LDMEs in ref.~\cite{Brambilla:2021abf}. 
We note that the polarization of $\Upsilon$ from decays of $\chi_b$ are almost
insensitive to the radial excitations; this happens because the $\chi_b$
production rate is dominated by the $^3S_1^{[8]}$ channel\footnote{
The pNRQCD results for the $\chi_b$ LDMEs in ref.~\cite{Brambilla:2021abf} provide
a natural explanation for the $^3S_1^{[8]}$ dominance in $\chi_b$ production:
since the scale-dependent gluonic correlator associated with the
$^3S_1^{[8]}$ LDME for the $\chi_b$ states grows with increasing factorization scale, 
the relative contribution from the $^3S_1^{[8]}$ channel is larger for $\chi_b$
compared to $\chi_c$.}, which yields similar values of $\lambda_\theta$ for 
$\Upsilon$ from decays of $\chi_{b1}$ and $\chi_{b2}$~\cite{Han:2014kxa}. 
Because of the feeddowns from $\chi_b$, the polarization parameter
$\lambda_\theta$ is smaller for inclusively produced $\Upsilon$, compared to direct production. 
The pNRQCD results for $\lambda_\theta$ of the inclusively produced
$\Upsilon(2S)$ and $\Upsilon(3S)$ are in good agreements with CMS data~\cite{CMS:2012bpf} at large $p_T$.
In figure~\ref{fig:upspol}, we also show our result for the $\Upsilon(1S)$ polarization compared to CMS data~\cite{CMS:2012bpf}, 
under the assumption that the pNRQCD analysis also applies to the $1S$ state.
Under this assumption, the direct polarization of $\Upsilon(1S)$ is the same as 
the one of $\Upsilon(2S)$ or $\Upsilon(3S)$, and we consider the effects of feeddowns
from $\Upsilon(2S)$, $\Upsilon(3S)$, and $\chi_b$. 
The result for $\Upsilon(1S)$ polarization is close to measurements, although 
the agreement with experiment is not as good as for $\Upsilon(2S)$ or $\Upsilon(3S)$. 

As we have argued previously, the value of $\lambda_\theta$ is larger for 
$\Upsilon$ compared to charmonium for comparable values of $p_T/m$, because the
correlator ${\cal E}_{10;10}$ takes a larger value at the scale of the bottom
quark mass compared to the charmonium case. This makes the cancellation between
the large positive transverse $^3S_1^{[8]}$ channel and the large negative 
transverse $^3P_J^{[8]}$ channel contributions not so strong as in the charmonium
case, cf. with figure~\ref{fig:channels}. 
That is, the pNRQCD analysis provides an explanation of the difference in the
behavior of $\lambda_\theta$ for charmonium and bottomonium.

\subsection[Photoproduction of $J/\psi$]
{\boldmath Photoproduction of $J/\psi$}
\label{sec:photo}

\begin{figure}[tbp]
\centering
\includegraphics[width=.8\textwidth]{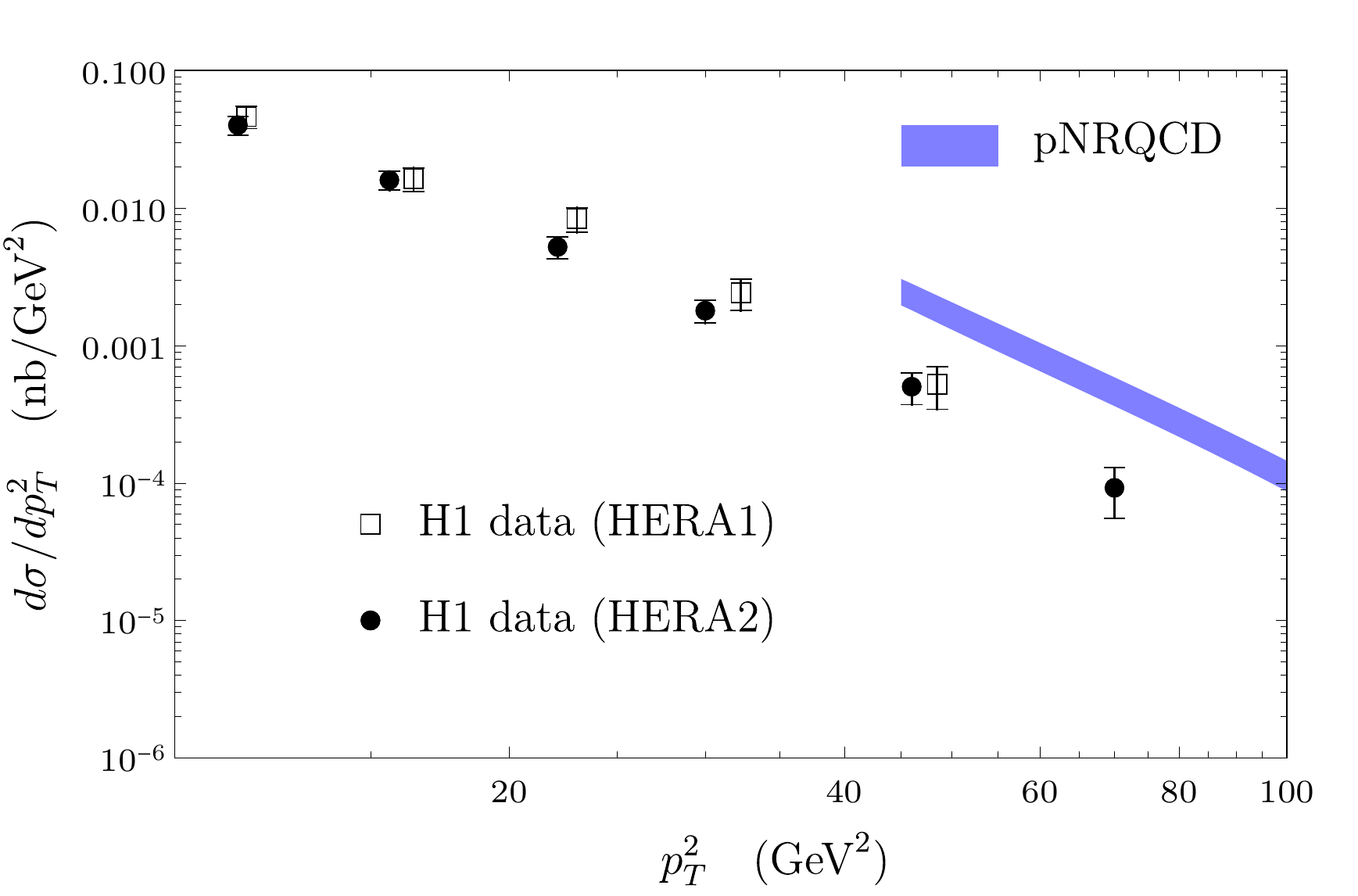}
\caption{\label{fig:jpsiphot}
Photoproduction cross section of $J/\psi$ compared to H1 data~\cite{H1:2002voc, H1:2010udv}. 
}
\end{figure}

In this section, we compute the $J/\psi$ production rate in $ep$ collisions. 
In order to compare with available data, we employ the kinematics used by the
H1 Collaboration for the measurement of the $p_T^2$-differential cross 
section~\cite{H1:2002voc, H1:2010udv}. 
That is, the center-of-mass energy of the $ep$ collision is 319~GeV, 
and kinematical cuts are made on the $\gamma p$ invariant mass $W = \sqrt{(p_\gamma+p_p)^2}$, 
elasticity $z = p_{J/\psi} \cdot p_p/p_\gamma \cdot p_p$, and the virtuality of
the photon $Q^2$ by setting $60\textrm{~GeV}<W<240\textrm{~GeV}$, 
$0.3<z<0.9$, and $Q^2 < 2.5$~GeV$^2$. 
Here the $p_\gamma$, $p_p$, and $p_{J/\psi}$ are the momentum of the photon
emitted by the electron, the momentum of the proton, and the momentum of the $J/\psi$, respectively. 
We employ the NLO short-distance coefficients computed in 
ref.~\cite{Butenschoen:2009zy}, which was
also adopted in ref.~\cite{Bodwin:2015yma}. 
The $p_T$-differential cross section of direct $J/\psi$ production computed by using our 
determination of LDMEs is shown in fig.~\ref{fig:jpsiphot} and compared with H1
data from refs.~\cite{H1:2002voc, H1:2010udv}. 
As we have done in the previous sections, the theoretical uncertainties
encompass the uncertainties in the LDMEs from the two $p_T$ regions. 
We see that our prediction for the direct $J/\psi$ cross section overshoots the measured
prompt cross section by more than a factor of 3 at the highest $p_T^2$ bin for
the H1 data from HERA 1, and more than a factor of 4 at the highest $p_T^2$ bin
for the H1 data from HERA 2. As it is expected that feeddown contributions 
will amount to about 15--20\% of the prompt cross section~\cite{H1:2010udv}, 
this discrepancy will increase once the effect of feeddowns are taken into account. 
In our case, the $^1S_0^{[8]}$ contribution is small, but the $^3P_J^{[8]}$
contribution is large and positive, because the short-distance coefficient 
for the $^3P_J^{[8]}$ channel is positive for photoproduction, unlike in the
hadroproduction case. 

It is worth noting that the kinematical constraints employed by the
experiments can make it difficult for NRQCD to give a precise description of
the photoproduction cross section: first, the $p_T$ of the $J/\psi$ is less than 10~GeV
at the highest $p_T^2$ bin, which is smaller than the $p_T$ cut that we have used
in our LDME determinations, so that nonperturbative effects that go beyond
next-to-leading power in the $m/p_T$ expansion and are unaccounted for in the
NRQCD factorization formula may become important. Second, the measurements
are made with kinematical cuts on the elasticity $z$, while in the calculation 
of the short-distance coefficients the elasticity is computed from 
the $Q \bar Q$ momentum instead of the $J/\psi$ momentum.
This introduces divergent distributions in $z$ that are strongly peaked near 
$z=1$ in the short-distance coefficients. Because for NRQCD factorization to
hold, the cross section must not depend strongly on small changes in $z$, 
NRQCD calculations are most reliable when the cross section is integrated over
a sufficiently inclusive region of $z$ that includes $z=1$. 
A kinematical cut on the maximum value of $z$ can make 
the cross section sensitive to changes in the $Q \bar Q$ momentum smaller than
the order of the heavy quark mass, and make the NRQCD calculation unreliable. 
This issue has already been pointed out in ref.~\cite{Beneke:1998re}.

\subsection[Hadroproduction of $\eta_c$]
{\boldmath Hadroproduction of $\eta_c$}
\label{sec:etac}

As we have shown in section~\ref{sec:HQSS}, our pNRQCD results for the LDMEs are compatible
with heavy quark spin symmetry, so that our determinations of the $J/\psi$ LDMEs
also lead to determinations of the $\eta_c$ LDMEs. By using heavy quark spin
symmetry, refs.~\cite{Butenschoen:2014dra, Han:2014jya, Zhang:2014ybe} 
employed the following NRQCD factorization formula 
\begin{eqnarray}
\label{eq:fac_etac}
\sigma_{\eta_c+X} &=&
\hat{\sigma}_{Q \bar{Q}(^1S_0^{[1]})}
\langle {\cal O}^{\eta_c}(^1S_0^{[1]}) \rangle
+ \hat{\sigma}_{Q \bar{Q}(^3S_1^{[8]})}
\langle {\cal O}^{\eta_c}(^3S_1^{[8]}) \rangle
\nonumber\\ &&
+ \hat{\sigma}_{Q \bar{Q}(^1S_0^{[8]})}
\langle {\cal O}^{\eta_c}(^1S_0^{[8]}) \rangle
+ \hat{\sigma}_{Q \bar{Q}(^1P_1^{[8]})}
\langle {\cal O}^{\eta_c}(^1P_1^{[8]}) \rangle, 
\end{eqnarray}
and the heavy-quark spin symmetry relations 
$\langle {\cal O}^{\eta_c}(^1S_0^{[1]}) \rangle
= \langle \Omega | {\cal O}^{J/\psi}(^3S_1^{[1]}) \rangle/3$, 
$\langle {\cal O}^{\eta_c}(^3S_1^{[8]}) \rangle$
$=$ $\langle {\cal O}^{J/\psi}(^1S_0^{[8]}) \rangle$, 
$\langle {\cal O}^{\eta_c}(^1S_0^{[8]})  \rangle
= \langle {\cal O}^{J/\psi}(^3S_1^{[8]}) \rangle/3$, 
and $\langle {\cal O}^{\eta_c}(^1P_1^{[8]}) \rangle
= 3 \times \langle {\cal O}^{J/\psi}(^3P_0^{[8]}) \rangle$ 
to compute the $\eta_c$ production rate from determinations of the $J/\psi$ LDMEs. 
An important caveat of this approach is that the factorization formula in 
eq.~(\ref{eq:fac_jpsi}) for $J/\psi$ production holds
when the color-singlet contribution from the $^3S_1^{[1]}$ channel is small,
compared to color-octet contributions. This does not necessarily hold for the
$\eta_c$ case: at values of $p_T$ where LHC measurements of the $\eta_c$ cross 
section are available, the short-distance coefficient for the $^1S_0^{[1]}$
channel is not suppressed compared to the $^1S_0^{[8]}$ and 
$^1P_1^{[8]}$ channels. While the short-distance coefficient for the 
$^3S_1^{[8]}$ channel is still enhanced by the gluon fragmentation 
contribution, $\langle {\cal O}^{\eta_c}(^1S_0^{[8]})  \rangle$ is
suppressed by powers of $v$ compared to the color-singlet LDME, 
so the $^3S_1^{[8]}$ contribution to the cross section is at best comparable 
to the color-singlet contribution. 
In this case, relativistic corrections to the color-singlet channel may be
important, similarly to what we see in NRQCD calculations of exclusive
production rates~\cite{Bodwin:2007ga, Sang:2009jc,
Fan:2012dy, Chung:2019ota}. 

Given the aforementioned limitations, we may still expect
eq.~(\ref{eq:fac_etac}) to give at least an estimate for the 
$\eta_c$ production rate at hadron colliders. 
We take the short-distance coefficients $\hat{\sigma}_{Q \bar{Q}(^1S_0^{[1]})}$
and $\hat{\sigma}_{Q \bar{Q}(^1P_1^{[8]})}$ given in ref.~\cite{Han:2014jya}, 
and use heavy-quark spin symmetry to compute the $\eta_c$ LDMEs from our
determinations of $J/\psi$ LDMEs. 
Because $\hat{\sigma}_{Q \bar{Q}(^1S_0^{[8]})}$ and 
$m^2 \hat{\sigma}_{Q \bar{Q}(^1P_1^{[8]})}$ are not enhanced compared to 
$\hat{\sigma}_{Q \bar{Q}(^1S_0^{[1]})}$, the contributions from the
$^1S_0^{[8]}$ and $^1P_1^{[8]}$ channels amount to less than 15\% of the
color-singlet contribution, which is smaller than the typical size of
relativistic corrections of relative order $v^2$ expected from 
velocity-scaling rules of NRQCD. 
Hence, the $\eta_c$ production rate computed from eq.~(\ref{eq:fac_etac}) is
dominated by the sum of the $^1S_0^{[1]}$ and $^3S_1^{[8]}$ contributions. 
As it has been pointed out in refs.~\cite{Butenschoen:2014dra, Han:2014jya,
Zhang:2014ybe}, the color-singlet contribution is already
comparable to the measured $p_T$-differential cross section. 
Because of this, LDME determinations where the $J/\psi$ production rate is
dominated by the $^1S_0^{[8]}$ channel, such as the results in 
refs.~\cite{Gong:2012ug, Bodwin:2014gia, Bodwin:2015iua}, 
give large positive $^3S_1^{[8]}$ contributions to the $\eta_c$ production
rate, which then lead to overestimations of the cross
section~\cite{Butenschoen:2014dra}.  In contrast, our LDME determinations give
small, or even negative values of $\langle {\cal O}^{J/\psi}(^1S_0^{[8]})\rangle$. 

\begin{figure}[ht]
\centering
\includegraphics[width=.48\textwidth]{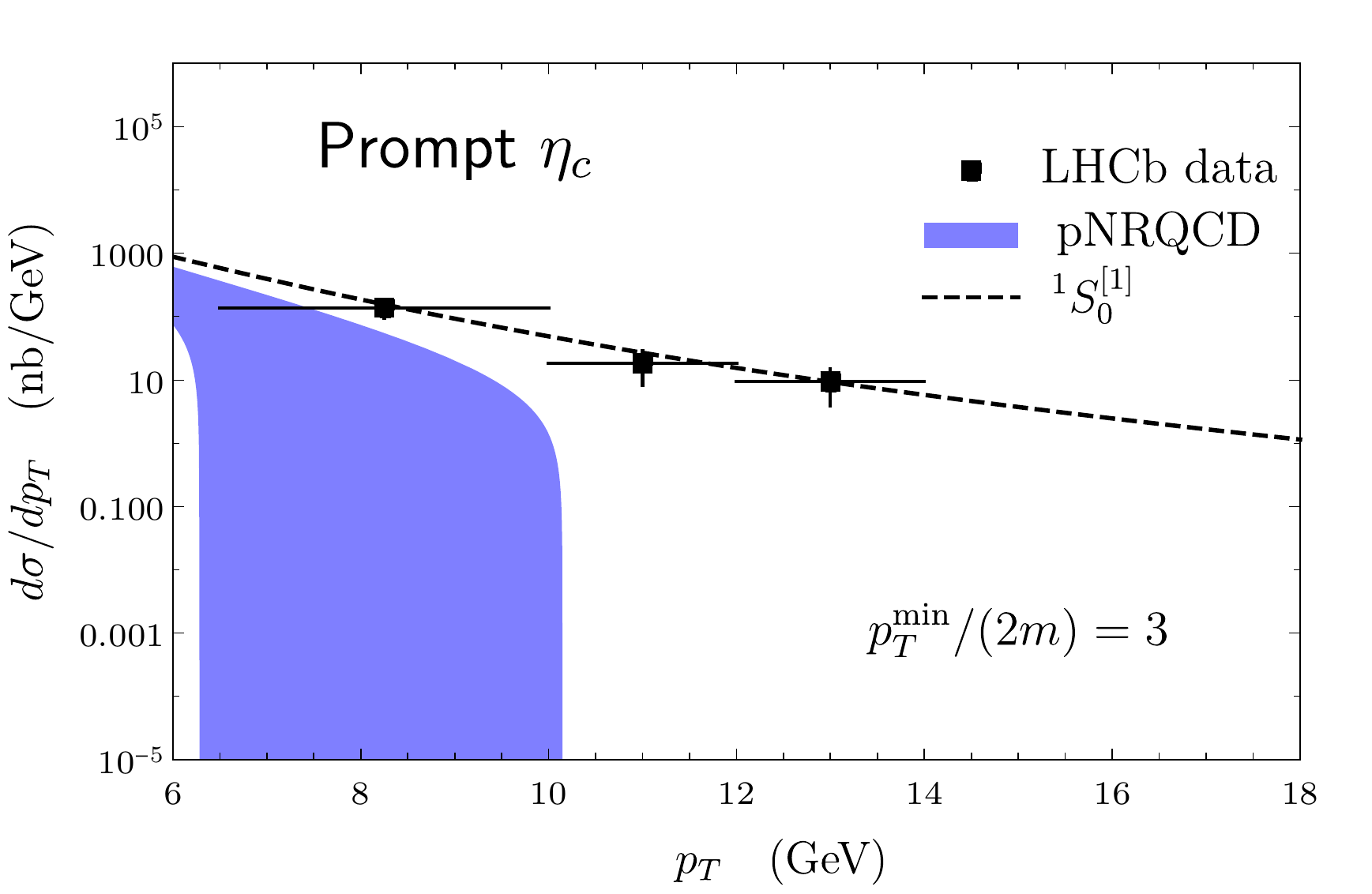}
\includegraphics[width=.48\textwidth]{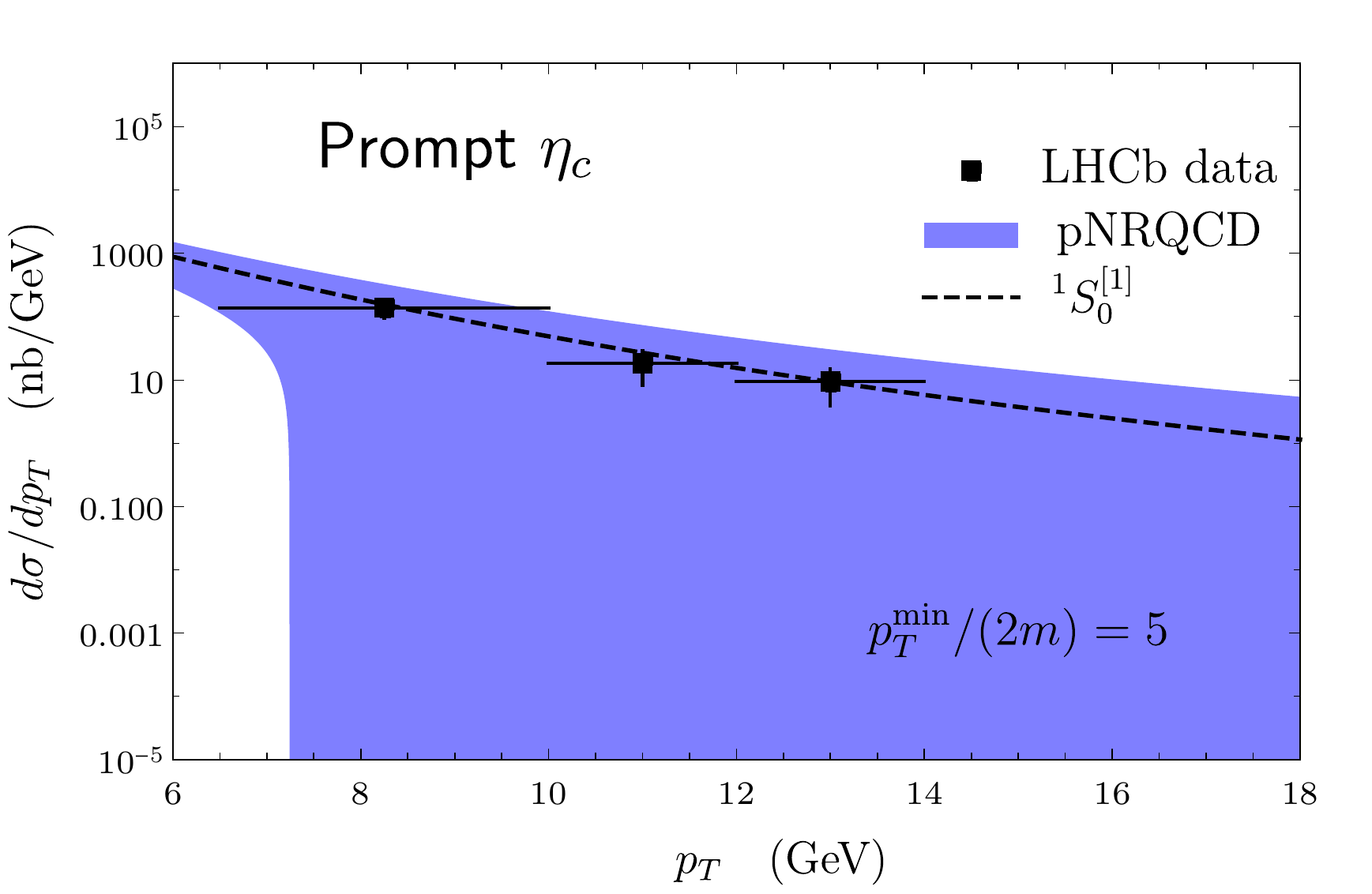}
\caption{\label{fig:etac}
Production rate of $\eta_c$ at the LHC center of mass energy $\sqrt{s}=7$~TeV in the rapidity range 
$2.0 < y < 4.5$ based on the heavy quark spin symmetry using the 
determinations of the $J/\psi$ LDMEs from fits with lower $p_T$ cuts 
$p_T/(2 m) > 3$ (left) and $p_T/(2 m) > 5$ (right), compared with LHCb
data~\cite{LHCb:2014oii}. 
The color-singlet contribution at leading order in $v$ is shown as black dashed lines. 
}
\end{figure}

We show our results for the $\eta_c$ production rate computed from the fits 
with $p_T$ cuts $p_T/(2 m) > 3$ and $p_T/(2 m) > 5$ compared to the LHCb
measurement~\cite{LHCb:2014oii} in figure~\ref{fig:etac}. 
The theoretical uncertainties come from the LDMEs. 
In the $p_T/(2 m) > 5$ case, the pNRQCD result for the $\eta_c$ cross section
is compatible with measurements, although the uncertainty is large due to the
uncertainty in our determination of $c_F^2 {\cal B}_{00}$. 
In the $p_T/(2 m) > 3$ case, the pNRQCD result undershoots the color-singlet 
contribution, and turns negative at large $p_T$. This may indicate that
a too negative value of  $c_F^2 {\cal B}_{00}$ (and $\langle {\cal O}^{J/\psi}(^1S_0^{[8]}) \rangle$),
as we obtain from small $p_T^{\rm min}$, is disfavored by $\eta_c$ data. 

Similarly to our calculations of the cross section ratios $\sigma_{\psi(2S)}/\sigma_{J/\psi}$
and $\sigma_{\Upsilon(3S)}/\sigma_{\Upsilon(2S)}$ in section~\ref{sec:ratios}, 
we also obtain a prediction for the ratio of $\eta_c(1S)$ and $\eta_c(2S)$ cross sections given by 
\begin{equation}
\label{eq:etacratio1}
\frac{\sigma_{\eta_c(2S)}^{\rm direct}}{\sigma^{\rm direct}_{\eta_c(1S)}}
= 
\frac{|R_{\eta_c(2S)}^{(0)} (0)|^2}{|R_{\eta_c(1S)}^{(0)} (0)|^2}
=
\frac{|R_{\psi(2S)}^{(0)} (0)|^2}{|R_{J/\psi}^{(0)} (0)|^2},
\end{equation}
where the last equality follows from the spin symmetry of the quarkonium
wavefunctions, which holds up to corrections of relative order $v^2$. 
For this result to be useful, the cross sections need to be multiplied by the
branching fractions into the $p \bar{p}$ final state that were employed by the
LHCb measurements~\cite{LHCb:2014oii, LHCb:2019zaj}.
While ${\rm B}_{\eta_c (1S) \to p \bar{p}} = (1.44 \pm 0.14) \times 10^{-3}$ is available
in ref.~\cite{ParticleDataGroup:2018ovx}, for the $\eta_c(2S)$, only the product 
${\rm B}_{B^+ \to \eta_c(2S) K^+} \times {\rm B}_{\eta_c(2S) \to p \bar{p}}
= (3.47 \pm 0.76 ) \times 10^{-8}$ has been reported in
ref.~\cite{LHCb:2016zqv}. 
By using the value ${\rm B}_{B^+ \to \eta_c(2S) K^+} = (4.4 \pm 1.0) \times
10^{-4}$ from ref.~\cite{ParticleDataGroup:2018ovx}, 
we obtain ${\rm B}_{\eta_c(2S) \to p \bar{p}}= (7.9^{+2.9}_{-2.3}) \times
10^{-5}$. These values of the branching fractions 
lead to the prediction 
\begin{equation}
\label{eq:etacratio2}
\frac{{\rm B}_{\eta_c(2S) \to p\bar{p}} \times\sigma_{\eta_c(2S)}^{\rm direct}}
{{\rm B}_{\eta_c (1S) \to p \bar{p}} \times \sigma^{\rm direct}_{\eta_c(1S)}}
= (\textrm{2 -- 5}) \times 10^{-2}, 
\end{equation}
which we expect to hold at values of $p_T$ much larger than the $\eta_c$
mass, independently of the rapidity or the center of mass energy.

\subsection[Production of $J/\psi +Z$ and $J/\psi +W$ at the LHC]
{\boldmath Production of $J/\psi +Z$ and $J/\psi +W$ at the LHC}
\label{sec:jpsiZW}

It has been suggested that associated production of a $J/\psi$ plus a gauge
boson would serve as a test of the $J/\psi$ LDMEs~\cite{Li:2014ava,
ATLAS:2014yjd,ATLAS:2014ofp, ATLAS:2019jzd, Butenschoen:2022wld}. 
The SDCs for the inclusive production of $J/\psi + \gamma$ have been computed in ref.~\cite{Li:2014ava},
and the $J/\psi + Z$ and $J/\psi + W$ production cross sections have been 
computed in ref.~\cite{Butenschoen:2022wld}. 
Experimentally, the $J/\psi + Z$ and $J/\psi + W$ production
rates at large $p_T^{J/\psi}$ have been measured by ATLAS~\cite{ATLAS:2014yjd,ATLAS:2014ofp, ATLAS:2019jzd}. 

We compute the $p_T$-differential prompt cross sections for $J/\psi + Z$ and 
$J/\psi + W$ at the LHC center of mass energy $\sqrt{s}=8$~TeV
by using the SDCs reported in ref.~\cite{Butenschoen:2022wld}, which were computed for the rapidity range 
$|y^{J/\psi}| < 2.1$ as used in the ATLAS measurements.  
We include the feeddown contributions from decays of $\psi(2S)$, 
and also the contribution from decays of $\chi_{c1}$ and $\chi_{c2}$,
computed from the pNRQCD determinations of the $\chi_c$ LDMEs in ref.~\cite{Brambilla:2021abf}. 
We consider the theoretical uncertainties coming from the gluonic correlators,
and we also consider uncertainties from uncalculated corrections of relative
order $v^2$, which we estimate to be 30\% of the central values. 
We add the uncertainties in quadrature.
Because the calculation in ref.~\cite{Butenschoen:2022wld} only includes the
contribution from single parton scattering (SPS), while the measurements in 
refs.~\cite{ATLAS:2014ofp, ATLAS:2019jzd} include both 
SPS and double parton scattering (DPS) contributions, 
following the analysis in ref.~\cite{Butenschoen:2022wld}
we subtract the estimated double parton scattering (DPS) contribution
from the measured SPS+DPS cross sections 
available from refs.~\cite{ATLAS:2014ofp, ATLAS:2019jzd} 
assuming the DPS effective area $\sigma_{\rm eff} = 15^{+5.8}_{-4.2}$~mb. 
We note that the estimated DPS contributions are generally smaller than the
uncertainties in the measured cross sections, and become negligible at very
large $p_T^{J/\psi}$, so that at the largest $p_T^{J/\psi}$ bins the estimated 
DPS contributions are only a fraction of a percent of the SPS+DPS cross
section. 
The measurements in refs.~\cite{ATLAS:2014ofp, ATLAS:2019jzd} are normalized to
the total cross sections $\sigma(pp \to Z+X)$ and $\sigma(pp \to W+X)$; 
to convert the data in refs.~\cite{ATLAS:2014ofp, ATLAS:2019jzd} to absolute
cross sections, we use $\sigma(pp \to Z+X) = 33.28\pm 1.19$~nb and 
$\sigma(pp \to W+X) = 112.43\pm 3.80$~nb based on the measurement in
ref.~\cite{CMS:2014pkt} and the analysis in ref.~\cite{Butenschoen:2022wld}. 

\begin{figure}[tbp]
\centering
\includegraphics[width=.48\textwidth]{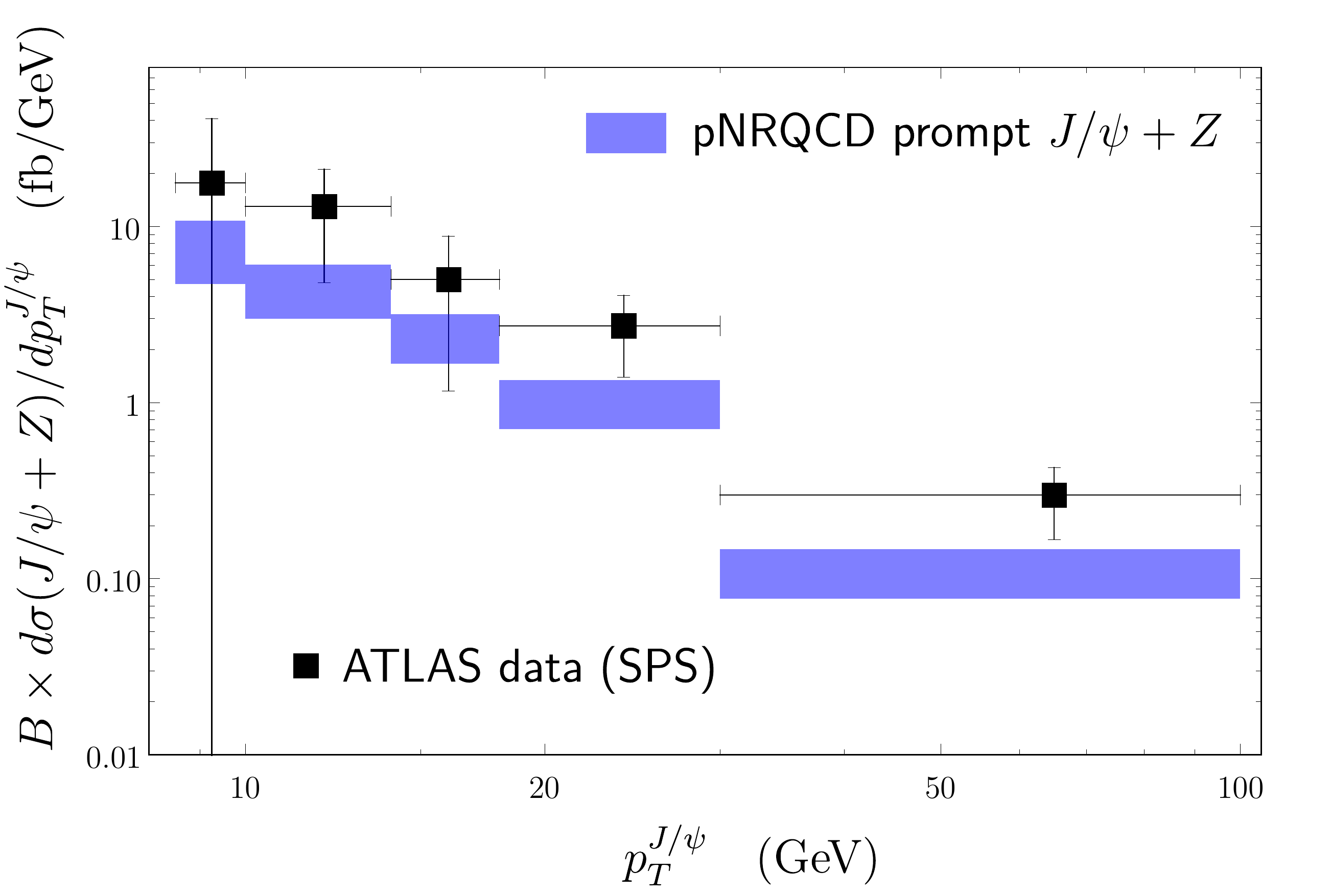}
\includegraphics[width=.48\textwidth]{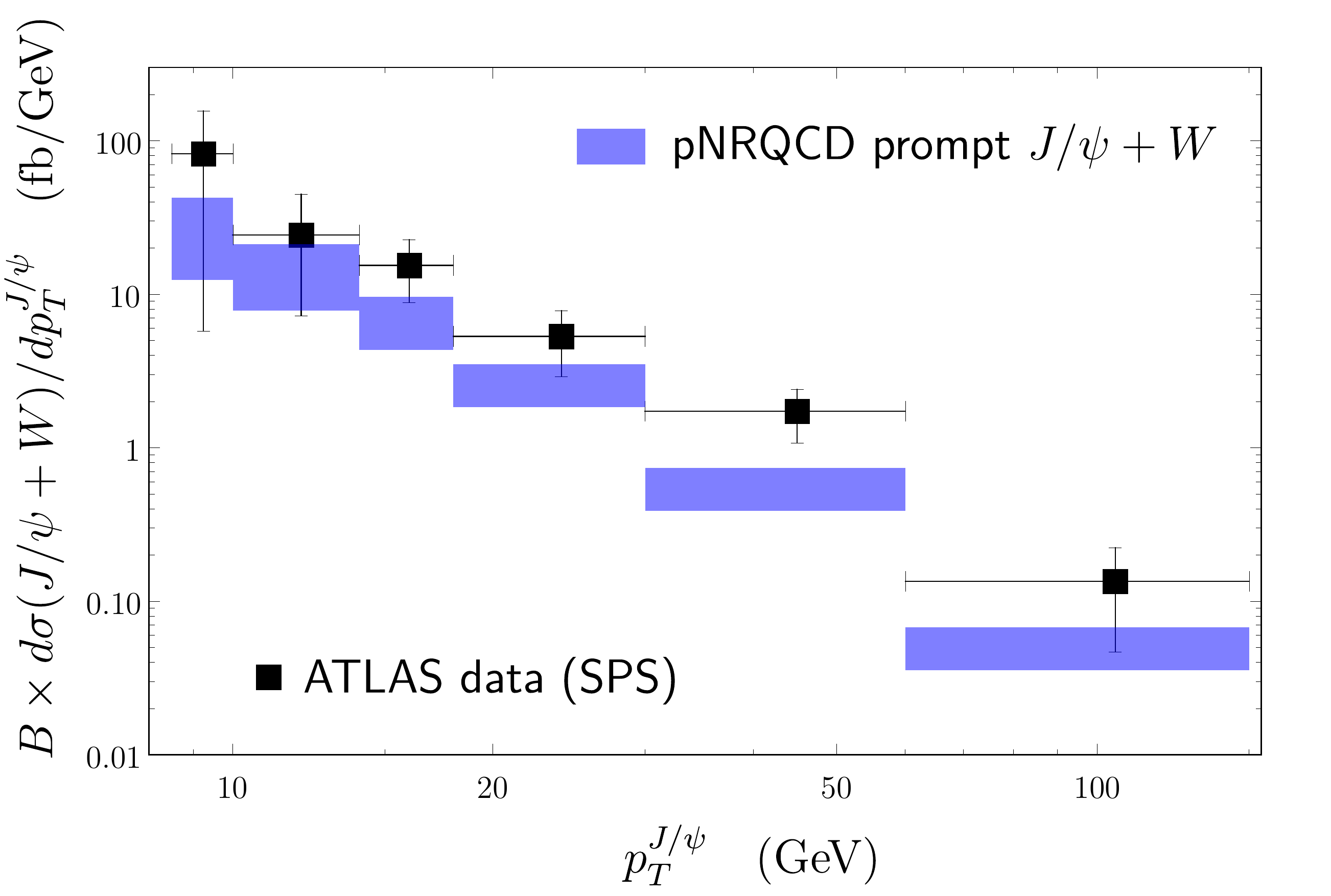}
\caption{\label{fig:jpsiZW}
Production cross sections of 
prompt $J/\psi + Z$ (left) and prompt $J/\psi+W$ (right) at the LHC center of mass energy $\sqrt{s}=8$~TeV
for $|y^{J/\psi}|<2.1$ in pNRQCD compared to ATLAS data~\cite{ATLAS:2014ofp, ATLAS:2019jzd}; 
$B$ is the dimuon branching fraction. 
}
\end{figure}

Our results for the $p_T$-differential $J/\psi + Z$ and $J/\psi + W$ 
cross sections at the LHC center of mass energy $\sqrt{s}=8$~TeV
compared to the ATLAS data in refs.~\cite{ATLAS:2014ofp,ATLAS:2019jzd} are shown in fig.~\ref{fig:jpsiZW}. 
As was reported in ref.~\cite{Butenschoen:2022wld}, the pNRQCD results for the
charmonium LDMEs lead to associated production cross sections that 
agree with measurements within uncertainties for the majority of the $p_T^{J/\psi}$ bins,
although the central values are systematically below the measured cross sections. 
Compared to the results in ref.~\cite{Butenschoen:2022wld} based on the
$J/\psi$ and $\psi(2S)$ LDMEs determined in ref.~\cite{Brambilla:2022rjd},
we have included the feeddown contributions from $P$-wave charmonia, and used
the updated $S$-wave charmonium LDMEs presented in sec.~\ref{sec:LDMEdeterminations}.

\subsection[Production of $J/\psi$ at the Electron-Ion Collider]
{\boldmath Production of $J/\psi$ at the Electron-Ion Collider}
\label{sec:EIC}

\begin{figure}[tbp]
\centering
\includegraphics[width=0.8\textwidth]{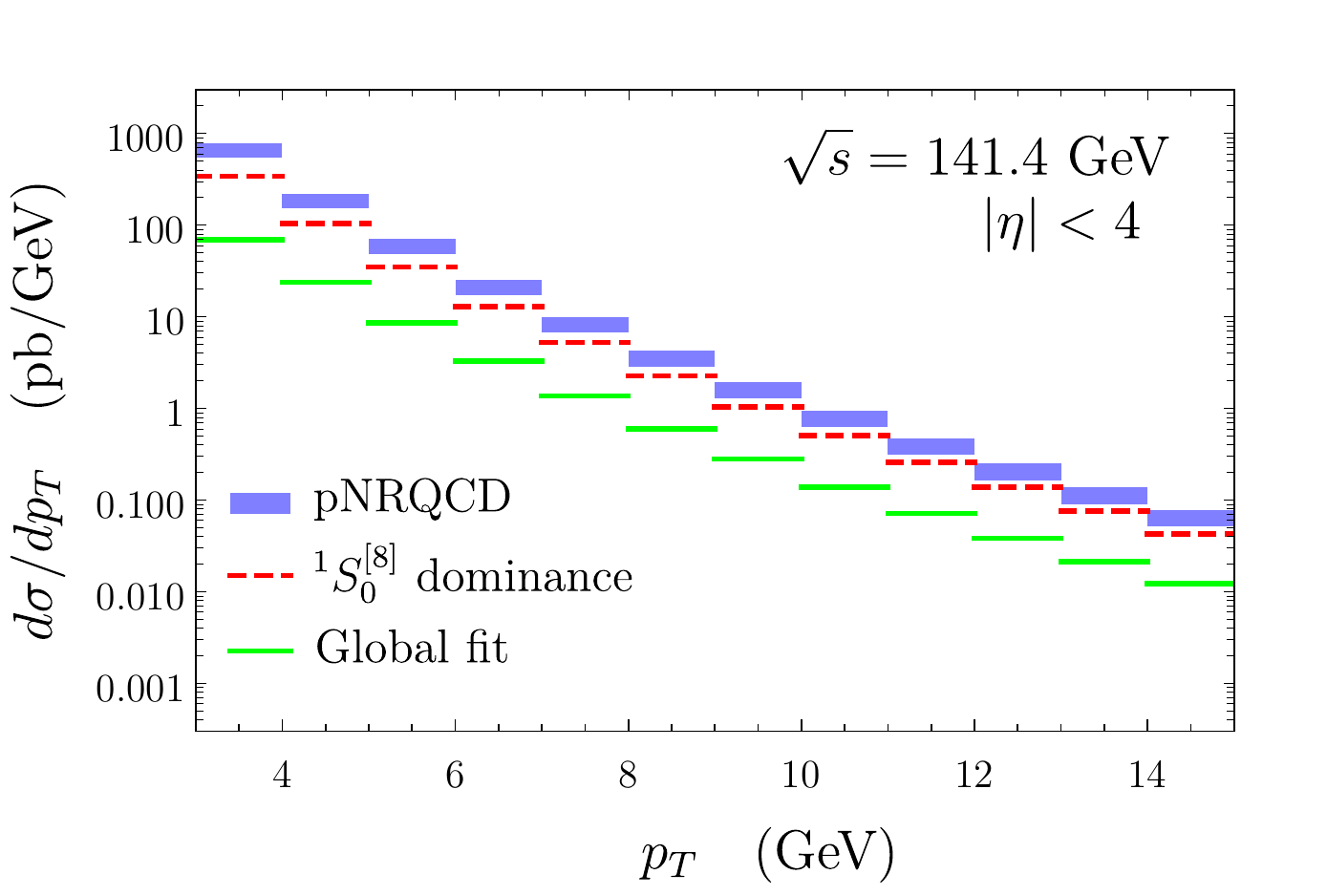}
\caption{\label{fig:EIC}
The pNRQCD prediction for the 
$p_T$-differential cross sections for $J/\psi$ from $ep$ collisions 
at the EIC with center of mass energy $\sqrt{s}=141.4$~GeV and pseudo-rapidity region $|\eta|<4$.
For comparison, predictions based on the $^1S_0^{[8]}$ dominance scenario in
ref.~\cite{Feng:2018ukp} and the global fit in ref.~\cite{Butenschoen:2011yh}
are also shown. 
}
\end{figure}

In ref.~\cite{Qiu:2020xum}, the authors propose to measure the $p_T$
distribution of the single inclusive $J/\psi$ production in the electron-hadron
rest frame at the Electron-Ion Collider (EIC) without tagging the
outgoing electron. As it is also pointed out in ref.~\cite{Liu:2021jfp}, the
inclusiveness of the final state electron helps to eliminate a major
uncertainty due to  QED radiative  corrections in semi-inclusive deep inelastic scattering.
Using collinear factorization for both QCD and QED initial
states and NRQCD factorization for the $J/\psi$ final state, within the
accuracy under our consideration, the inclusive $p_T$ differential cross
section of $J/\psi$ at the EIC is expressed as~\cite{Qiu:2020xum}
\begin{equation}
\label{eq:collfac}
d\sigma_{eh\to J/\psi + X} = \sum_{a,b,n} f_{a/e}(x_a,\mu_f^2) \otimes 
f_{b/h}(x_b,\mu_f^2) 
 \otimes \hat{\sigma}_{ab \to c\bar c[n] + X}(x_a, x_b, p_T, \eta ,m_c, \mu_f^2) \langle \mathcal{O}^{J/\psi}(n) \rangle, 
\end{equation}
where, $\eta$ is the pseudorapidity of $J/\psi$,
$\mu_f$ is the factorization scale,
$a=e, \gamma$ and $b = q, \bar{q}, g$ under our considerations,
$f_{a/e}$ is the collinear distribution of finding an electron and a photon from the colliding electron,
$f_{b/h}$ is the parton distribution function of the colliding hadron $h$,
and $\hat{\sigma}_{ab \to c\bar c[n] +X }$ is the partonic cross section with $n=  {}^1S_{0}^{[8]}, {}^3P_{J}^{[8]}$ at LO in
the strong coupling and $n = {}^3S_{1}^{[1]}, {}^3S_{1}^{[8]}, {}^1S_{0}^{[8]}, {}^3P_{J}^{[8]}$ at NLO in the strong coupling.
Since  at LO in the strong coupling, only the ${}^1S_{0}^{[8]}$ and ${}^3P_{J}^{[8]}$ channels contribute, the
observable $d\sigma_{eh\to J/\psi +X}$ in the electron-hadron rest frame has the advantage to provide better information on
$\langle \mathcal{O}^{J/\psi}({}^1S_{0}^{[8]}) \rangle$ and $\langle \mathcal{O}^{J/\psi}({}^3P_{0}^{[8]}) \rangle$.
Combing the NLO SDCs calculated in ref.~\cite{Qiu:2020xum} with our fitting results of the $J/\psi$ LDMEs, we
plot our prediction for the $p_T$ distribution of the single inclusive $J/\psi$
production in the electron-proton rest frame at the EIC in figure~\ref{fig:EIC}.
The theory uncertainties are determined so that they encompass the uncertainties in the correlators in both $p_T$ regions.
For comparison, we also show in figure~\ref{fig:EIC} the prediction based on the $^1S_0^{[8]}$ dominance scenario
by using the $J/\psi$ LDMEs determined in ref.~\cite{Feng:2018ukp},
and the prediction from the global fit in ref.~\cite{Butenschoen:2011yh}. 

\section{Summary and outlook}
\label{sec:summary}

In this work, we have presented a calculation of NRQCD long-distance matrix elements
that appear in the NRQCD factorization formula for inclusive production of a 
spin-triplet $S$-wave heavy quarkonium, based on the strongly coupled pNRQCD
formalism developed in refs.~\cite{Brambilla:2020ojz, Brambilla:2021abf}. 
In the pNRQCD formalism, the three color-octet long-distance matrix elements that appear
in the factorization formula, corresponding to the $^3S_1^{[8]}$, $^1S_0^{[8]}$,
and $^3P_J^{[8]}$ channel contributions, are given by quarkonium wavefunctions
at the origin and three universal gluonic correlators.
The results of this calculation have been first reported in
ref.~\cite{Brambilla:2022rjd}, and in this paper we show the 
technical details for the derivations in section~\ref{sec:LDMEs}.
The results are displayed in eqs.~(\ref{eq:LDMEs_summary}).
The universality of the gluonic correlators give rise to universal relations
between color-octet long-distance matrix elements for different $S$-wave
quarkonium states shown in eqs.~(\ref{eq:LDMEs_universal_relations}). 
These relations, together with the evolution equations of
the gluonic correlators, see eqs.~(\ref{eq:RGsolution}),
give strong constraints on the long-distance matrix elements in phenomenological analyses. 

We have presented phenomenological results for production of $J/\psi$,
$\psi(2S)$, and $\Upsilon$ states in section~\ref{sec:pheno}. These include
cross section ratios, cross sections and polarizations at the LHC and 
photoproduction cross sections at DESY HERA. 
Furthermore we have presented the hadroproduction rates of $\eta_c$ at the LHC based on the heavy-quark spin symmetry relations,
and predictions for the associated production of $J/\psi + W$ and $J/\psi + Z$, as well as the 
production rate of $J/\psi$ at the Electron-Ion Collider. 
In particular, the direct cross section ratio of $J/\psi$ and $\psi(2S)$, and
the ratio of $\Upsilon(3S)$ and $\Upsilon(2S)$ do not depend on the specific
values of the color-octet long-distance matrix elements, thanks to the
universal relations in eqs.~(\ref{eq:LDMEs_universal_relations}). 
By using only the quarkonium wavefunctions at the origin, and the measured
values of feeddown and branching fractions, we computed the cross section
ratio of prompt $J/\psi$ and $\psi(2S)$ production, and the ratio of inclusive
$\Upsilon(2S)$ and $\Upsilon(3S)$ production in section~\ref{sec:ratios}. 
These results agree with experiments within
uncertainties at large $p_T$, which supports the validity of the pNRQCD
approach. In order to compute absolute cross sections and polarizations, we
determined the color-octet long-distance matrix elements in
section~\ref{sec:LDMEdeterminations}, by using large-$p_T$ cross sections
measured at the LHC. Because in pNRQCD the color-octet long-distance matrix
elements are given by wavefunctions at the origin times universal gluonic
correlators, the number of nonperturbative unknowns are greatly reduced, 
which leads to stronger constraints on the phenomenological determinations of
the long-distance matrix elements compared to alternative approaches. 
Based on the long-distance matrix elements determined in
section~\ref{sec:LDMEdeterminations}, we computed $p_T$-differential 
cross sections of $J/\psi$, $\psi(2S)$, and $\Upsilon$ states at the
LHC center of mass energy $\sqrt{s}=7$~TeV in section~\ref{sec:crosssection}. 
In section~\ref{sec:crosssection} we also show results for the cross sections
of $^3S_1$ heavy quarkonia computed from our predictions of the long-distance matrix elements obtained
without using cross sections measurements for that specific quarkonium state,
which have never been possible without the pNRQCD formalism. 
The results agree well with data at large $p_T$. 
We also computed the polarizations of $J/\psi$, $\psi(2S)$, and $\Upsilon$ 
at the LHC in section~\ref{sec:polarization}, which agree with measurements. 
The results for absolute cross sections and polarizations at the LHC shown in
this paper update and supersede the previous analysis in
ref.~\cite{Brambilla:2022rjd}. 
On the other hand, our determinations of long-distance matrix elements lead to
an overestimation of photoproduction cross section of $J/\psi$ at DESY HERA; 
we note that the kinematical cuts employed in the photoproduction cross section 
measurements can make it difficult for NRQCD to give a satisfactory description of the production rate. 
By using the heavy-quark spin symmetry relations, which we reproduce explicitly
by using the pNRQCD calculations of the long-distance matrix elements in
section~\ref{sec:HQSS}, we have also computed the hadroproduction rate of
$\eta_c$ at the LHC in section~\ref{sec:etac}. 
Although the uncertainties in the cross sections that we
obtain are much larger compared to the results from previous works that used 
the measured $\eta_c$ production rates as inputs, we found that our
determination of the long-distance matrix elements with a large $p_T$ cut is
compatible with the measured $\eta_c$ cross section. 
In section~\ref{sec:jpsiZW}, we computed the associated production cross
sections of $J/\psi + Z$ and $J/\psi + W$ at the LHC using the recent results
for the short-distance coefficients in ref.~\cite{Butenschoen:2022wld}, 
and found fair agreements with ATLAS measurements~\cite{ATLAS:2014ofp,
ATLAS:2019jzd}. Finally, we made
predictions for $J/\psi$ production rate at the Electron-Ion Collider in
section~\ref{sec:EIC}. 

As we have mentioned in section~\ref{sec:NRQCDfac}, arguments for the validity
of the NRQCD factorization have been made in the expansion in powers of $m/p_T$, up
to next-to-leading power (relative order $m^2/p_T^2$). 
Hence, we expect NRQCD factorization to hold for values of $p_T$ much larger than the quarkonium mass. 
For values of $p_T$ similar to or smaller than the
quarkonium mass, the production rates can be strongly affected by unsuppressed
nonperturbative effects. For example, soft gluons emitted in the evolution of a
color-octet $Q \bar Q$ into a quarkonium can interact nonperturbatively with
initial and final states; at values of $p_T$ much larger than the quarkonium
mass, such contributions are expected to cancel, or to be absorbed into
nonperturbative matrix elements, based on general arguments in collinear
factorization~\cite{Nayak:2005rt, Nayak:2005rw, Nayak:2006fm, Kang:2014tta}. 
The same arguments cannot be made if $p_T$ is of the order of the quarkonium mass or smaller.  

Our phenomenological results also seem to support the
validity of NRQCD factorization at large $p_T$. The quality of the fits (section~\ref{sec:LDMEdeterminations}), 
as well as the theoretical descriptions of 
the cross section ratios (section~\ref{sec:ratios}), 
absolute cross sections (section~\ref{sec:crosssection}), 
polarizations (section~\ref{sec:polarization}), 
and $\eta_c$ hadroproduction (section~\ref{sec:etac}) 
all improve with increasing $p_T$.
Concerning the lower $p_T$ cut $p_T^{\rm min}$, a small $p_T^{\rm min}$ improves the NRQCD description of the
total inclusive cross sections, which are dominated by contributions from $p_T$
of the order of the quarkonium mass or smaller.
This is also the case for $\sigma(e^+ e^- \to J/\psi +X)$ at the $B$ factories,
whose prediction with our long-distance matrix elements or 
in other large-$p_T$ hadroproduction-based approaches, when using the 
short-distance coefficients computed to next-to-leading order accuracy in ref.~\cite{Zhang:2009ym}, 
far exceeds the Belle measurement~\cite{Belle:2009bxr}.
Since the dominant color-octet contribution to 
$\sigma(e^+ e^- \to J/\psi +X)$ is given by a linear combination of 
the $^1S_0^{[8]}$ and $^3P_0^{[8]}$ long-distance matrix elements with positive
coefficients, the discrepancy diminishes, however, if we decrease $p_T^{\rm min}$ so that
the $^1S_0^{[8]}$ contribution becomes more negative\footnote{It has however 
been argued in ref.~\cite{Butenschoen:2012qr} that the Belle measurement in 
ref.~\cite{Belle:2009bxr} should be interpreted as a lower bound, because it
was obtained from a data sample with the multiplicity of charged tracks larger than four,
and no corrections for this limitation were included.}.  
Nevertheless it should be recalled that reducing the lower $p_T$ cut $p_T^{\rm min}$
and making the $^1S_0^{[8]}$ long-distance matrix element even more negative
makes the $\eta_c$ hadroproduction cross section turn negative at even smaller $p_T$.
The apparent disparity between low and high-$p_T$ behaviors suggests that one
needs to be cautious when applying results of large-$p_T$ analysis to small-$p_T$
or $p_T$-integrated observables. 

The pNRQCD analysis presented in this paper suggests a noticeable pattern in
the production mechanism of spin-triplet $S$-wave heavy quarkonia at very large
$p_T$: large cancellations occur in the sum of the $^3S_1^{[8]}$ and 
$^3P_J^{[8]}$ channel contributions, which mix due to renormalization of the
long-distance matrix elements, and the remnant of this cancellation makes up
for the bulk of cross section. This is similar to the case of  $P$-wave 
production, where the cross section at leading order in $v$ is given by the sum
of color-singlet $P$-wave and color-octet $S$-wave contributions, which also
mix due to renormalization of the long-distance matrix elements, and large
cancellations occur in the sum at large $p_T$. 
This pattern emerges because we obtain positive values for both $^3S_1^{[8]}$ 
and $^3P_0^{[8]}$ long-distance matrix elements. 
We note that similar scenarios for $J/\psi$ production have been suggested in
phenomenological analyses based on $J/\psi$ and $\eta_c$ hadroproduction data
in refs.~\cite{Zhang:2014ybe, Han:2014jya}. Interestingly, a similar
configuration of long-distance matrix elements have been obtained for the
$\psi(2S)$ state in the global fit analysis in ref.~\cite{Butenschoen:2022orc}
when the $p_T$ cut $p_T > 7$~GeV was used. 
A caveat of this scenario is that large cancellations can be affected by
radiative corrections, so that inclusion of corrections of higher orders in
$\alpha_s$ may bring sizable changes in the phenomenologically obtained values 
of the long-distance matrix elements. However, in the pNRQCD analysis, we expect
the $^3P_0^{[8]}$ long-distance matrix elements to be less
susceptible to radiative corrections, because their values are also 
constrained by the evolution equation~(\ref{eq:RG_E1010})
and the universality of the gluonic correlators. 
As we have shown in section~\ref{sec:pheno}, the production mechanism for
spin-triplet $S$-wave heavy quarkonia suggested by the pNRQCD analysis 
leads to large-$p_T$ production rates that agree well with measurements at
the LHC.
It would be interesting to see how the heavy quarkonium production 
mechanism 
presented in this work
will test against upcoming measurements and future experiments 
such as those planned at the Electron-Ion Collider.

\acknowledgments 
We thank Jian-Xiong Wang and Yu Feng for their support while using the FDCHQHP package.  We thank Julian Mayer-Steudte and Viljami Leino for their instructions on using C2PAP.
The work of N. B. and  X.-P. W. is supported by the DFG (Deutsche
Forschungsgemeinschaft, 
German Research Foundation) Grant No. BR 4058/2-2. N.~B., H.~S.~C., A.~V.
and X.-P.W. 
acknowledge support from the DFG cluster of excellence ``ORIGINS'' under
Germany's Excellence Strategy - EXC-2094 - 390783311. 
The simulations have been carried out on the computing facilities of the
Computational Center for Particle and Astrophysics (C2PAP). 
The work of H.~S.~C is supported by the National Research Foundation of Korea
(NRF) Grant funded by the Korea government (MSIT) under Contract No.
NRF-2020R1A2C3009918 and by a Korea University grant.
The work of A.~V. is funded by the DFG Project-ID 196253076 - TRR 110.

\bibliography{hadropro-Swave_long.bib}
\bibliographystyle{JHEP}

\end{document}